\documentclass[%
 reprint,
nofootinbib,
 amsmath,amssymb,
 aps,
]{revtex4-2}
\usepackage[normalem]{ulem}
\usepackage[nolist]{acronym}
\usepackage{xcolor}
\usepackage{graphicx}
\usepackage{dcolumn}
\usepackage{bm}
\usepackage[breaklinks, plainpages=false, colorlinks=true, anchorcolor=cyan, linkcolor=magenta, citecolor=cyan, urlcolor=purple, bookmarks=false]{hyperref}
\usepackage{multirow}
\usepackage[caption=false]{subfig}

\newcommand{\bilby}{\texttt{Bilby}}
\newcommand{\parallelbilby}{\texttt{Parallel Bilby}}
\newcommand{\dynesty}{\texttt{dynesty}}

\begin{document}

\preprint{APS/123-QED}

\title[Impact of Prior Choices]{Impact of Bayesian Priors on the Inferred Masses of Quasi-Circular Intermediate-Mass Black Hole Binaries}

\author{Koustav Chandra}
    \email{koustavchandra@gmail.com}
    \affiliation{Department of Physics, Indian Institute of Technology Bombay, Powai, Mumbai 400 076, India}
    \affiliation{Institute for Gravitation and the Cosmos, Department of Physics, Pennsylvania State University, University Park, PA 16802, USA}
\author{Archana~Pai}
    \email{archanap@iitb.ac.in}
    \affiliation{Department of Physics, Indian Institute of Technology Bombay, Powai, Mumbai 400 076, India}
\author{Samson H. W. Leong}
    \email{samson.leong@link.cuhk.edu.hk}
    \affiliation{Department of Physics, The Chinese University of Hong Kong, Shatin, N.T., Hong Kong}
\author{Juan Calder\'on~Bustillo}
 	\email{juan.calderon.bustillo@gmail.com}
	\affiliation{Instituto Galego de F\'{i}sica de Altas Enerx\'{i}as, Universidade de
Santiago de Compostela, 15782 Santiago de Compostela, Galicia, Spain}
\affiliation{Department of Physics, The Chinese University of Hong Kong, Shatin, N.T., Hong Kong}

\date{\today}

\begin{abstract}
   Observation of gravitational waves from inspiralling binary black holes has offered a unique opportunity to study the physical parameters of the component black holes. To infer these parameters, Bayesian methods are employed in conjunction with general relativistic waveform models that describe the source's inspiral, merger, and ringdown. The results depend not only on the accuracy of the waveform models but also on the underlying fiducial prior distribution used for the analysis. In particular, when the pre-merger phase of the signal is barely observable within the detectors' bandwidth, as is currently the case with intermediate-mass black hole binary signals in ground-based gravitational wave detectors, different prior assumptions can lead to different interpretations. In this study, we utilise the gravitational-wave inference library, \parallelbilby, to evaluate the impact of mass prior choices on the parameter estimation of intermediate-mass black hole binary signals. While previous studies focused primarily on analysing event data, we offer a broader, more controlled study by using simulations. Our findings suggest that the posteriors in total mass, mass ratio and luminosity distance are contingent on the assumed mass prior distribution used during the inference process. This is especially true when the signal lacks sufficient pre-merger information and/or has inadequate power in the higher-order radiation multipoles. In conclusion, our study underscores the importance of thoroughly investigating similarly heavy events in current detector sensitivity using a diverse choice of priors. Absent such an approach, adopting a flat prior on the binary's redshifted total mass and mass ratio emerges as a reasonable choice, preventing biases in the detector-frame mass posteriors.
\end{abstract}
\maketitle

\acrodef{IGWN}[IGWN]{International Gravitational-Wave Observatory Network}
\acrodef{LVK}[LVK]{LIGO Scientific, Virgo and KAGRA}
\acrodef{CI}[CI]{credible intervals}
\acrodef{IMBH}[IMBH]{intermediate-mass black hole}

\section{Introduction} 
\label{sec:intro}

The field of gravitational-wave astronomy is experiencing significant advancements owing to the frequent upgrades made to the \acf{IGWN}~\citep{LIGOScientific:2014pky, VIRGO:2014yos}. Since their inception, the Advanced LIGO and the Advanced Virgo detectors have observed $\gtrsim 100$ signals, most of which are attributed to binary black hole mergers~\citep{LIGOScientific:2018mvr, Nitz:2018imz, Venumadhav:2019lyq, Nitz:2020oeq, LIGOScientific:2020ibl, Nitz:2021uxj, LIGOScientific:2021djp, Nitz:2021zwj, Olsen:2022pin}. This growing catalogue has enabled precision strong-field tests of general relativity and has shed light on the astrophysical origin and the nature of the black hole population in the local Universe~\citep{LIGOScientific:2016lio, LIGOScientific:2018jsj, LIGOScientific:2019fpa, Roulet:2020wyq, KAGRA:2021duu, LIGOScientific:2020tif, LIGOScientific:2021sio}.

Quite a few of these results rely heavily on Bayesian parameter inference. For instance, the inspiral-merger-ringdown consistency test examines the consistency of the signal's low and high-frequency components by independently inferring the mass and spin of the remnant object from each portion~\citep{Ghosh:2016qgn, Ghosh:2017gfp}. Similarly, the residual test measures the coherent residual signal-to-noise ratio in the data after subtracting the best-fit waveform to ensure overall signal consistency~\citep{LIGOScientific:2016aoc}. Likewise, studying secondary parameters, such as the remnant object's recoil velocity, requires accurate inference of the binary's primary parameters, such as their masses and spins~\citep{Bekenstein:1973zz, Gonzalez:2006md, Lousto:2007db, Lousto:2012su, Lousto:2012gt, CalderonBustillo:2018zuq, Varma:2020nbm, Varma:2022pld, CalderonBustillo:2022ldv}. Also, to determine the shape of the black hole mass or spin spectrum, we must first accurately obtain the astrophysical properties of the observed binary black holes. Therefore, addressing the potential issues in Bayesian parameter inference is essential. 

One such issue is the impact of prior choices in signal parameter estimation. In gravitational-wave astronomy, Bayesian parameter inference involves analysing signal-containing data segments from a detector network with a prior distribution to obtain posterior samples of binary parameters. If the analysed data lacks sufficient information, prior assumptions will significantly influence the resulting posteriors. For example, GW190521 has had multiple interpretations due to its barely observable pre-merger phase~\citep{LIGOScientific:2020iuh, Fishbach:2020qag, Nitz:2020mga, Romero-Shaw:2020thy, Gayathri:2020coq, CalderonBustillo:2020fyi, CalderonBustillo:2020xms, Gamba:2021gap, Estelles:2021jnz, Olsen:2021qin}. Even under the quasi-spherical signal hypothesis, the binary constituents have a non-negligible probability of occupying the pair-instability gap or straddling it, depending on the prior choice. 

This paper studies the effect of competing mass prior choices on synthetic intermediate-mass black hole binary signals. This work is motivated by similar previous studies~\citep{Estelles:2021jnz, Olsen:2021qin, bustillo2022searching} which highlighted the potential limitations of certain prior choices in exploring the high-likelihood regions of parameter space. However, these previous studies primarily focused on analysing data segments containing the GW190521 event (or other comparatively low-significance events). In contrast, our study offers a broader and more controlled perspective as we use simulations.

Our study shows that the posteriors in binary's redshifted total mass, mass ratio, and luminosity distance critically depend on the mass prior choice. This effect is particularly accentuated when the signal has a barely observable inspiral phase and/or has insufficient power in the higher-order radiation multipoles. The amplitudes and phases of these higher-order radiation multipoles relative to the dominant quadrupole mode are determined by the binary properties, such as mass ratio and inclination and therefore encode them. We, therefore, advocate mapping the likelihood of similarly high-mass events with diverse mass prior choices whenever possible, although reweighting may be enough~(See Appendix~\ref{appx:reweigh} for details).

The remainder of this paper is structured as follows. Section~\ref{sec:basics} briefly describes Bayesian inference and summarises the notations and conventions of the binary parameters used in this article. The following section, Section~\ref{sec:methods}, summarises our simulation setup and our prior choices. The results of our analysis are in Section~\ref{sec:results}, and a summary and conclusion of our study are in Section~\ref{sec:summary}. 

\section{Preliminaries}
    \label{sec:basics}

\subsection{Bayesian inference in a nutshell}

Bayesian parameter inference libraries, like \bilby, model the calibrated detector output $d(t)$ as 
\begin{equation}
    d(t) = n(t) + s(t\mid \boldsymbol{\theta})
\end{equation}
where $n(t)$ is the noise, and $s(t | \boldsymbol{\theta})$ is the signal strain~\citep{Ashton:2018jfp, Smith:2019ucc, thrane2019introduction, Romero-Shaw:2020owr, Christensen:2022bxb}. These libraries compute the posterior probability distribution of the signal parameters $\boldsymbol{\theta}$ by using the Bayes theorem:
\begin{equation}\label{eq:PE-1}
    p(\boldsymbol{\theta} \mid d, \mathcal{M}) = \frac{\mathcal{L}(d \mid \boldsymbol{\theta}, \mathcal{M}) \pi(\boldsymbol{\theta} \mid \mathcal{M})}{\mathcal{Z}_\mathcal{M}}~.
\end{equation}
Here, $\mathcal{L}(d | \boldsymbol{\theta}, \mathcal{M})$ is the likelihood function, and $\pi(\boldsymbol{\theta} | \mathcal{M} )$ is the prior probability distribution which is typically chosen to avoid imprinting astrophysical assumptions on the results. $\mathcal{Z}_\mathcal{M} =\int d\boldsymbol{\theta}\mathcal{L}(d | \boldsymbol{\theta}, \mathcal{M}) \pi(\boldsymbol{\theta}| \mathcal{M})$ is the probability of observing the data given the model $\mathcal{M}$, otherwise known as the model evidence. The likelihood function is defined as follows:
\begin{equation}\label{eq:signal-likelihood}
    \mathcal{L}\left(d \mid \boldsymbol{\theta}, S_n \right) \propto \exp \left[-\sum_i \frac{2\left|\tilde{d}\left(f_i\right)-\tilde{h}\left(f_i \mid \boldsymbol{\theta}\right)\right|^2}{T S_n\left(f_i\right)}\right]~.
\end{equation}
assuming wide-sense stationary Gaussian noise and involves the data segment duration $T$, the noise power spectrum $S_n(f_i)$ at each frequency $f_i$, the Fourier representations of the data $\tilde{d}(f_i)$ and the template $\tilde{h}(f_i \mid \boldsymbol{\theta})$. When using data from multiple detectors, \bilby~ obtains the joint likelihood by multiplying individual likelihoods for each detector $j$ while requiring the template to be coherent across the $N_\mathrm{ifo}$ detector network:
\begin{equation}
        \mathcal{L} (\{d\} \mid \boldsymbol{\theta}, \{S_n \}) = \prod_j^{N_\mathrm{ifo}}
\mathcal{L}\left(d_j \mid \boldsymbol{\theta}, S_n \right)~.
\end{equation}

To compute this likelihood function and estimate the posterior probability distribution described in Eq.~\eqref{eq:PE-1}, Bayesian inference algorithms rely on off-the-shelf stochastic samplers. Given the dimensionality involved, this can be computationally expensive for some waveform models and prior choices. Therefore, it is preferable to use marginalised likelihood over specific parameters whenever possible. The marginalised parameters can be retrieved later using post-processing techniques.

Posterior sampling generates a multi-dimensional probability density function that describes our joint inference about the signal parameters. Understanding and/or visualising the full posterior probability distribution is tricky. So, we often focus on just one or two parameters and marginalise the others. For example, if we are interested in the joint posterior for component masses $m_1,~m_2$ of an event, we integrate over the other parameters as follows:
\begin{equation}
    p\left(m_1, m_2 \mid \{d\} \right)=\sum_{\boldsymbol{\theta}_{\left(-m_1,-m_2\right)}} p\left(\boldsymbol{\theta} \mid \{d\} \right)
\end{equation}
This marginalised posterior indicates what we can infer about the component masses when the uncertainties in the omitted parameters are considered. Once obtained, the marginal posterior samples can be visualised using histograms or kernel density plots. We can also calculate its summary statistics, such as the median and estimate the uncertainty in the parameter(s) of interest by constructing symmetric \acf{CI} defined as follows:
\begin{equation}
\sum_{\theta_k = -\infty}^{a} p\left(\theta_k  \mid \{d\} \right) = \frac{1 - X}{2} = \sum_{\theta_k = b}^{\infty} p\left(\theta_k  \mid \{d\} \right)~.
\end{equation}
Here, the interval $(a,b)$ is a symmetric 100X\% \ac{CI} for parameter $\theta_k$. In this article, we choose $X=0.68, 0.9$  and construct two-dimensional marginal posteriors to consider the potential influence of one parameter on the other. The 68\% \ac{CI} corresponds to approximately one standard deviation from the mean of a normal distribution. Conversely, the 90\% \ac{CI} provides a wider interval, encompassing a higher degree of uncertainty and has traditionally been employed in gravitational-wave data analysis.

Finally, when multiple models are available, it is natural to quantify how much one model, say $A$, is preferred to another, say $B$. Within the Bayesian framework, the Bayes factor is the primary tool for comparing the preferences of two competing models. The log Bayes’ factor for $A$ over model $B$ is calculated by subtracting the log model evidence: $\ln \mathrm{BF}^{A}_{B}=\ln \mathcal{Z}_A - \ln \mathcal{Z}_B$. A $\ln \mathrm{BF}^{A}_{B} > 1$ indicates positive support for model $A$ over model $B$~\citep{bayesfactor}.

    \subsection{Quasi-circular black hole binary parameters: Notations and Conventions}
For quasi-circular or non-precessing black hole binaries~\footnote{Precessing binaries are quasi-spherical as the orbital orientation evolves with time.}, $\boldsymbol{\theta}$ represents four intrinsic and seven extrinsic parameters. The intrinsic parameters include the component's masses $m_i$ and the dimensionless spins $\chi_i$ aligned with each other and the orbital angular momentum. The extrinsic parameters comprise the source's two-dimensional sky location $(\alpha, \delta)$, the luminosity distance $D_L$ (or equivalently the redshift $z$ to the source), the source's inclination $\iota$, the polarisation angle $\psi$, the azimuth $\varphi$ and the merger time $t_c$. 

It is possible to capture the dominant effect of spin on the inspiral rate using the mass-weighted effective aligned spin parameter $\chi_\mathrm{eff}$. At a given reference frequency, say $f_\mathrm{ref}$, during the binary's inspiral, it is defined as follows~\citep{Racine:2008qv, Ajith:2009bn, Santamaria:2010yb}:
\begin{equation}
    \chi_\mathrm{eff} = \frac{\chi_1 + Q \chi_2}{1+Q}~,
\end{equation}
using the binary's spins and the mass ratio $Q = m_2/m_1 \leq 1$. Aligned spin binaries are characterised by positive values of $\chi_\mathrm{eff}$ whereas anti-aligned spin binaries have negative $\chi_\mathrm{eff}$. Low-spin binaries have $\chi_\mathrm{eff} \rightarrow 0$.

The redshifted (detector-frame) total mass $M_T(1+z)$ is crucial for detection, as it determines the signal's overall strength and frequency content. In agreement with other works, we express it and the individual component masses in solar mass units.

    \section{Methodology}
    \label{sec:methods}

    \subsection{Data}
    \label{sec:data}

\begin{table}[t]
    \centering
    \begin{tabular}{lll}
        \hline \hline Parameter & Symbol & Value \\
        \hline Detector-frame total mass & $M_T(1+z)$ & $268.83 M_{\odot}$ \\
        Azimuth & $\varphi$ & 0.004 \\
        Polarization angle & $\psi$ & 2.38 \\
        Coalescence GPS time & $t_c$ & $1242442967.41 \mathrm{~s}$ \\
        Right ascension & $\alpha$ & 0.16 \\
        Declination & $\delta$ & -1.14 \\
        \hline \hline
    \end{tabular}
    \caption{True parameter values that are shared by both the simulation sets. They represent the maximum likelihood point of the GW190521 posterior samples that LIGO-Virgo has publicly released~\citep{gw190521-public}. All angles are in radians.}
    \label{table:common}
\end{table}

We aim to explore the differences between posteriors sampled under different mass priors for systems with detector frame masses similar to GW190521. We, therefore, create two sets of simulated quasi-circular intermediate-mass black hole binary signals for our analysis. The first set uses the phenomenological waveform model IMRPhenomXAS, which models just the dominant waveform harmonic, $(2,2)$ of a quasi-circular black hole binary~\citep{Pratten:2020fqn}. The second set of signals uses the phenomenological waveform model IMRPhenomXHM, which includes the modes $(2,1),(3,3),(3,2),(4,4)$ in addition to the $(2,2)$ mode~\citep{Garcia-Quiros:2020qpx}. 

\begin{table}[h]
    \centering
    \begin{tabular}{lll}
        \hline \hline Parameter & Symbol & Value \\
        \hline Mass ratio & $Q$ & 0.82, 0.25 \\
        Inclination & $\iota$ &  $28.6^\circ$, $54.4^\circ$ \\
        Primary Spin & $\chi_1$ & 0.054, 0.7, -0.7 \\
        Secondary Spin & $\chi_2$ & 0.144, 0.4 \\
        \hline \hline
    \end{tabular}
    \caption{True parameter values for the twelve simulated binaries from the first simulation set.}
    \label{table:set-1}
\end{table}

These simulated signals share a set of common parameter values, as shown in Table~\ref{table:common}. However, they differ in their $Q$, $\iota$, and $\chi_\mathrm{eff}$ values. Even though we perform our study with several combinations of these varying parameters, we only report the results for simulations with specific combinations to summarise the features. These combinations are listed in Table~\ref{table:set-1} and Table~\ref{table:set-2} for simulation set-1 (quadrupole-only signal) and simulation set-2 (multipole signal), respectively. 

\begin{table}[htb]
    \centering
    \begin{tabular}{lll}
        \hline \hline Parameter & Symbol & Value \\
        \hline Mass ratio & $Q$ & 0.82, 0.25, 0.11 \\
        Inclination & $\iota$ & $28.6^\circ$, $54.4^\circ$, $74.5^\circ$ \\
        Primary Spin & $\chi_1$ & 0.054 \\
        Secondary Spin & $\chi_2$ & 0.144 \\
        \hline \hline
    \end{tabular}
    \caption{True parameter values for the nine simulated binaries from the second simulation set.}
    \label{table:set-2}
\end{table}

We use a three-detector network comprising advanced LIGO and advanced Virgo interferometers, operating at the anticipated fourth observing run sensitivity~\citep{noise}. We do not add Gaussian noise to the simulated data in each detector for simulation set-2, thus enabling their analysis in zero-noise~\citep{Nissanke:2009kt}. Therefore, statistically, these results are equivalent to the average results obtained from a large sample of runs on Gaussian noise~\citep{noise}. The distance to each system is adjusted to fix the network optimal SNR at 20. Though we show results for a network SNR of 20, our analysis has produced consistent results for optimal network SNR values of 15 and 30.

Increasing the binary’s mass asymmetry and inclination decreases the system’s intrinsic luminosity, thus placing it at a relatively closer distance than mass-symmetric systems with a face-on orientation. However, the former systems will have significantly larger contributions from the non-quadrupole modes, particularly for simulation set 2. This can help reduce the correlation between certain signal parameters, such as source inclination and distance~\citep{Graff:2015bba, Usman:2018imj}.

Given that the primary objective of our work is to provide a preliminary understanding of the measurability of mass parameters based on prior choices, we have restricted our discussion primarily to quasi-circular black hole binaries in the main text. Relativistic precession introduces amplitude and phase modulations in the signal inspiral, which is limited for the heavy systems considered in this study. Consequently, we anticipate that the fundamental qualitative attributes of our findings would remain unaltered even if we use quasi-spherical signals. A demonstration of this assertion is provided in Appendix~\ref{appx:quasi-spherical} where we use a quasi-spherical binary at two different inclinations, finding that it doesn't qualitatively change our results. Nonetheless, it is essential to underscore that relativistic precession, particularly for certain specific spin configurations, can significantly suppress the inspiral phase before the merger, affecting measurements~\citep{Boyle:2019kee, Healy:2019jyf}.

    \subsection{Inference Settings}

To obtain the posterior samples, we use the parallelised version of the parameter estimation package \bilby~\citep{Ashton:2018jfp,Smith:2019ucc}~ with the nested sampler \dynesty~\citep{Speagle:2019ivv}. In our analysis, we fix the number of $\texttt{walks}$ and $\texttt{maxmcmc}$ at 200 and 15000, respectively. Additionally, we set the number of live points, denoted as $\texttt{nlive}$, to 1024, and the number of autocorrelation times, denoted as $\texttt{nact}$, to 30. Other \dynesty~attributes, such as $\texttt{naccept}$, $\texttt{proposals}$, etc., are set to their default values. 

To expedite convergence, we downsample our simulated data to 1024 Hz and utilise marginalised likelihood with marginalisation over distance and time, which is performed numerically following the approach of \citet{willfarr2014} and \citet{Singer:2015ema}. Further, while using the IMRPhenomXAS waveform model, we use the likelihood function that is analytically marginalised over the signal phase~\citep{marginalise2013}. We compute all spin posteriors at $f_\mathrm{ref}=11$Hz and use an identical lower-frequency cutoff of 11Hz for likelihood integration. The high-frequency cutoff is set to the Nyquist frequency, 512Hz, corresponding to the data's sampling rate of 1024Hz.

    \subsection{Priors}

\begin{figure}
    \begin{center}
    \includegraphics[width=\columnwidth]{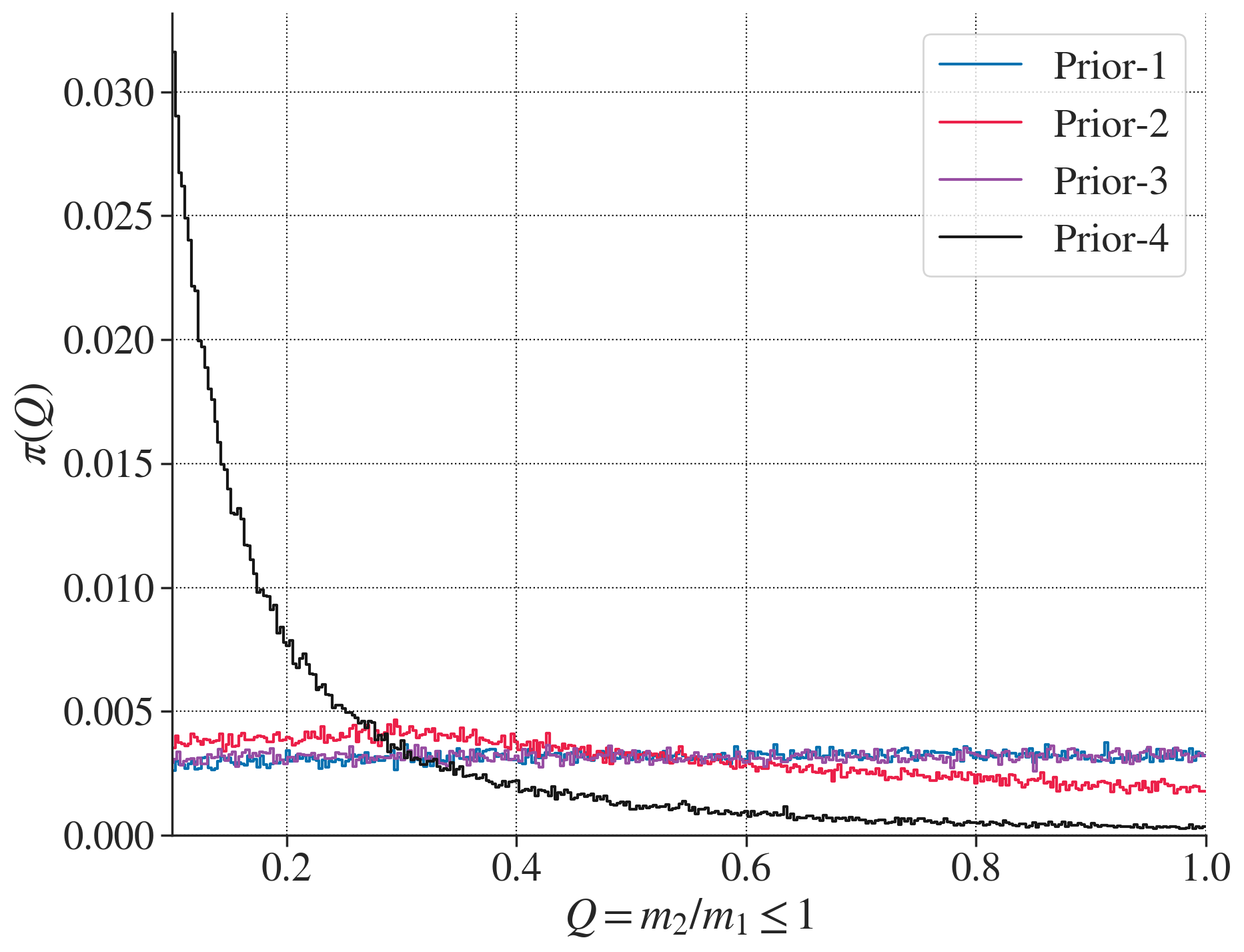}
    \caption{The figure compares mass ratios $Q$ resulting from four different prior choices used in our analysis. Unlike the other three priors, Prior-4 distinctly favours systems with unequal component masses. Also, as evident, Prior-2 effectively imposes a flat prior on $Q$.}
    \label{fig:mass-ratio-prior}
    \end{center}
\end{figure}

Except for the mass priors, we adopt default prior choices for our quasi-circular binary black hole analysis, with some adjustments based on our injections~\citep{Veitch:2014wba, Romero-Shaw:2020owr, LIGOScientific:2020ibl}. Table~\ref{Table:default-priors-x} summarises these settings for the signal parameters. The maximum spin values are set to 0.99 to ensure consistency with the employed waveform models. We employ the \texttt{UniformSourceFrame} luminosity distance prior, which implements a uniform prior in comoving volume $V_c$ with an initial factor $1/1+z$, assuming the Planck15 cosmology --- the factor of $1/1+z$ accounts for time dilation due to the expansion of the Universe~\citep{Romero-Shaw:2020owr}.  
\begin{table}[h]
    \centering
    \begin{tabular}{cccc}
    \hline Parameter & Shape & Limits & Boundary \\
    \hline \hline
    $\chi_1, \chi_2$ & Uniform & (0,~0.99) & - \\
    $\iota$ & Sinusoidal & $(0,~\pi)$ & - \\
    $\psi$ & Uniform & $(0,~\pi)$ & Periodic \\
    $\varphi$ & Uniform & $(0,~2 \pi)$ & Periodic \\
    $\alpha$ & Uniform & $(0,~2 \pi)$ & Periodic \\
    $\delta$ & Cosinusoidal & $(-\pi / 2,~\pi / 2)$ & - \\
    $D_L$ & \texttt{UniformSourceFrame} & (0.1, 10)Gpc & - \\
    \hline
    \end{tabular}
    \caption{Default prior settings of signal parameters to study binary black hole signals.}
    \label{Table:default-priors-x}
\end{table}
 
As for the mass priors, we explore the following four choices for our analysis. \emph{Prior-1} adopts a flat prior on the detector frame chirp mass $M_c(1+z)$ in the range of $[30,300]M_\odot$ and $Q$ in the range of $[0.1,1]$ while constraining the individual masses $m_i(1+z)$ to lie within $[5,300]M_\odot$. \emph{Prior-2} aligns with the prior used in LIGO/Virgo's GW190521 analysis, setting flat priors on the component masses $m_i(1+z)$ in the range of $[5,300]M_\odot$ while ensuring that the mass ratio and the detector frame total mass range lie within $Q\in[0.1,1]$ and $M_T(1+z)\in[80,400]M_\odot$~\footnote{Please note that \citet{LIGOScientific:2020iuh} implemented a prior which is $\pi(d_L) \propto d_L^2$ which distribute mergers uniformly throughout Euclidean Universe. Throughout our study, we adopt the \texttt{UniformSourceFrame} as our distance prior.}. \emph{Prior-3} imposes a flat prior on $Q\in[0.1,1]$ and $M_T(1+z)\in[80,400]M_\odot$ while constraining the individual masses $m_i(1+z)$ to lie within $[5,300]M_\odot$. Finally, \emph{Prior-4}, following \citet{Nitz:2020mga}, is flat in $1/Q \in [1,10]$ and $M_T(1+z)\in[80,400]M_\odot$. Figure~\ref{fig:mass-ratio-prior} compares the mass ratio distribution of our prior choices. Prior-4 favours unequal mass ratio systems, while the first three have no such preference. As meticulously elucidated in~\citet{Estelles:2021jnz, Olsen:2021qin, bustillo2022searching}, when examining the GW190521 event under the quasi-spherical signal hypothesis, such prior choices can notably impact the signal inference.

Irrespective of our prior choice, we sample in $M_c(1+z)$ and $Q$ space following \bilby's default setting. \bilby~accomplishes this by utilising appropriate transformations between the chosen parameterisation and the $M_c(1+z)-Q$ representation, thereby folding the chosen prior into $M_c(1+z)-Q$ space.

Finally, to assess the similarity/dissimilarity between posterior probability distributions obtained using different prior choices, we measure the Jensen-Shannon divergence (JSD)  defined as follows~\citep{Lin1991}:
\begin{align*}
    \mathrm{JSD}(p_1,p_2) &= \frac{1}{2} \sum_{k} p_1(\theta_k \mid d) \ln \left(\frac{2p_1(\theta_k \mid d)}{p_1(\theta_k \mid d) + p_2(\theta_k \mid d)}\right) \\
    &\quad+ \frac{1}{2} \sum_{k} p_2(\theta_k \mid d) \ln \left(\frac{2p_2(\theta_k \mid d)}{p_1(\theta_k \mid d) + p_2(\theta_k \mid d)}\right)~.
\end{align*}
Here, $p_1(\theta_i \mid d)$ and $p_2(\theta_i \mid d)$ represents the posterior of parameter $\theta_i$ for two different prior choices. A value of 0 nats~\footnote{``nat'' is a unit of information based on natural logarithms~\citep{mackay2005information}.} indicates that no additional information is gained when transitioning from one posterior distribution to another, implying that the two distributions are similar. On the other hand, a value of $\ln (2) = 0.69$ nat represents the maximum divergence between the distributions.

    \section{Results}
    \label{sec:results}

    \subsection{Simulation Set-I}

\begin{figure*}[htb]
    \centering
        \subfloat[For simulated systems at an inclination $\iota=28.6^\circ$]{\label{fig:iota-0.5}
        \centering
        \includegraphics[width=0.32\textwidth]{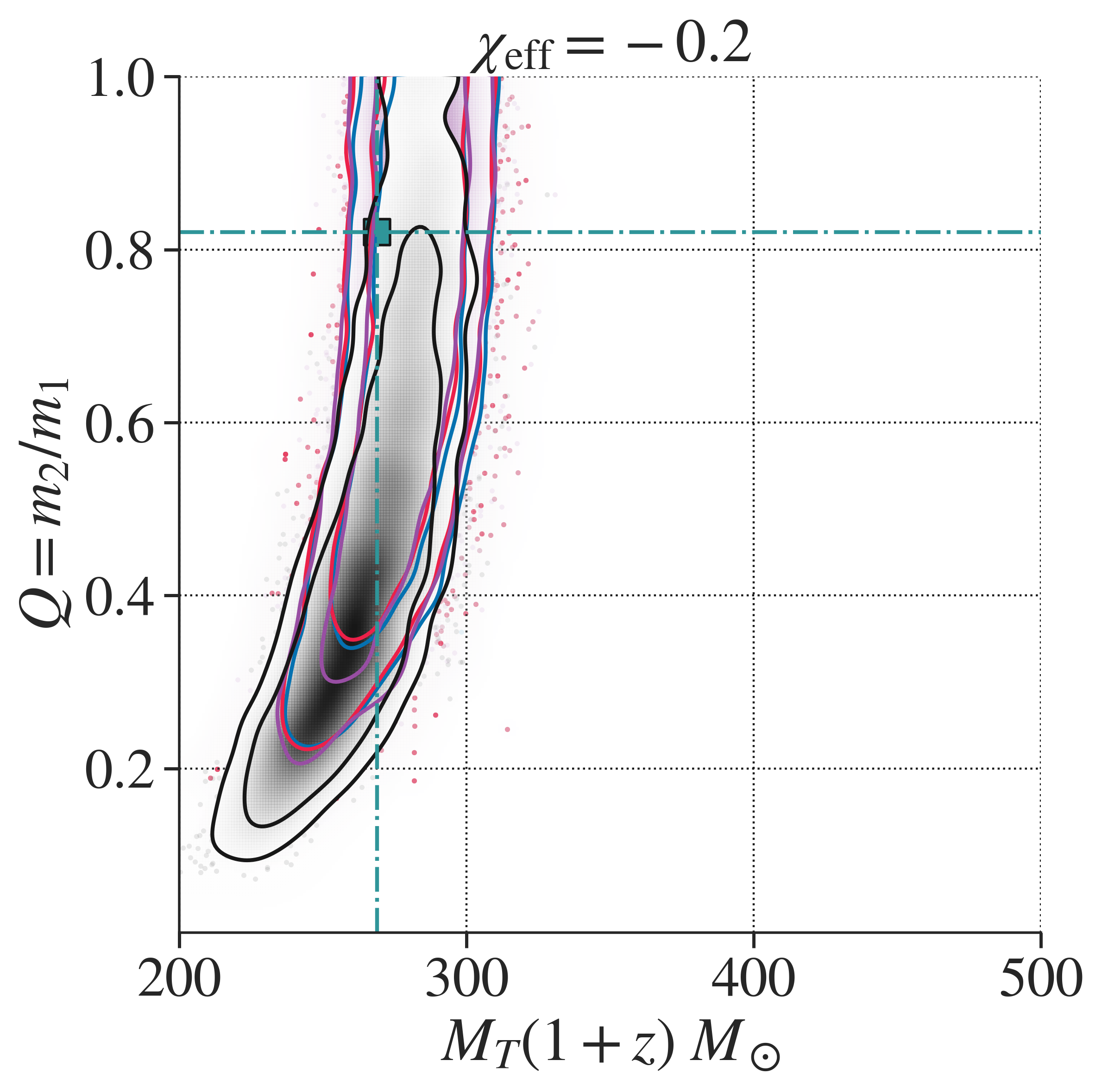}
        \includegraphics[width=0.32\textwidth]{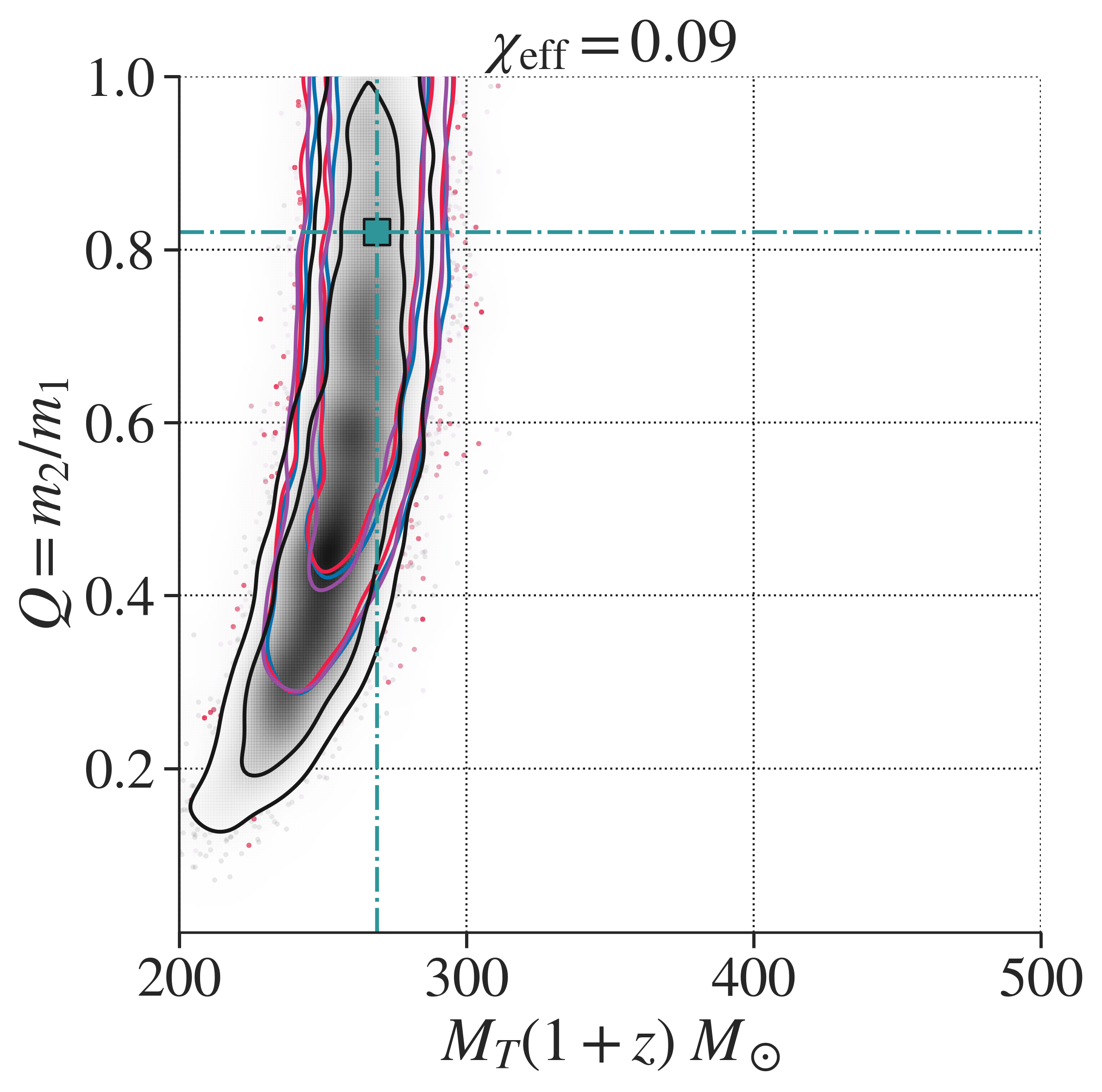}
        \includegraphics[width=0.32\textwidth]{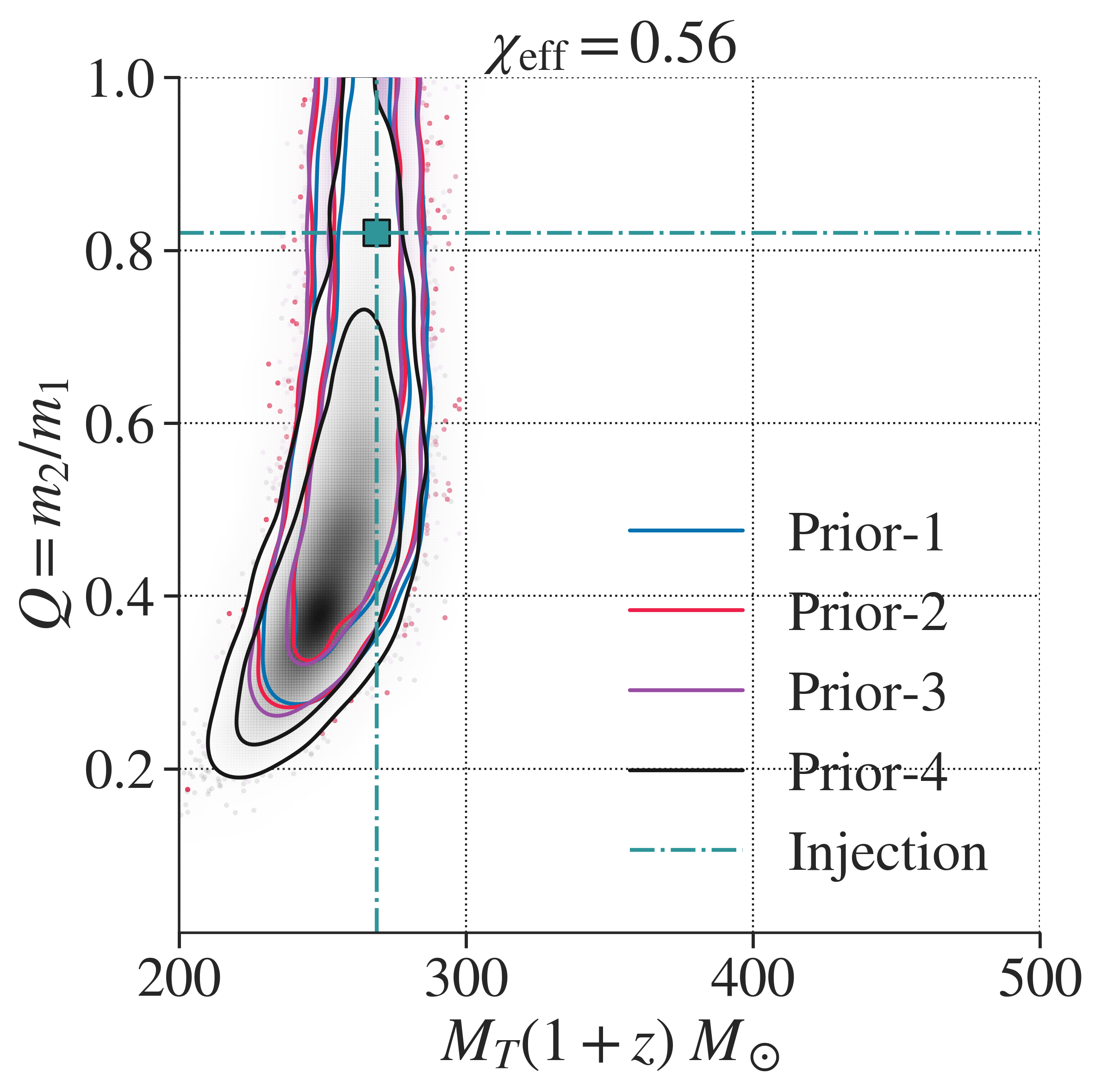}
    }\\
    \subfloat[For simulated systems at an inclination $\iota=54.4^\circ$]{\label{fig:iota-0.95}
        \centering
        \includegraphics[width=0.32\textwidth]{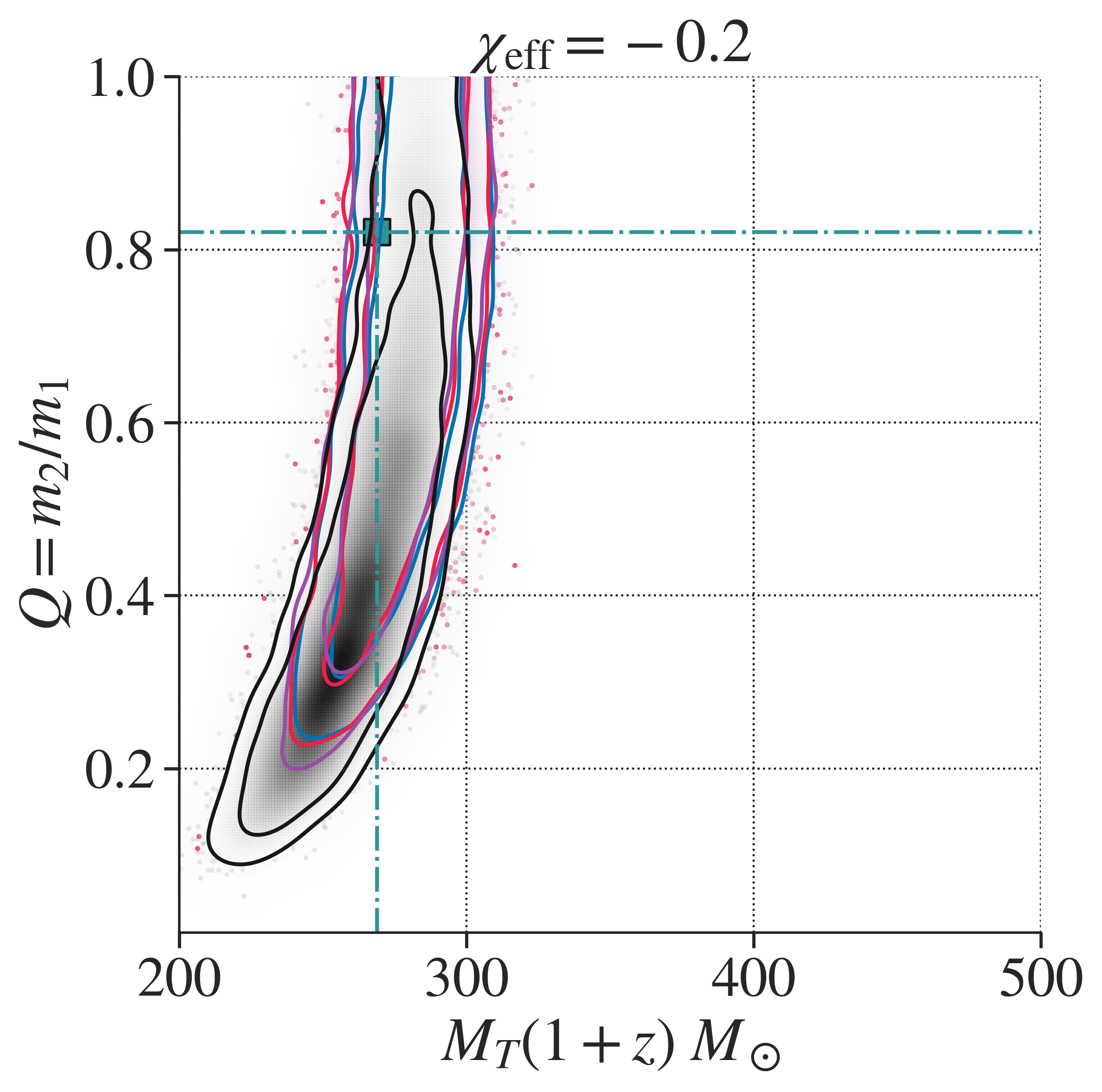}
        \includegraphics[width=0.32\textwidth]{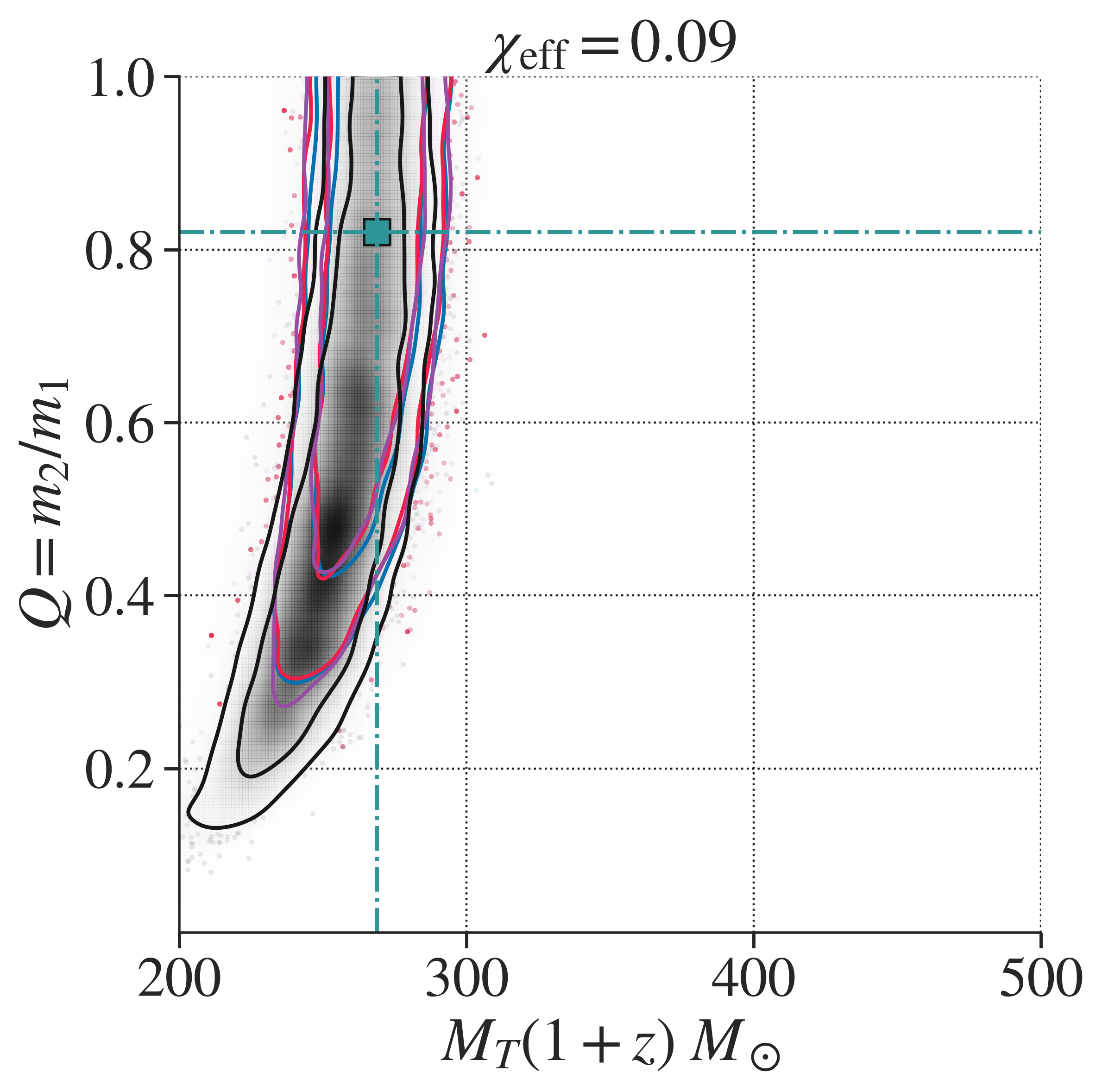}
        \includegraphics[width=0.32\textwidth]{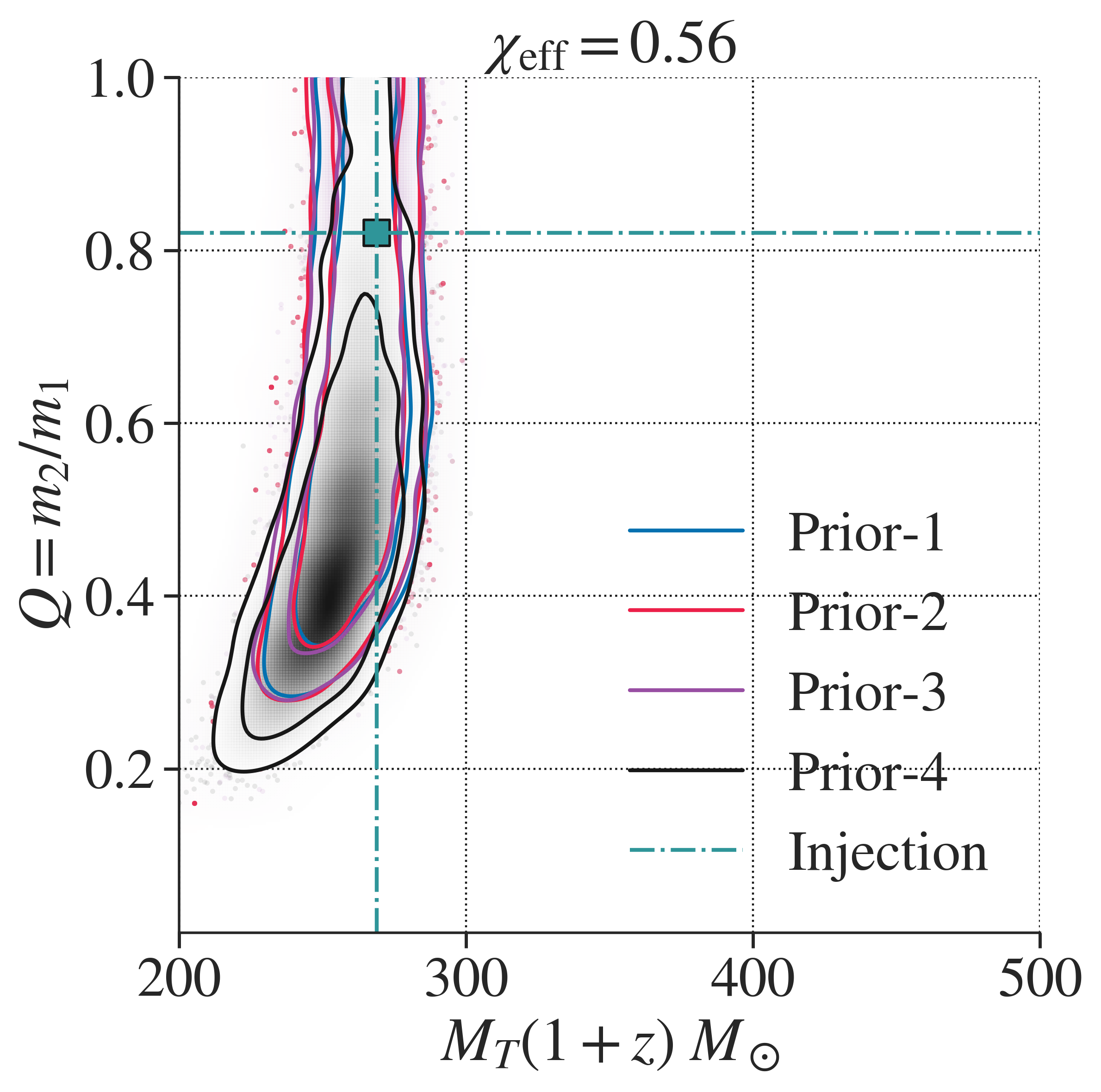}
    }
    \caption{Comparison of joint posterior distributions for $M_T(1+z)$ and $Q$ for different simulated systems with $Q = 0.82$ and varying $\chi_\mathrm{eff}$ as summarised in Table~\ref{table:set-1}. The contours represent the \acf{CI}, with the smaller contour indicating the 68\% \ac{CI} and the larger contour representing the 90\% \ac{CI}. The different colours in the figure represent the various mass prior choices used in the analysis, while the dotted line corresponds to the true mass parameters of the injected signals.}
    \label{fig:low-mass-ratio}
\end{figure*}

\begin{figure*}[ht]
    \begin{center}
\subfloat[For simulated systems at an inclination $\iota=28.6^\circ$]{\label{fig:iota-0.5-1}
        \centering
        \includegraphics[width=0.32\textwidth]{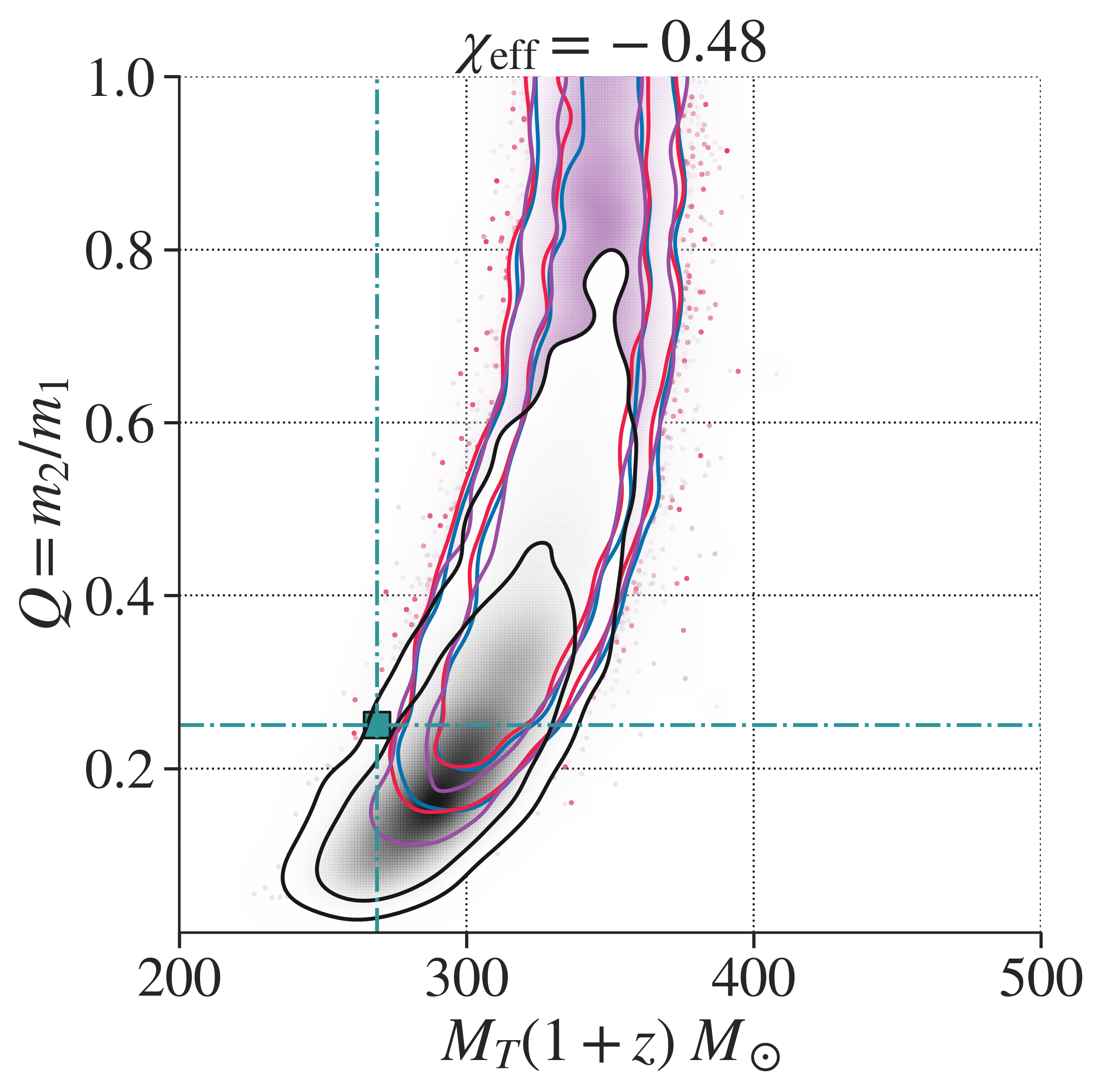}
        \includegraphics[width=0.32\textwidth]{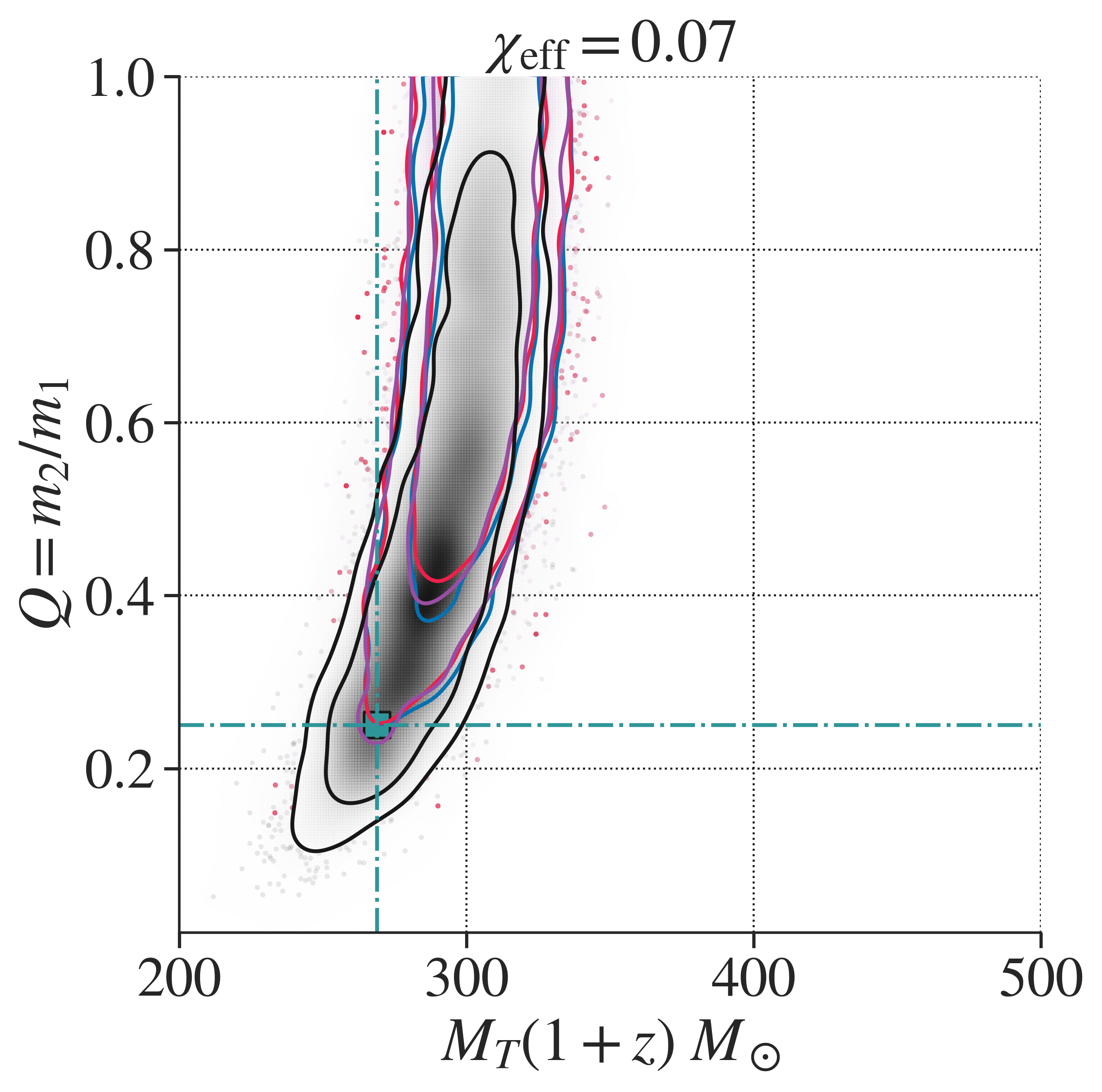}
        \includegraphics[width=0.32\textwidth]{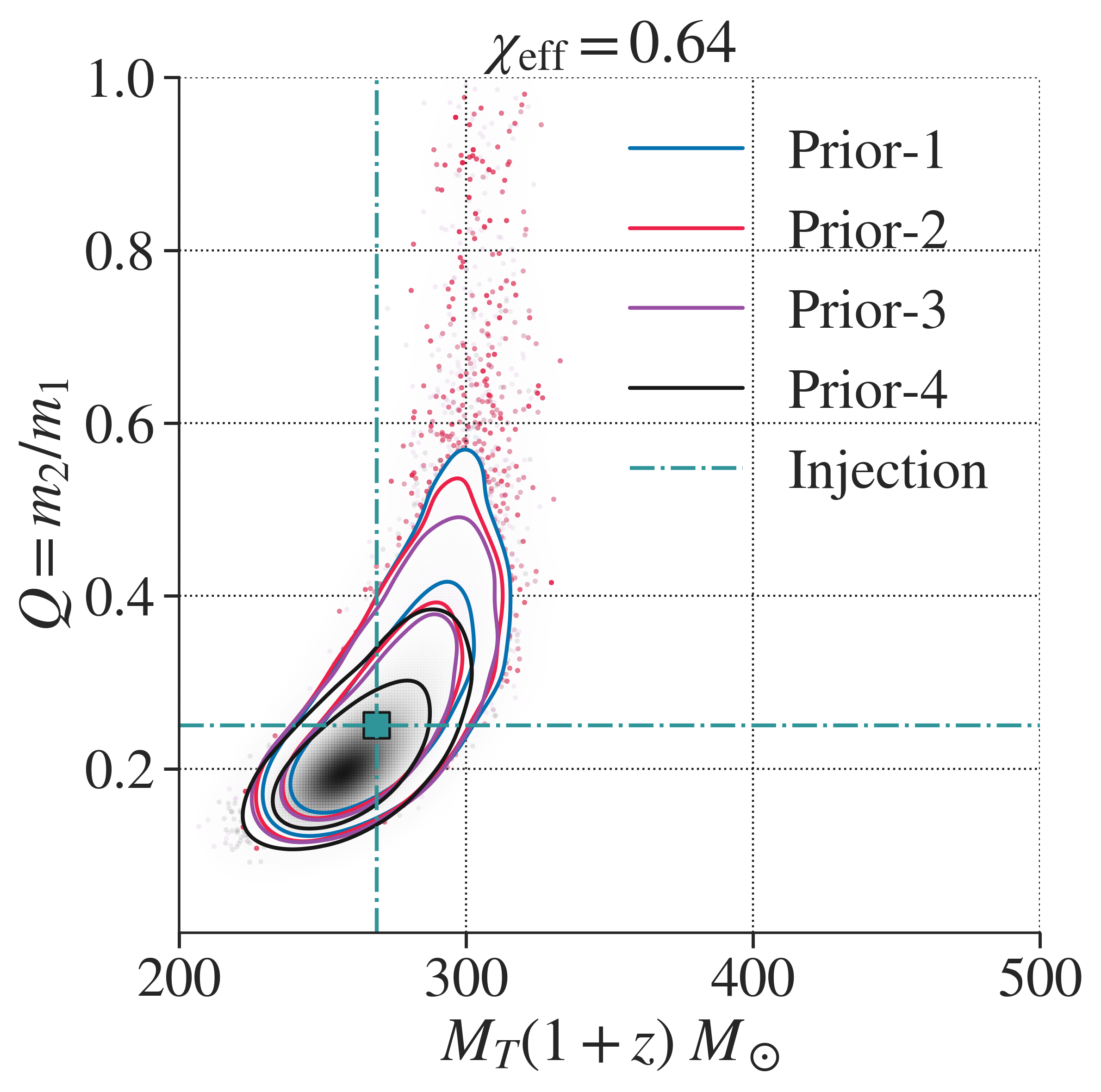}
    }\\
    \subfloat[For simulated systems at an inclination $\iota=54.4^\circ$]{\label{fig:iota-0.95-1}
        \centering
        \includegraphics[width=0.32\textwidth]{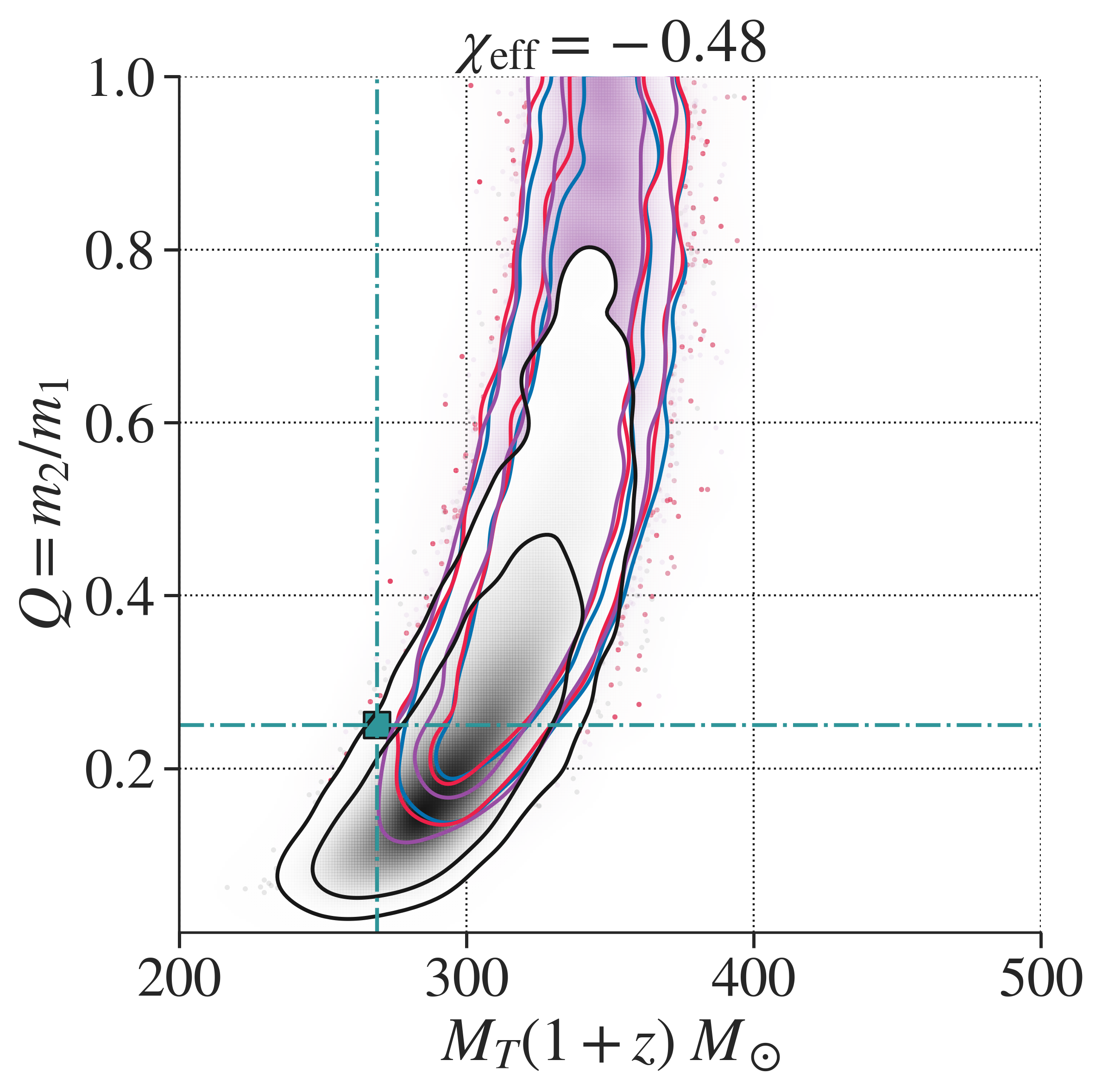}
        \includegraphics[width=0.32\textwidth]{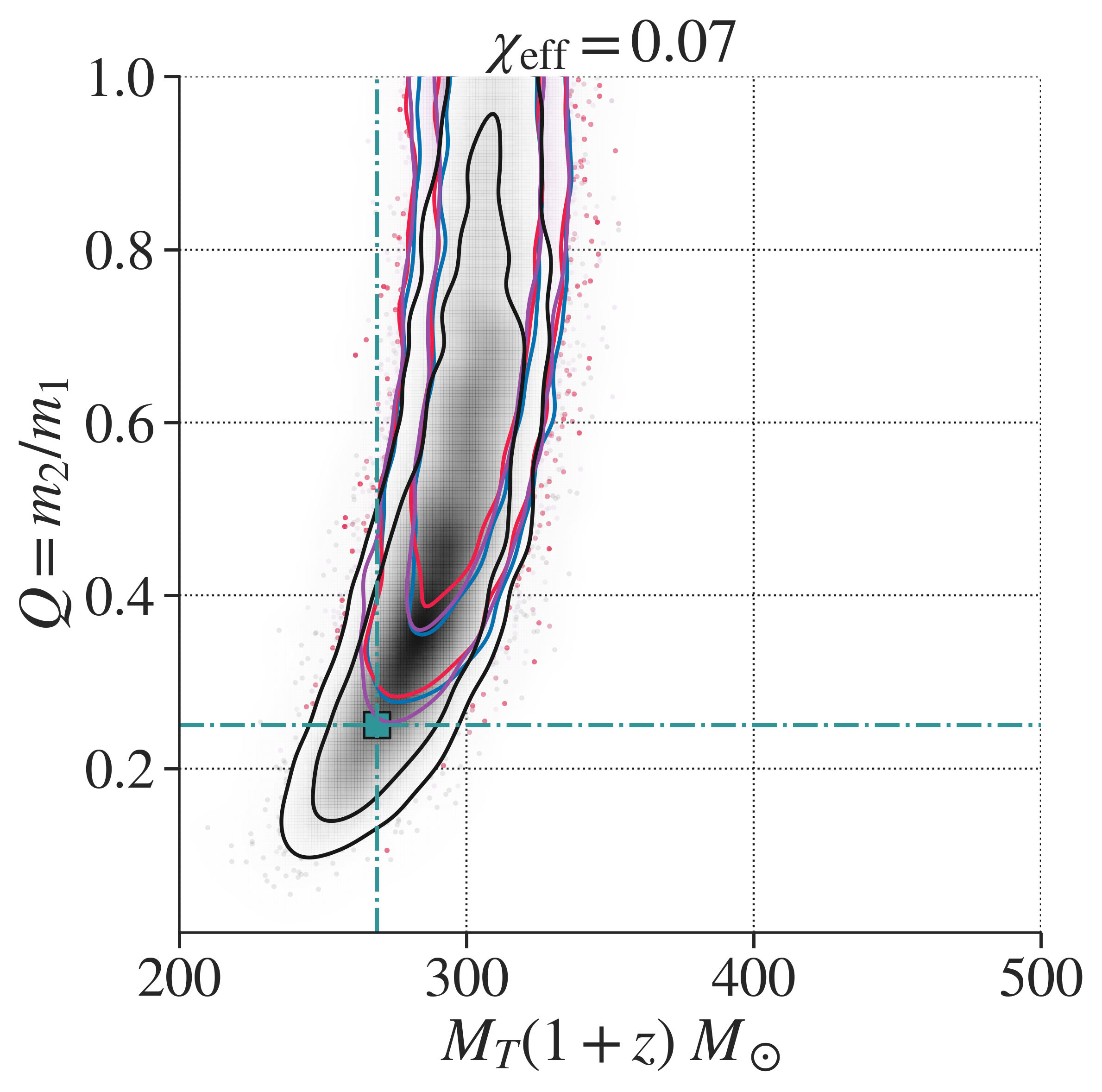}
        \includegraphics[width=0.32\textwidth]{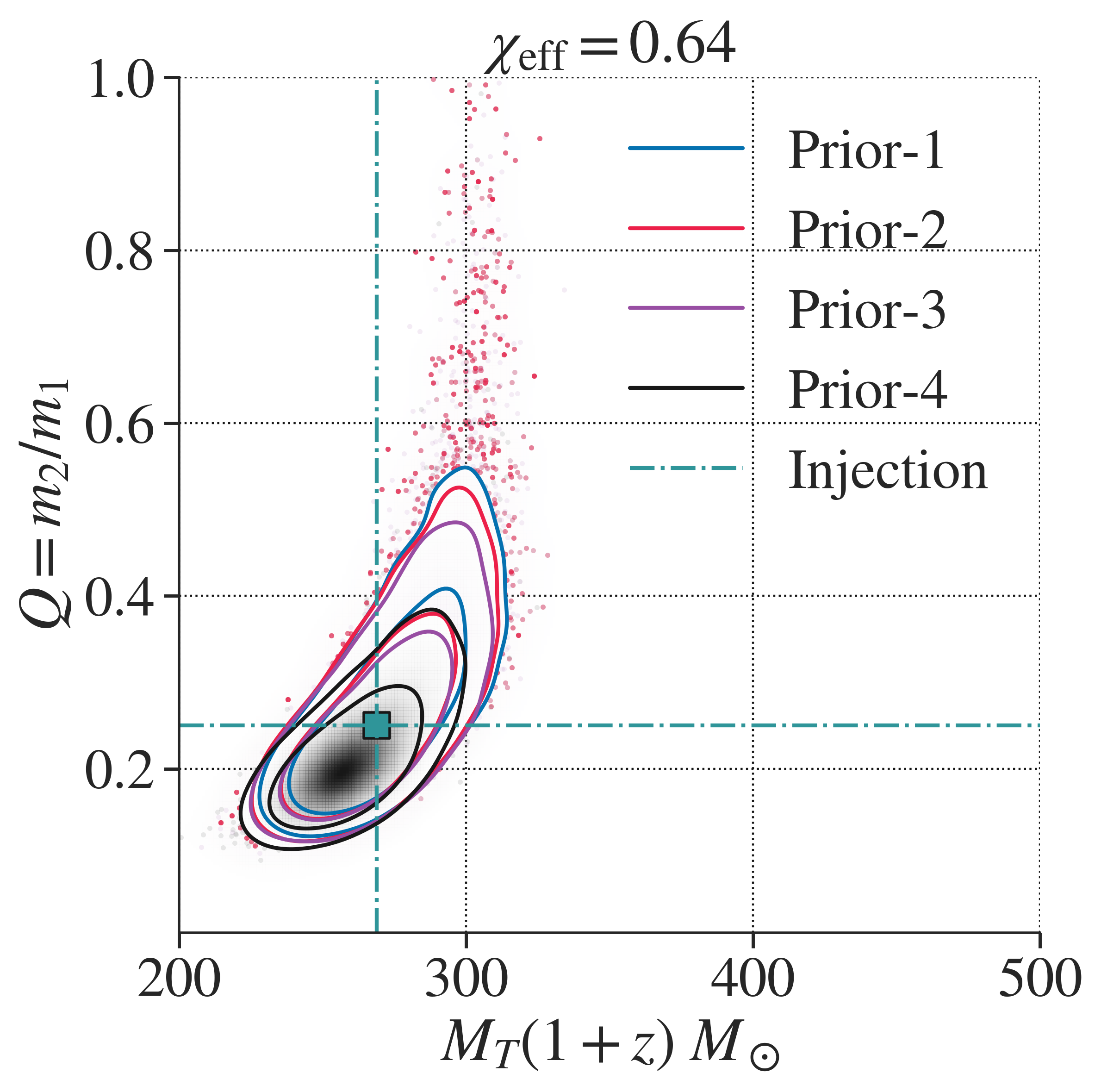}
    }
    \caption{Comparison of joint posterior distributions for $M_T(1+z)$ and $Q$ for different simulated systems with $Q = 0.25$ and varying $\chi_\mathrm{eff}$ as summarised in Table~\ref{table:set-1}. }
    \label{fig:higher-mass-ratio}
    \end{center}
\end{figure*}

We first present the results obtained using quasi-circular dominant waveform harmonics. As mentioned in Sec.~\ref{sec:data}, we simulate and recover set-1, detailed in Table~\ref{table:set-1}, using the waveform approximant IMRPhenomXAS to prevent any systematic errors caused by a disagreement between waveform models. Therefore, the effects discussed in this section are entirely due to different prior assumptions made for the analysis. We focus on the system's redshifted total mass and mass ratio since these parameters are better constrained for the heavy binaries discussed. We note that most of the signal SNR comes from the merger and ringdown phases for all our systems. Consequently, during likelihood evaluation and thus posterior sampling, the sampler prefers waveforms that better describe the post-inspiral phase. 

\subsubsection*{Nearly equal-mass binaries}

Figure~\ref{fig:low-mass-ratio} shows the two-dimensional posterior distribution of $M_T(1+z)$ and $Q$ for systems with injected $Q=0.82$ at varying inclinations and aligned spin values. We find that the JSD value for $M_T(1+z)$ and $Q$ posteriors lie in the range [0.0008, 0.0100] for Prior-1, 2 and 3, indicating that within statistical uncertainties, the samples drawn using Prior-1, 2 and 3 have an identical distribution. Further, the figure illustrates that the posterior's 68\% \ac{CI} consistently encompasses the injected value across inclinations and spin values for Prior-1 through Prior-3. However, the behaviour differs for Prior-4 due to its preference towards unequal mass systems. Notably, at $\chi_\mathrm{eff}=-0.2$ and $\chi_\mathrm{eff}=0.56$, the injected value lies outside the posterior's 68\% \ac{CI} for this prior choice. Also, this prior choice leads to a comparatively broader posterior distribution of masses for all spin values and inclinations, with a notable impact on mass ratios.

The spread of the posterior for all prior choices is least when the systems have $\chi_\mathrm{eff} = 0.56$ and most for $\chi_\mathrm{eff}=-0.2$~\footnote{For $\chi_\mathrm{eff}=-0.2$ system, $Q=0.43^{+0.44}_{-0.25}$ and for $\chi_\mathrm{eff}=0.09$ system, $Q=0.52^{+0.40}_{-0.31}$ (These measurements are reported as symmetric 90\% \ac{CI} around the median of the marginalized posterior distribution).}. This is due to the orbital-hangup effect, which results in a delay (or acceleration) in the merger of aligned (anti-aligned) spin systems relative to systems with lower spin ($\chi_\mathrm{eff}=0.09$) ~\citep{Campanelli:2006uy}. Consequently, our set of aligned spin systems spends $\sim 0.05$ seconds more within the detector bandwidth than anti-aligned systems, thus providing more information about the binary components and reducing measurement uncertainties. This trend holds irrespective of the binary's inclination. Additionally, altering parameters such as the number of live points (e.g., \texttt{nlive}=2048) or the network optimal SNR of the system to 30 does not alter the qualitative nature of these findings.

\subsubsection*{Asymmetric-mass binaries}

Similar to $Q=0.82$ systems, we observe that for $Q=0.25$ systems, the JSD values of $M_T(1+z)$ and $Q$ posteriors lie within the range of $[0.0024, 0.0099]$ for Prior-1, 2, and 3, indicating a coherence between their distributions for these specific prior selections. However, the overall results for these systems are quite different from $Q=0.82$ systems. As illustrated in Figure~\ref{fig:higher-mass-ratio}, the accuracy of the inference for the first three priors is unsatisfactory for anti-aligned and low-spin systems, which contrasts with the previously observed consistent agreement. The inferred masses' 90\% \ac{CI} (indicated by lighter contours) fail to adequately enclose the injected parameter values. This stems from the flat nature of these priors in the $Q$ parameter and our chosen prior on luminosity distance, which together favours intrinsically luminous systems --- systems that are mass-symmetric and located further away from us.

Conversely, Prior-4, as depicted in Figure~\ref{fig:mass-ratio-prior}, exhibits a propensity for asymmetric mass systems, resulting in a better recovery of the injected value. The injected parameter value lies within the 90\% \ac{CI} and the 68\% \ac{CI} for anti-aligned and low spin systems, respectively, when using Prior-4. As for aligned spin systems, the maximum a posteriori estimate of $M_T(1+z)$ and $Q$ is close to the injected parameter values for all prior choices, irrespective of inclination angle choice as evident from Figure~\ref{fig:higher-mass-ratio}.

Also, asymmetric systems tend to have a larger total mass posterior than mass-symmetric binaries. This is expected as an increase in mass asymmetry corresponds to a decrease in intrinsic loudness of the gravitational wave signal for a fixed total mass. Consequently, to account for this diminished signal strength, the resulting posterior distribution for the total mass naturally extends toward larger values of $M_T(1+z)$.

Therefore, in summary, owing to a flat prior on $Q$, the $M_T(1+z)$ and $Q$ posteriors obtained using Prior-1, 2 and 3 remain identical for different mass ratios, inclinations, and spins. However, due to their preference towards symmetric mass systems, their capability to effectively retrieve higher mass ratio systems with low spins or anti-aligned configurations is limited for the quasi-circular quadrupole case.


\begin{figure*}[htb]
            \begin{center}
        \subfloat[For simulated systems with $Q=0.82$]{\label{fig:q-1.219-1}
            \includegraphics[width=0.32\textwidth]{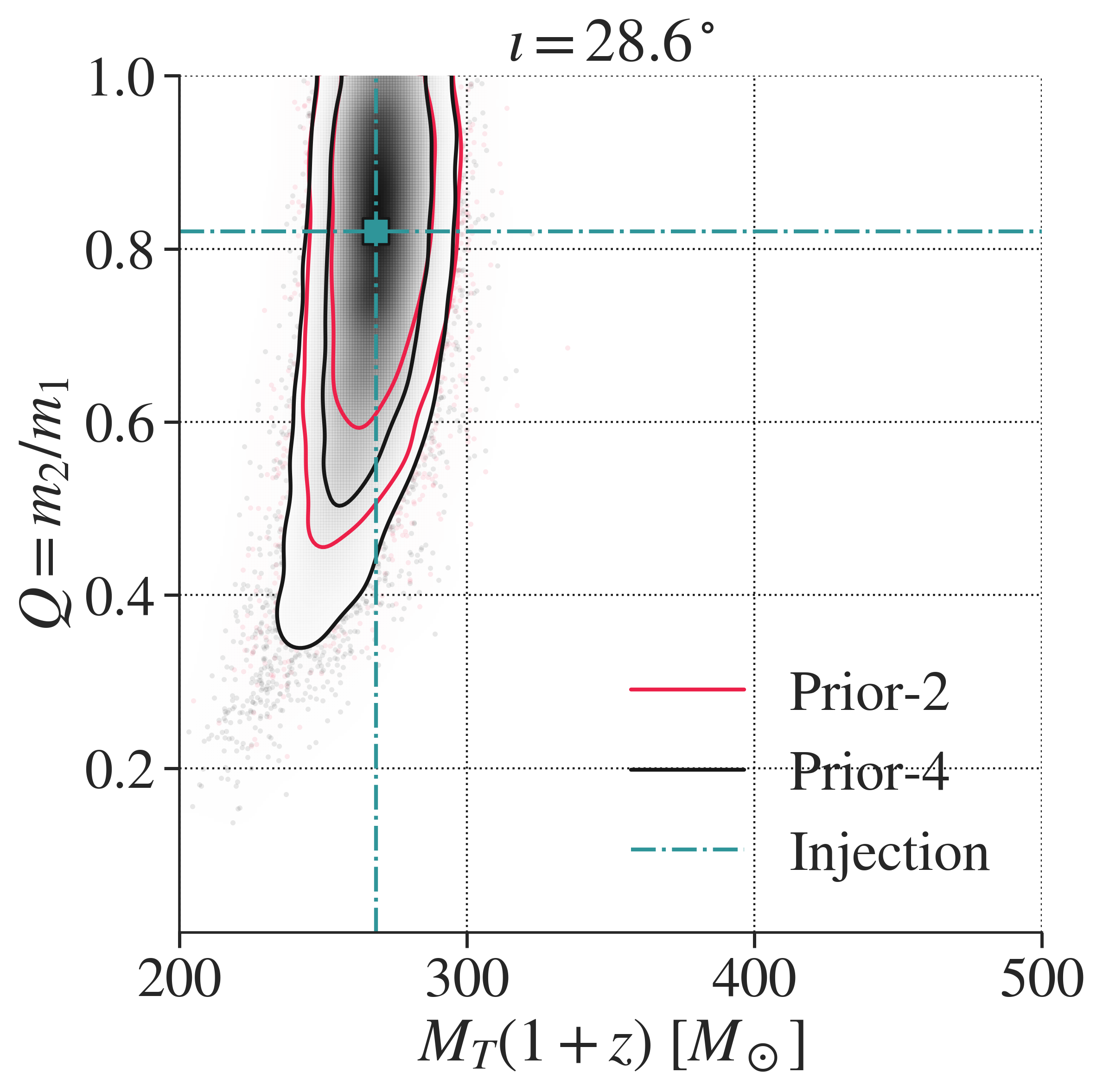}
            \includegraphics[width=0.32\textwidth]{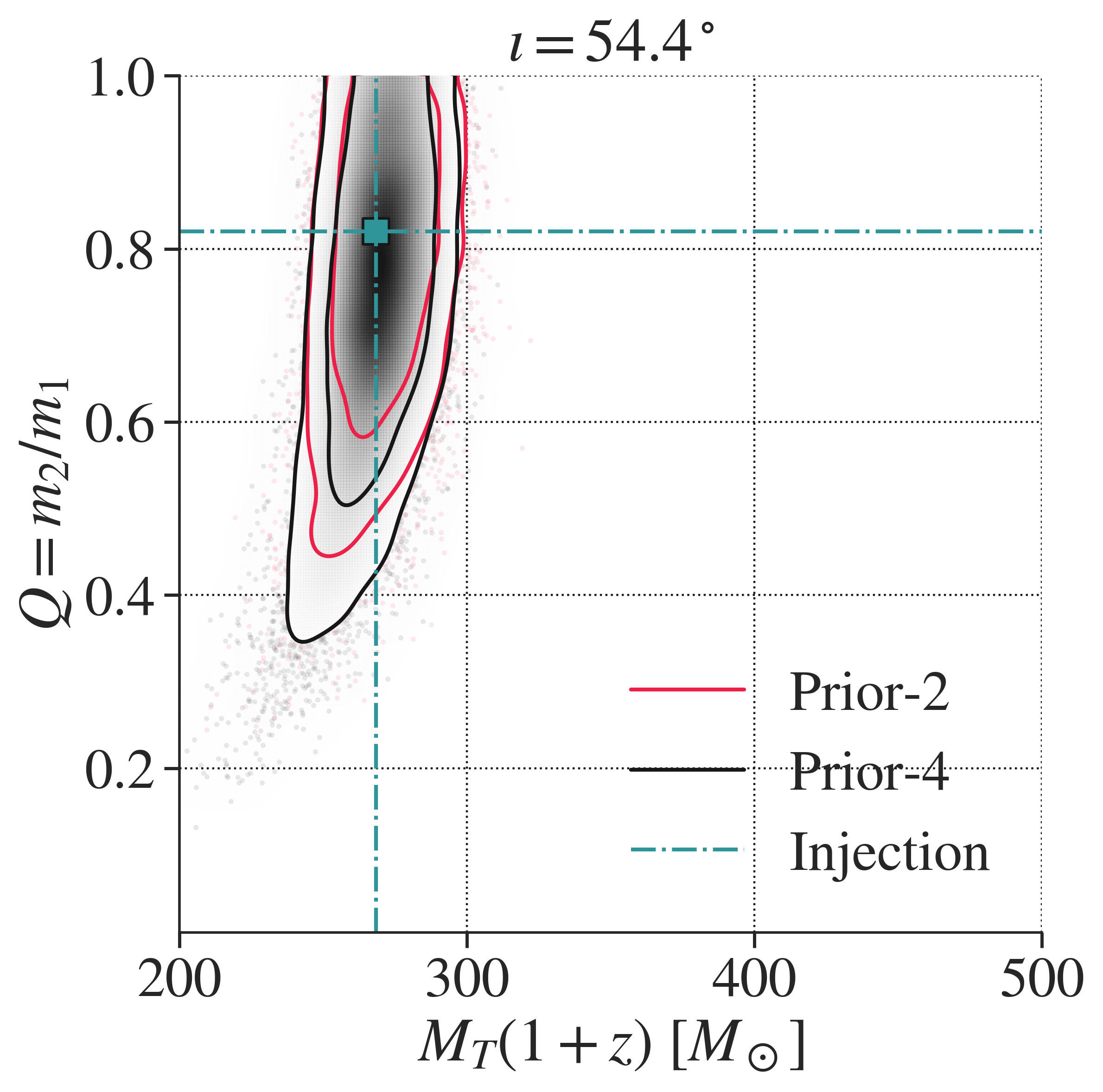}
            \includegraphics[width=0.32\textwidth]{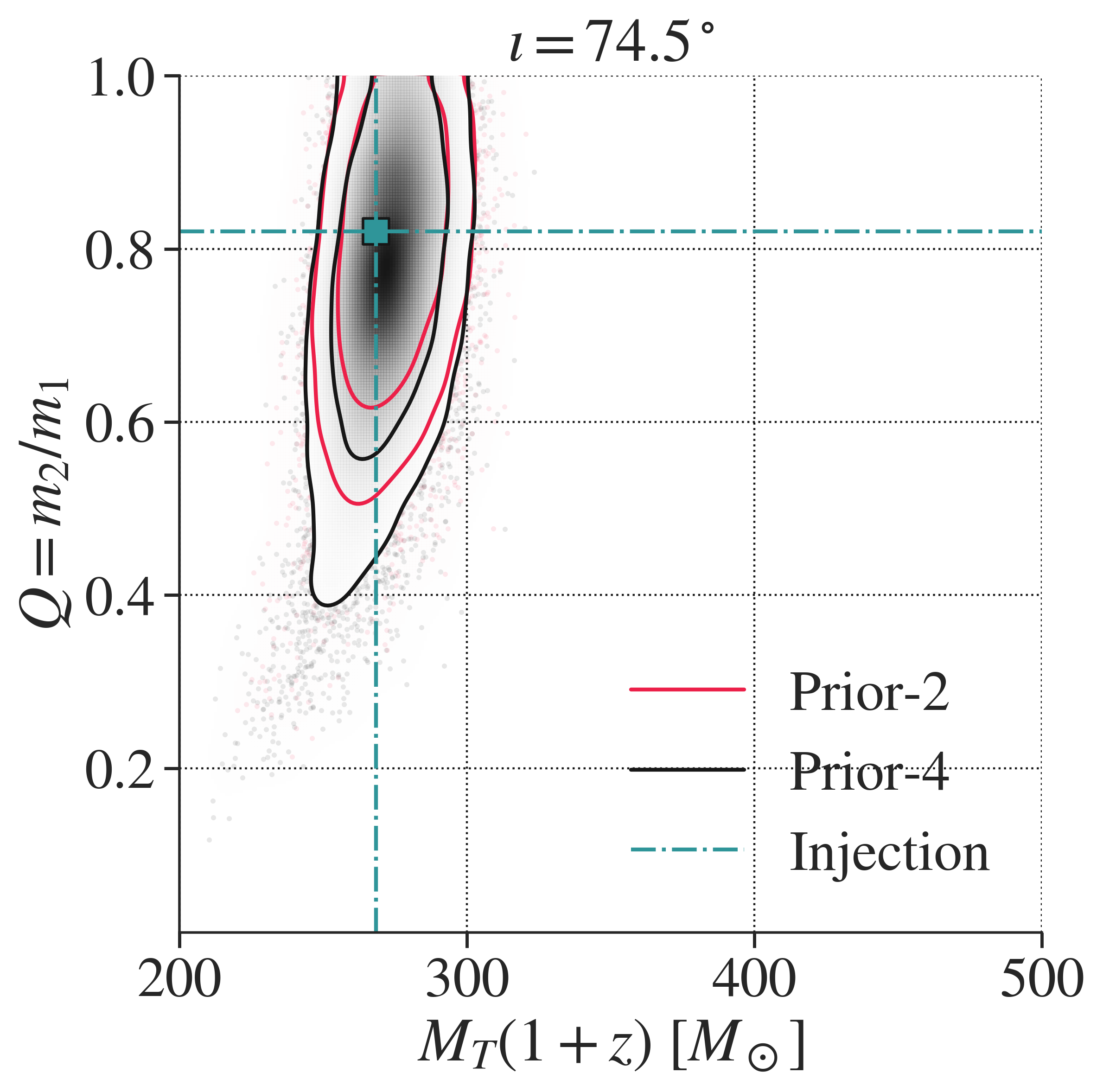}
        }\\
        \subfloat[For simulated systems with $Q=0.25$]{\label{fig:q-4-1}
            \includegraphics[width=0.32\textwidth]{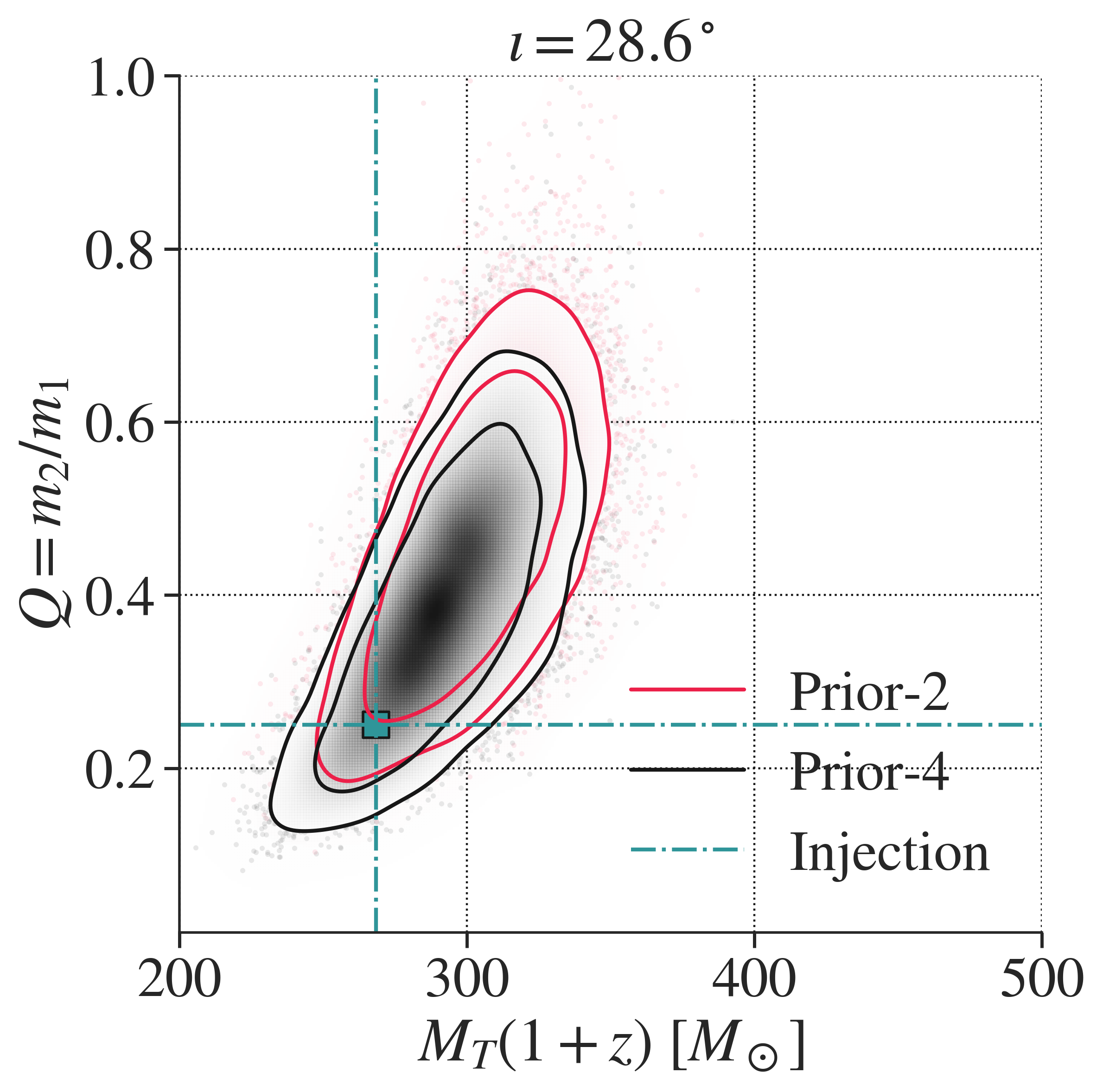}
            \includegraphics[width=0.32\textwidth]{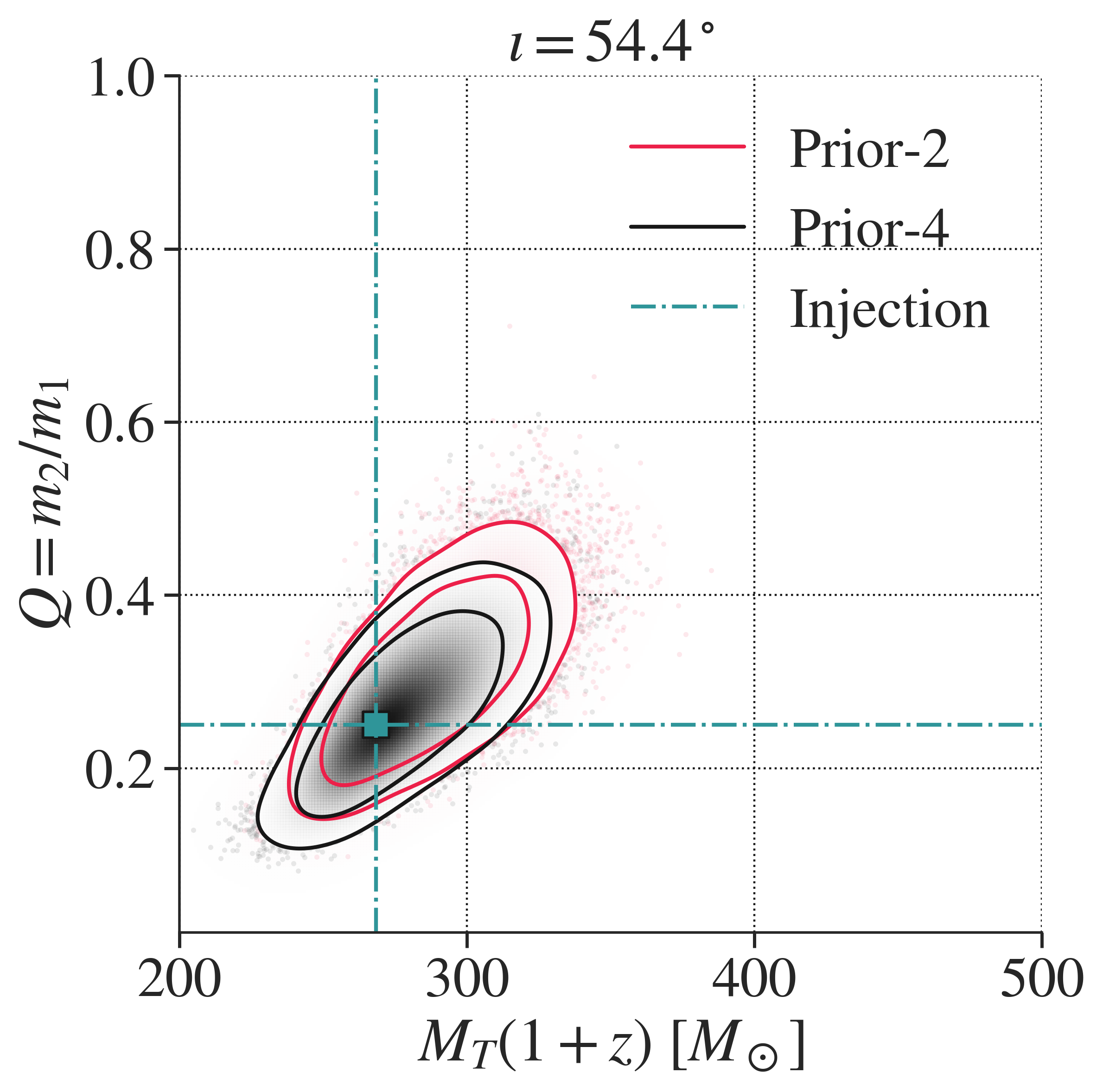}
            \includegraphics[width=0.32\textwidth]{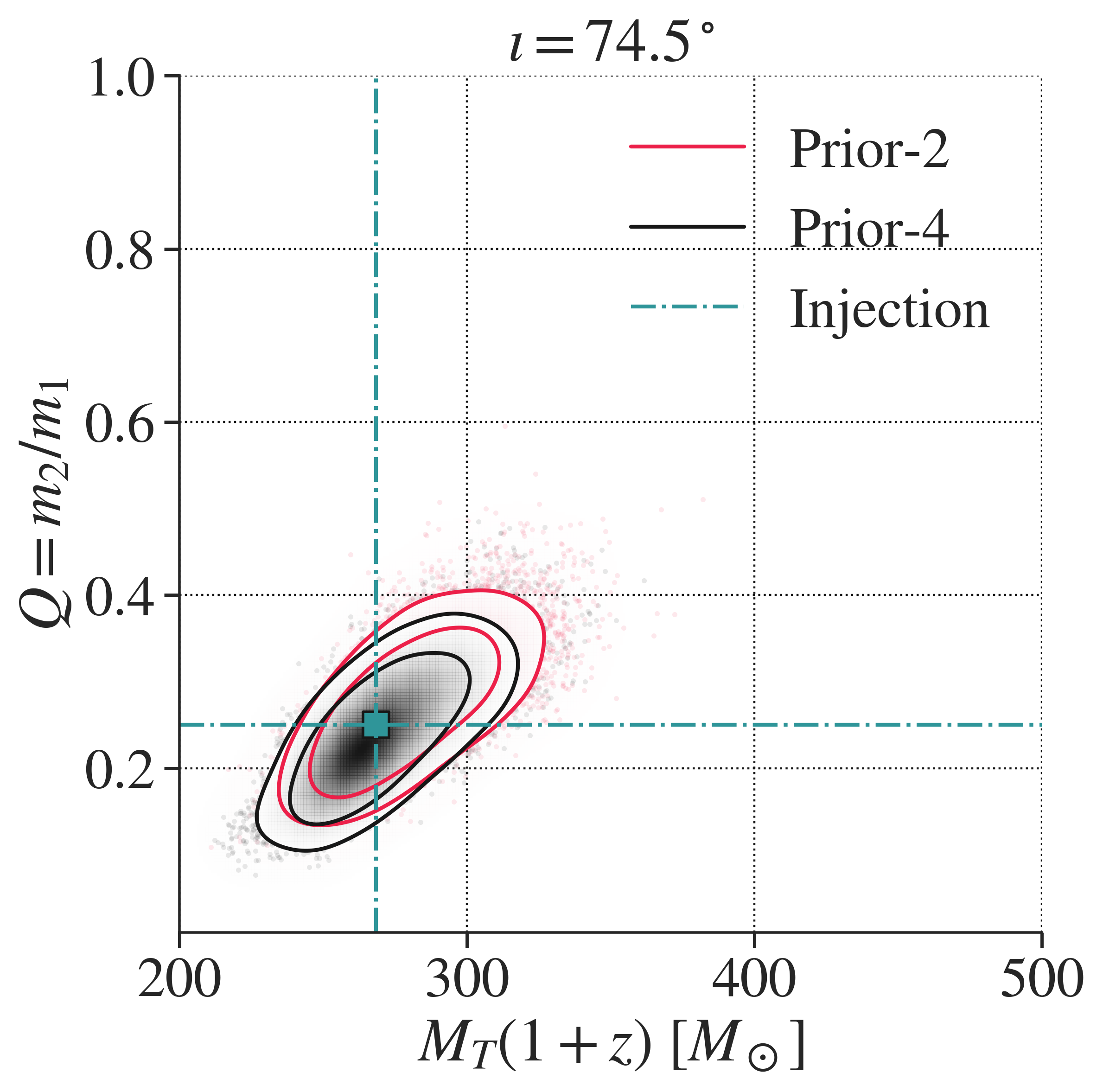}
        }\\
        \subfloat[For simulated systems with $Q=0.11$]{\label{fig:q-9-1}
            \includegraphics[width=0.32\textwidth]{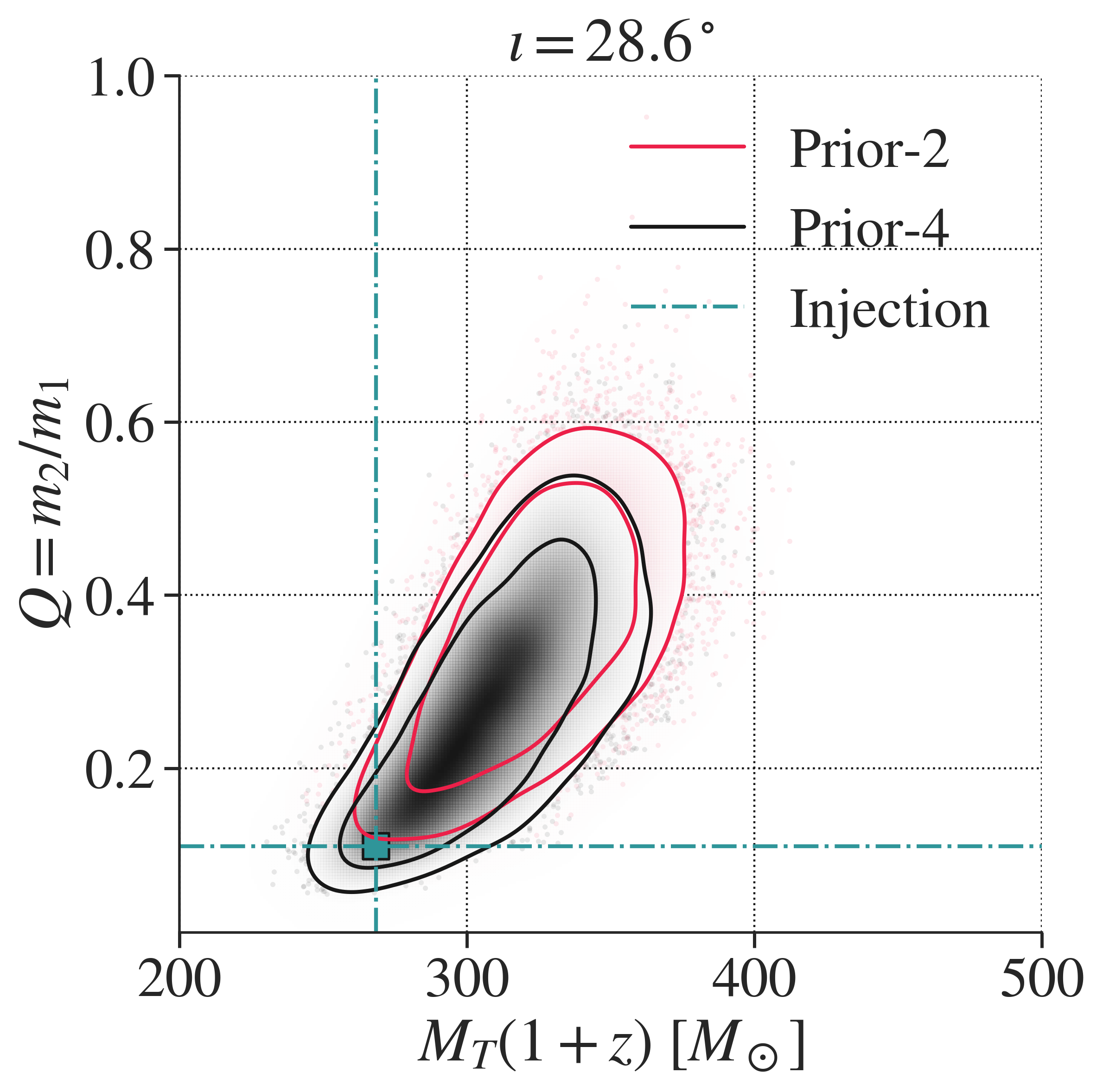}
            \includegraphics[width=0.32\textwidth]{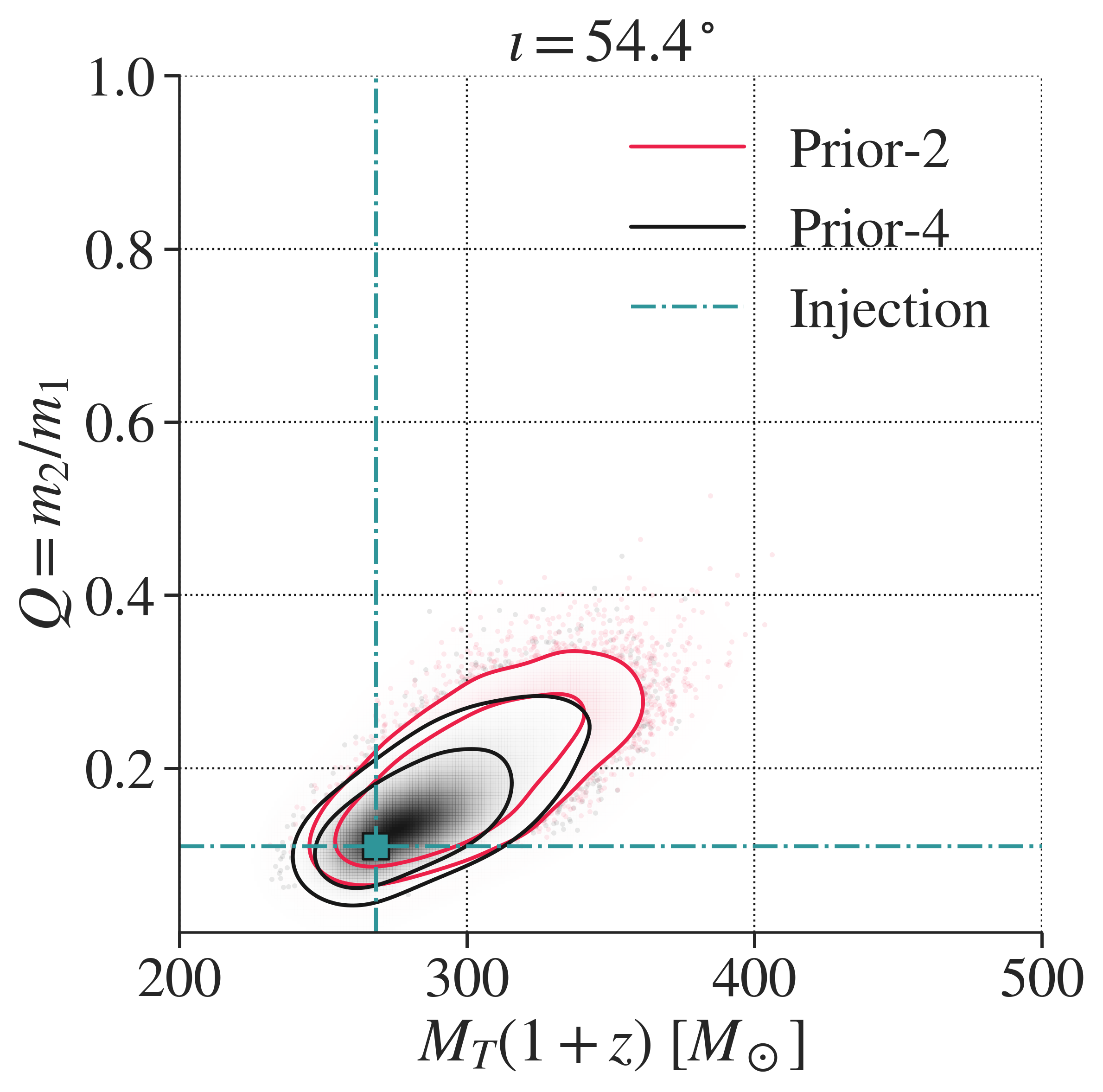}
            \includegraphics[width=0.32\textwidth]{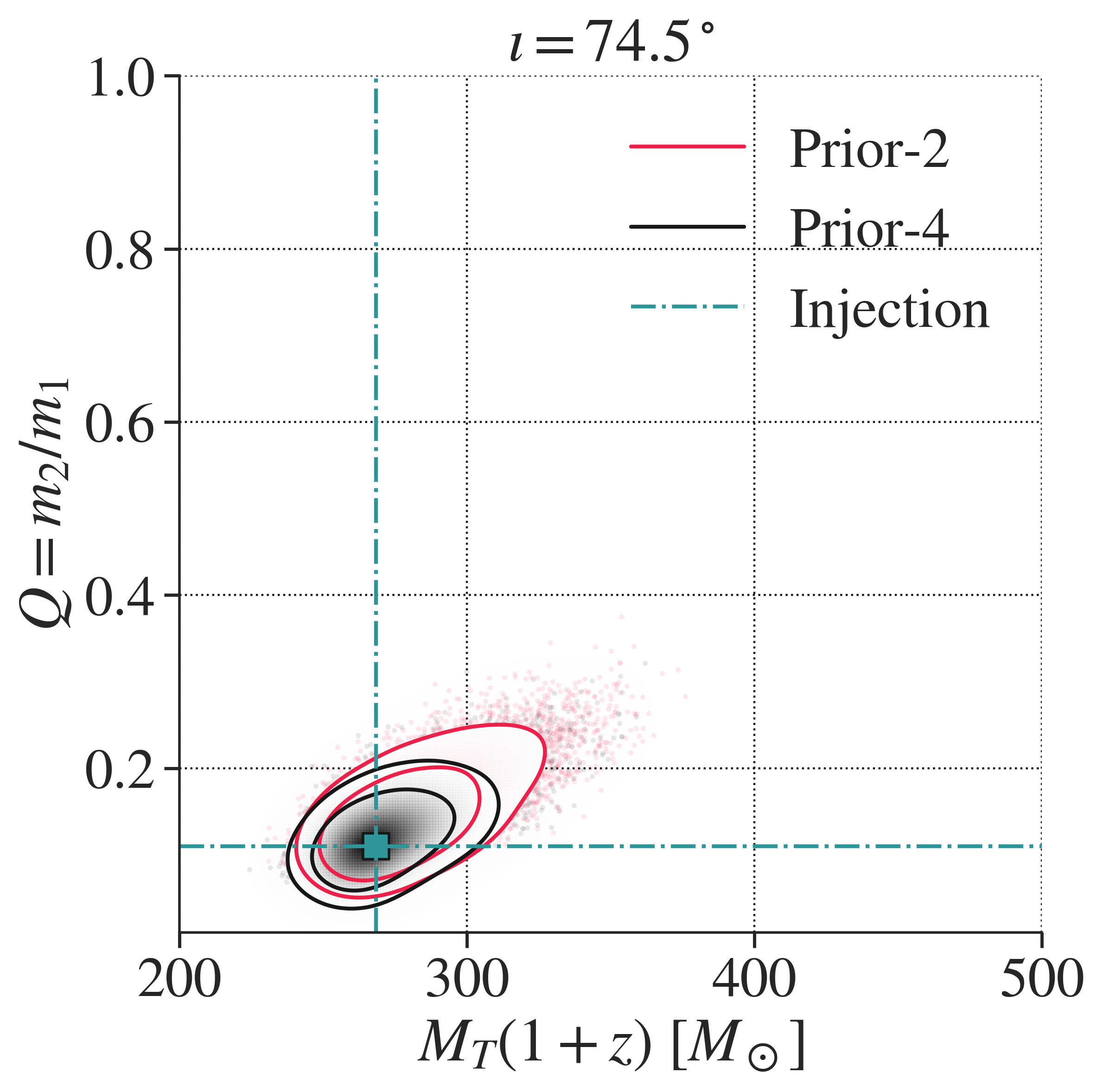}
        }\\    
        \caption{Joint posterior distributions for $M_T(1+z)$ and $Q$ of simulated systems with parameters summarized in Table~\ref{table:set-2}.}
        \label{fig:hm}
        \end{center}
\end{figure*}

    \subsection{Simulation Set-2}

We now examine the results obtained when analysing the injections described in Table~\ref{table:set-2} with IMRPhenomXHM, zero-noise realisation and different prior choices. This set of simulated binary systems has low spin and is oriented at three different inclination angles. Given that the posterior distributions obtained using Prior-1, Prior-2, and Prior-3 yield similar results (with a JSD value of $M_T(1+z)$ and $Q$ falling within the range of [0.0016, 0.0099]), we only consider the results for Prior-2 (LVK-like prior) and Prior-4 presented in Fig~\ref{fig:hm}. The area inside the $M_T(1+z)-Q$ contour decreases compared to the quadrupole analysis performed with simulation set-1, indicating a lower measurement uncertainty~. This is expected as the presence of higher harmonics in the signal facilitates a more accurate determination of the binary parameters in contrast to systems dominated by the quadrupole harmonic alone. Furthermore, the correlation between the $M_T(1+z)$ and $Q$ for a fixed mass ratio injection decreases as the inclination angles increase from lower to higher values. Similarly, the correlation becomes less from higher to lower mass ratios for a fixed inclination angle. This behaviour arises due to the increasing strength of the higher harmonics relative to the dominant quadrupole harmonic as the inclination and mass asymmetry rise, reducing the correlation between these two parameters of the binary system.

However, for Prior-4, the estimate's uncertainty decreases as we move from low ($\iota=28.6^\circ$) to moderate inclination ($\iota=54.4^\circ$) but grows as we move from moderate to high inclination ($\iota=74.5^\circ$) except for $Q=0.82$ system.

Lastly, it is essential to note that for the injection with a low mass ratio and low inclination ($Q=0.11, \iota = 28.6^\circ$), the actual value of the parameters, as indicated by the blue lines, lies outside the 90\% \ac{CI} of Prior-2 but inside the 68\% \ac{CI} of Prior-4, indicating that \textit{the prior distribution, rather than the higher harmonic content, is more crucial} in such cases. As the injection's inclination increases, the actual parameter value lies inside the 68\% \ac{CI} for all prior choices, thus illustrating the opposite. 

\begin{figure*}[htb]
    \begin{center}
    \subfloat[For simulated systems with $Q=0.82$]{
        \includegraphics[width=0.32\textwidth]{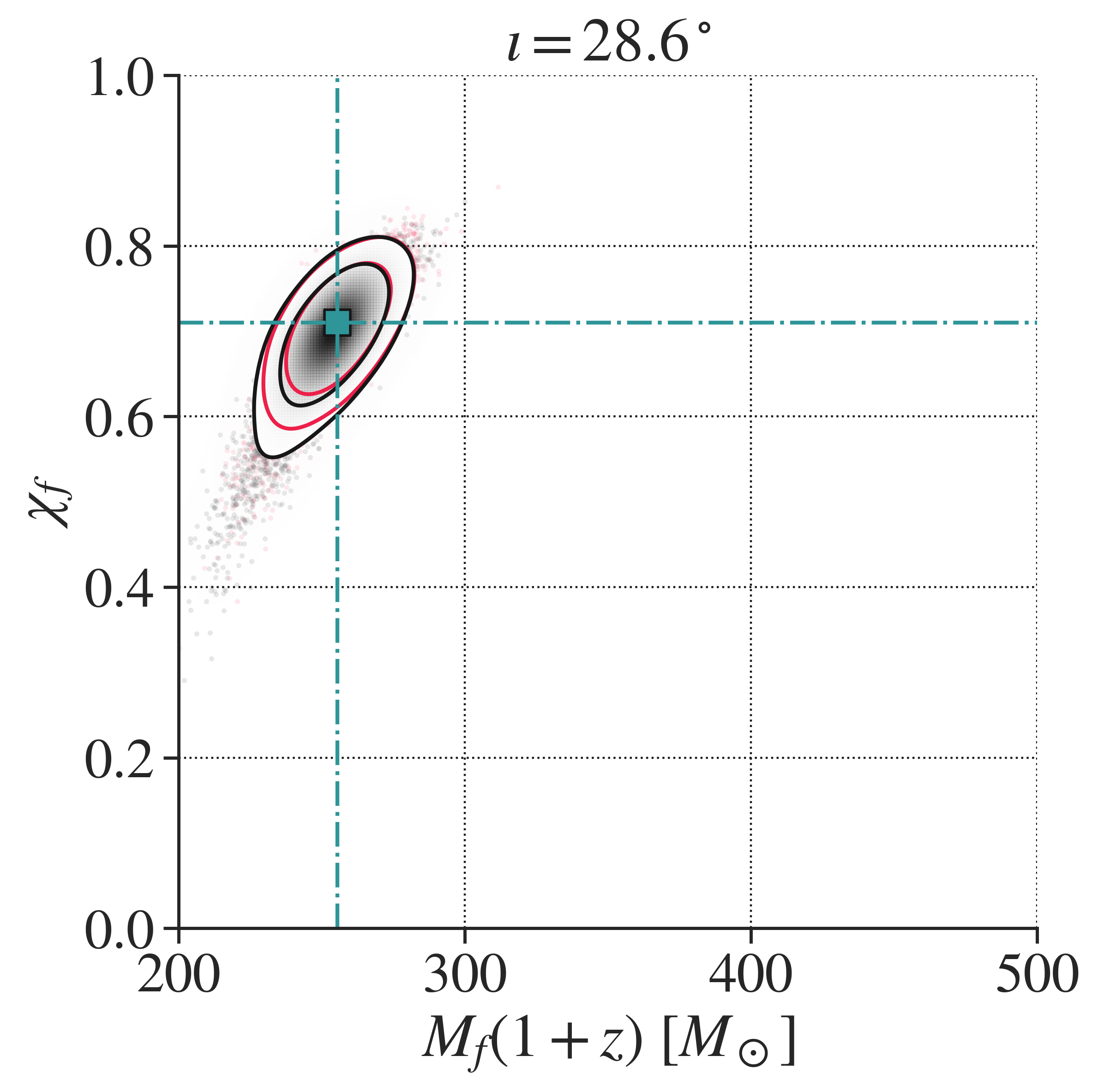}
        \includegraphics[width=0.32\textwidth]{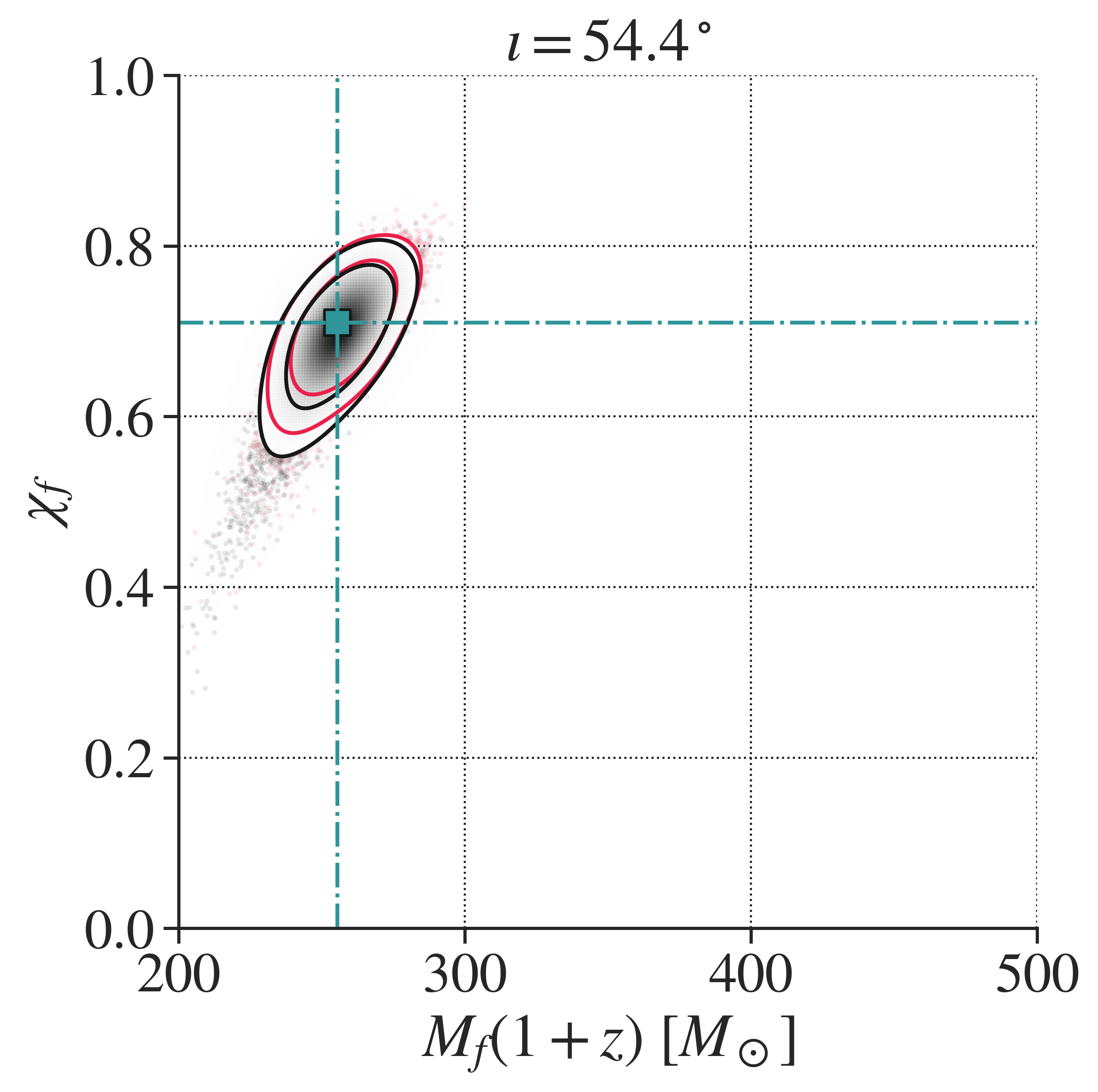}
        \includegraphics[width=0.32\textwidth]{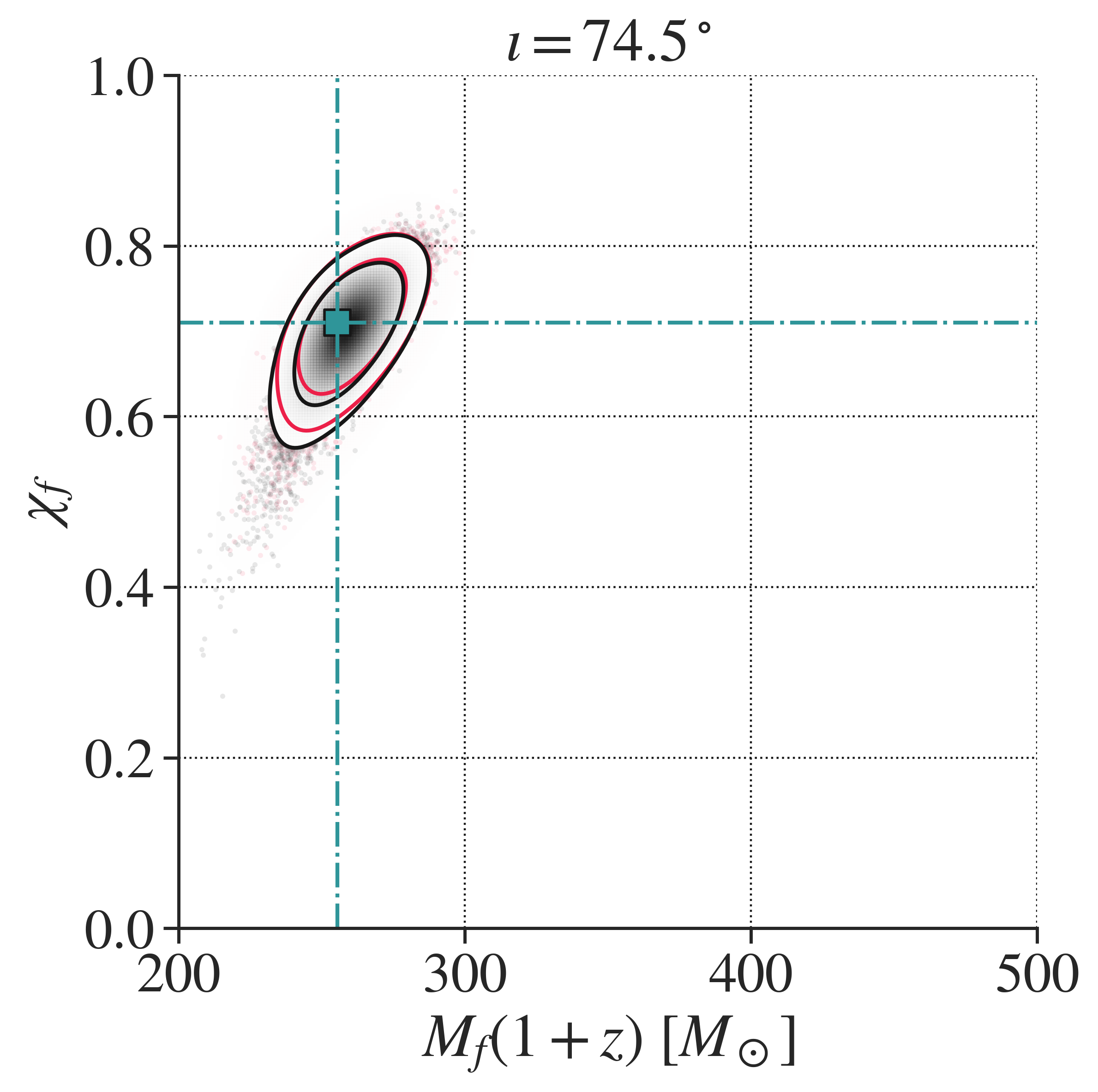}} \\
    \subfloat[For simulated systems with $Q=0.25$]{
        \includegraphics[width=0.32\textwidth]{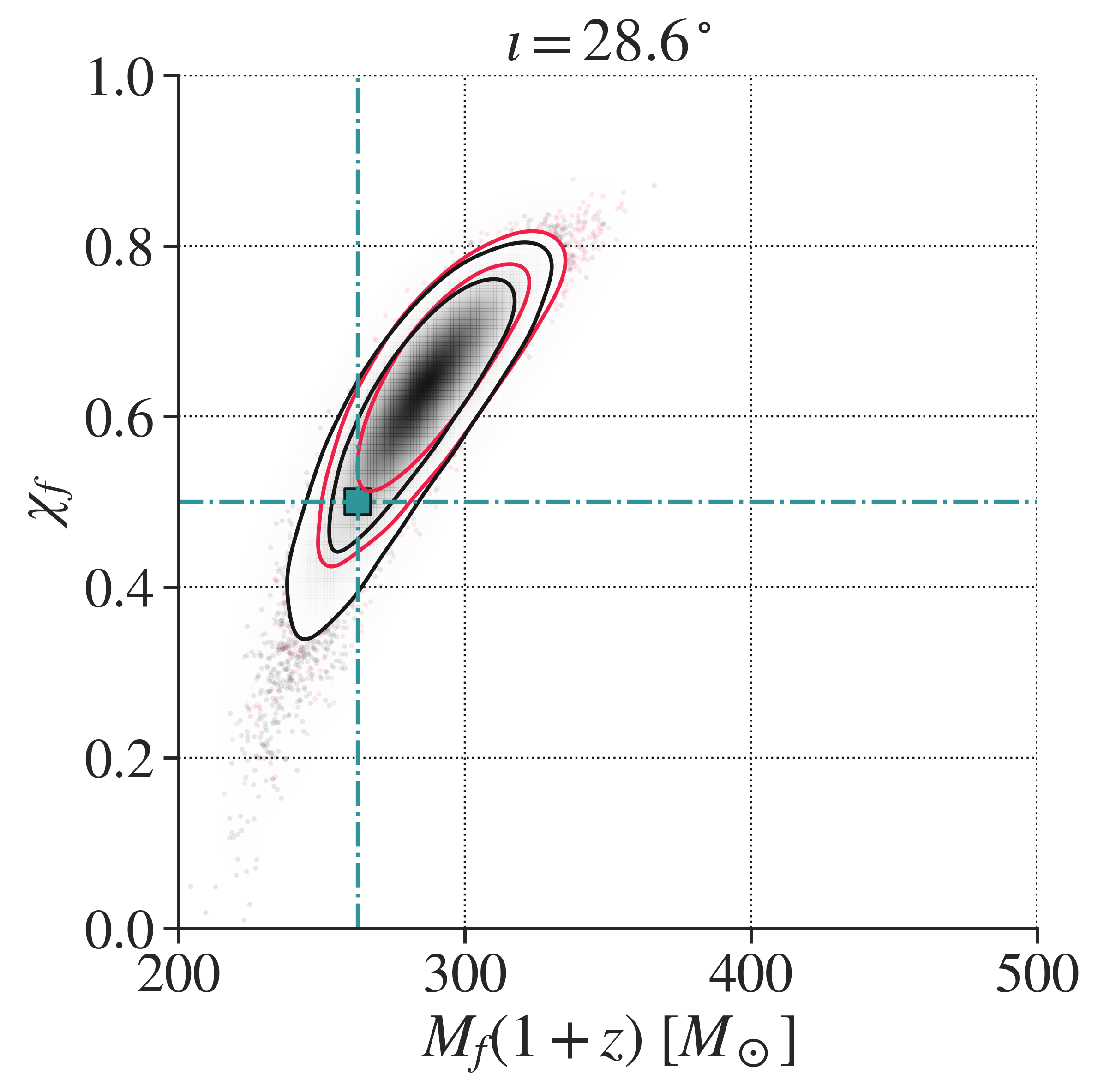}
        \includegraphics[width=0.32\textwidth]{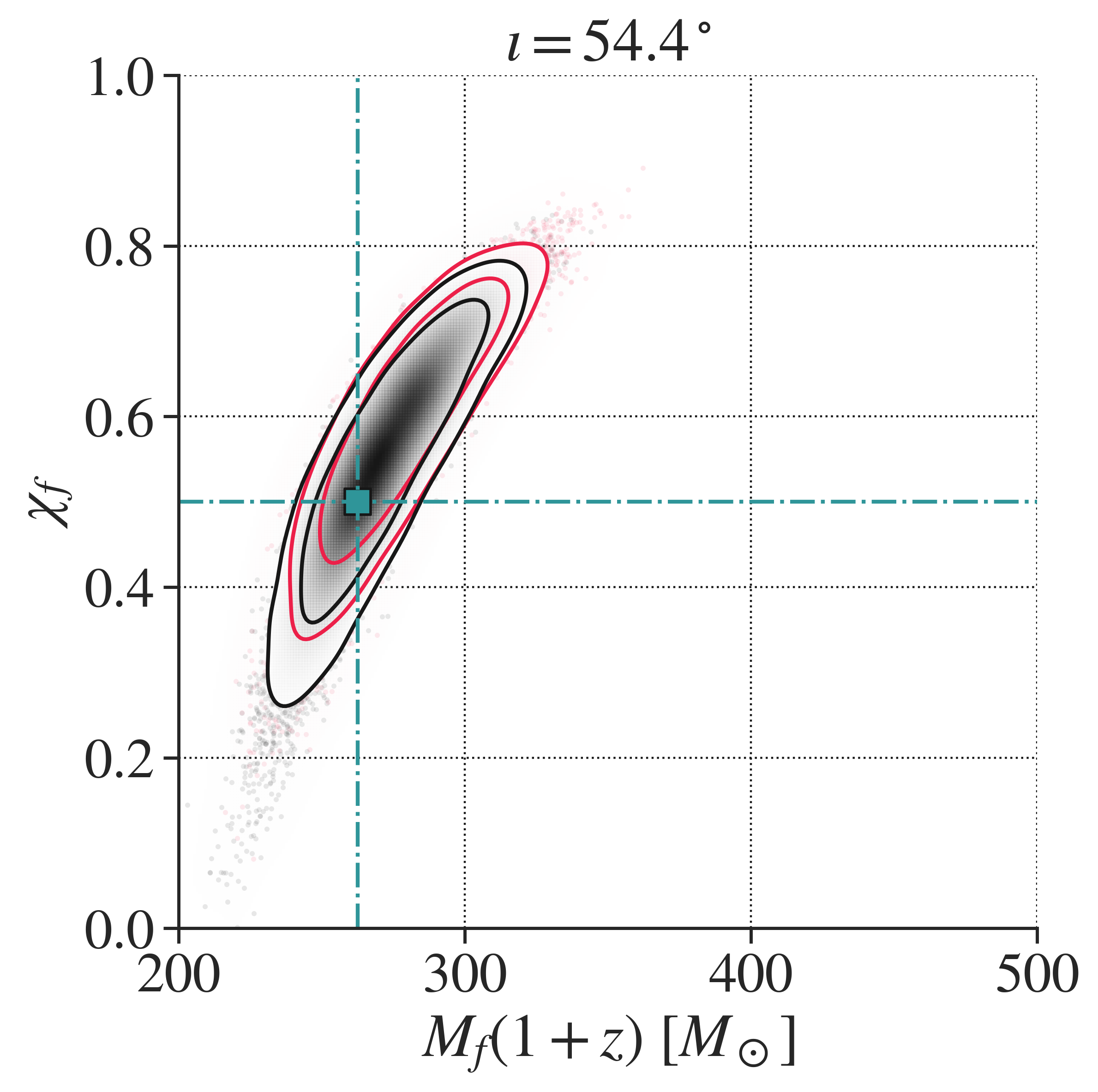}
        \includegraphics[width=0.32\textwidth]{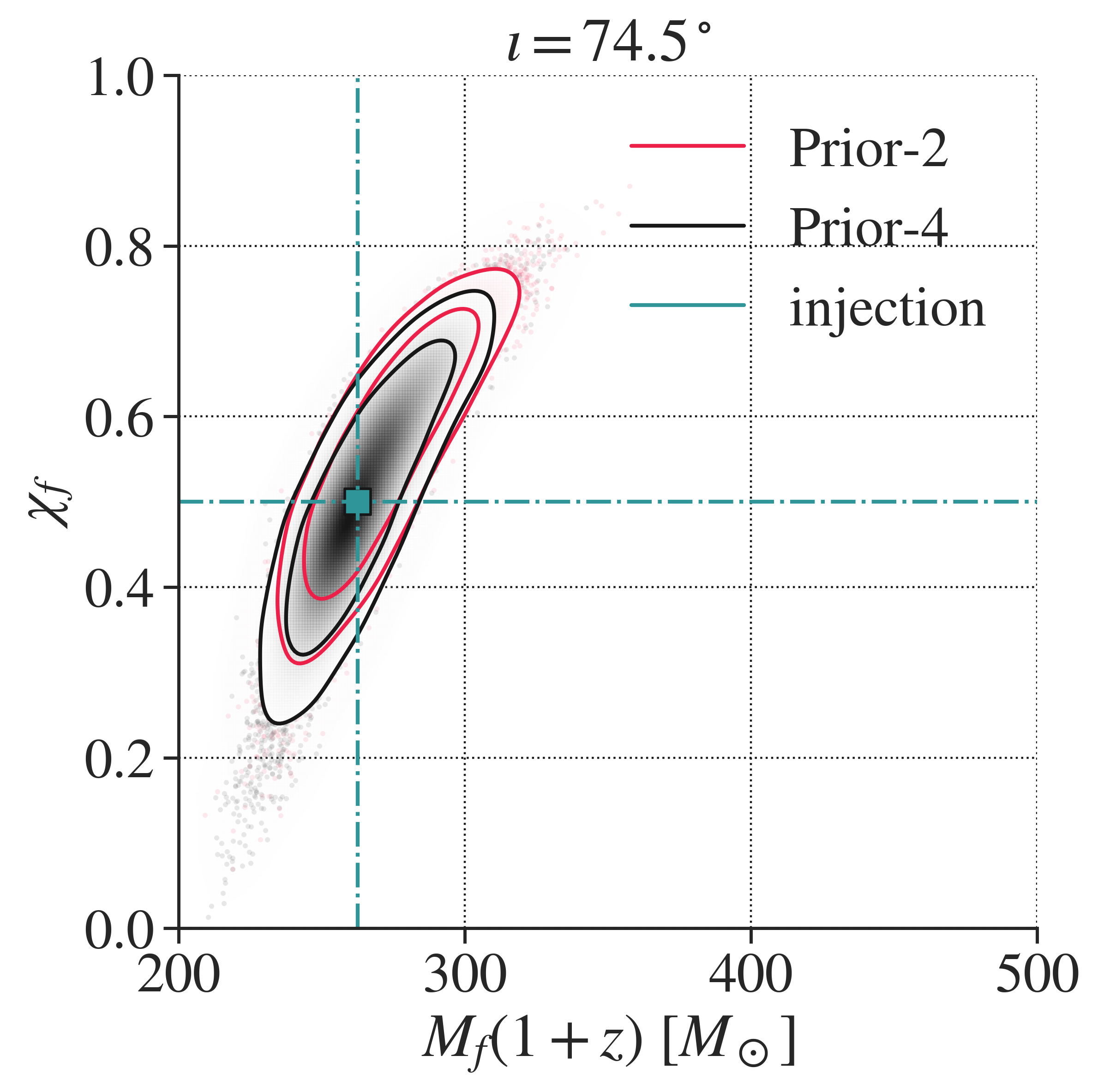}} \\
    \subfloat[For simulated systems with $Q=0.11$]{        
        \includegraphics[width=0.32\textwidth]{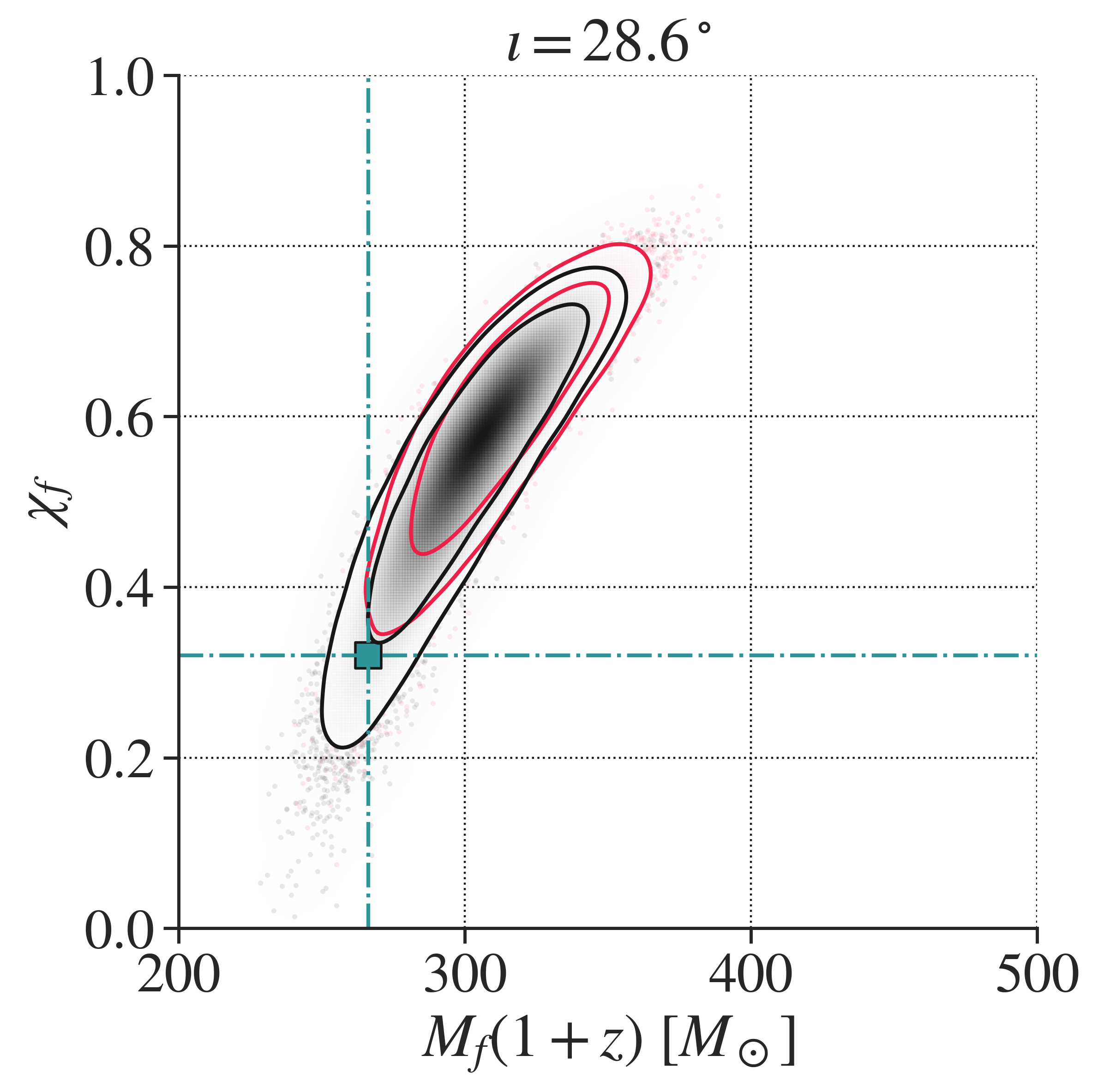}
        \includegraphics[width=0.32\textwidth]{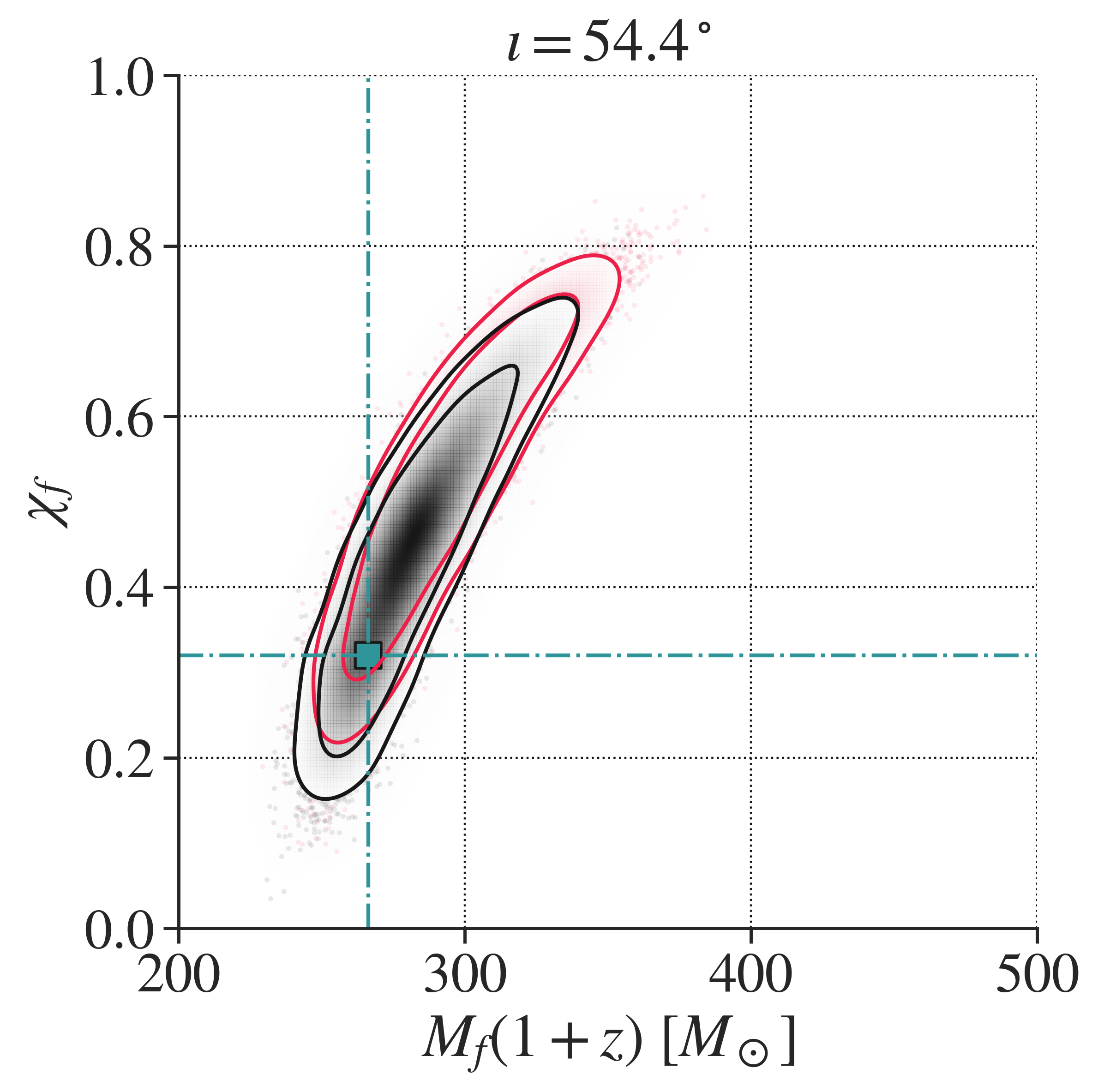}
        \includegraphics[width=0.32\textwidth]{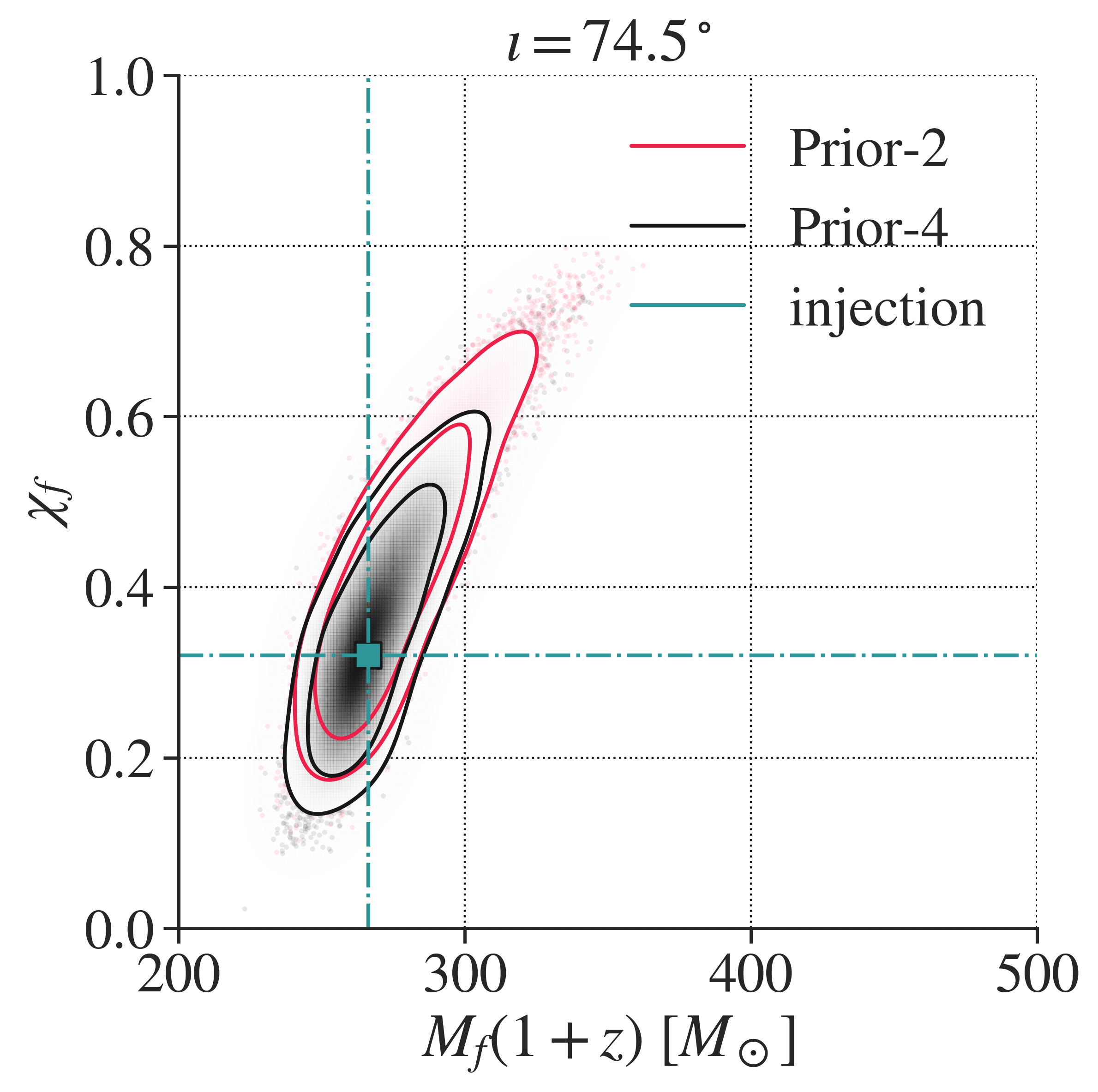}} \\
    \caption{The 2-dimensional posterior distribution for simulated systems' redshifted final mass and final spin.}
    \label{fig:hm-remnant}
    \end{center}
\end{figure*}

    \subsection{Impact of mass prior choices on remnant properties}

\begin{figure}
    \begin{center}
\includegraphics[width=\columnwidth]{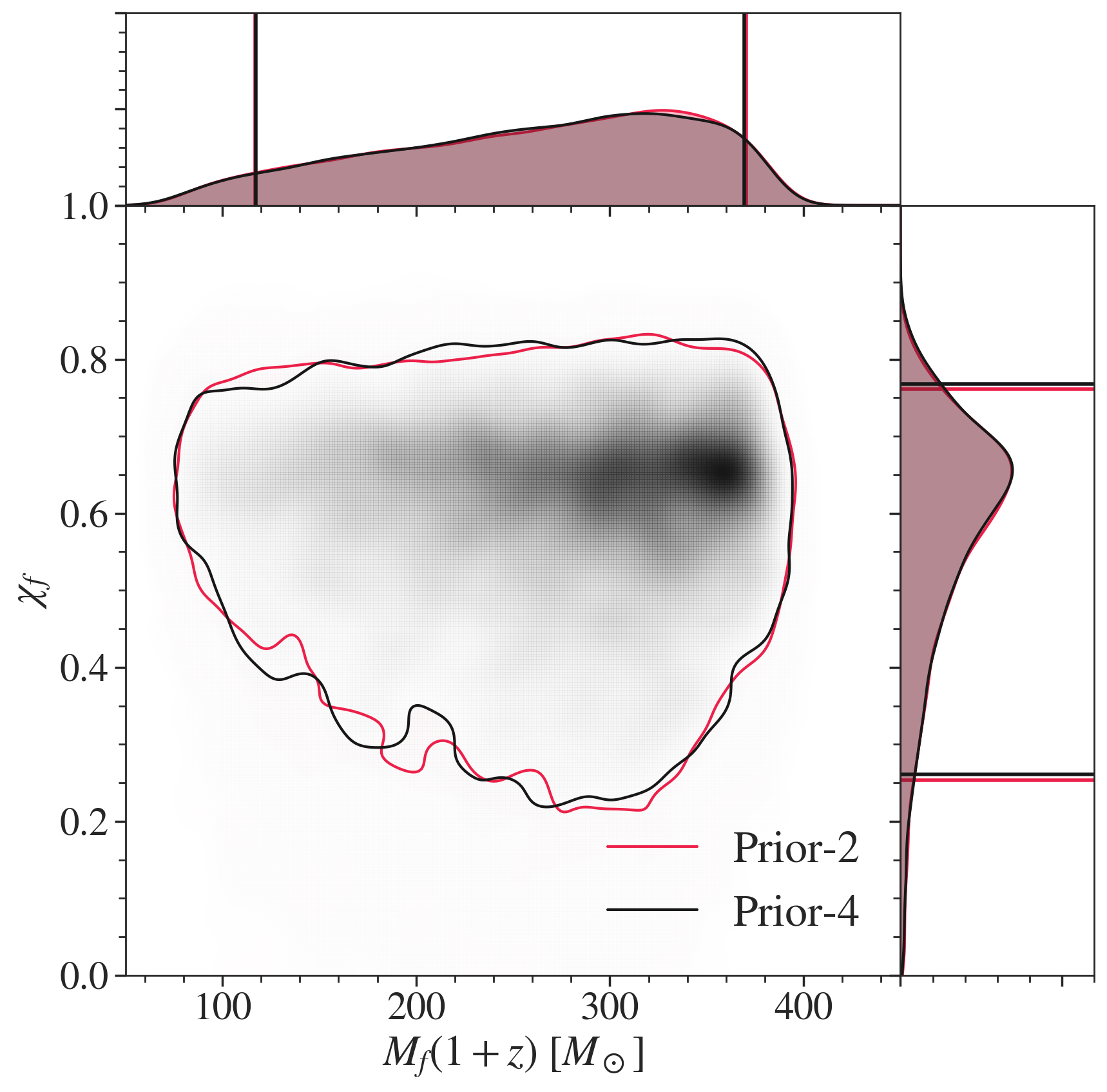}
    \caption{The figure compares the priors on the binary's redshifted final mass and final spin. Both the prior choices strongly prefer remnants with final spin $\sim 0.6-0.7$.}
    \label{fig:remnant-prior}
    \end{center}
\end{figure}

The properties and configuration of an inspiralling black hole binary fully specify the remnant's mass, spin and kick. Therefore, in practice, we need numerical relativity simulations to compute the remnant properties for arbitrary binary mass and spin sets. However in this article, we use surrogate fits from~\citet{Healy:2016lce, Hofmann:2016yih, Jimenez-Forteza:2016oae} to infer the final redshifted mass $M_f(1+z)$ and final spin $\chi_f$.

Fig~\ref{fig:hm-remnant} shows that as the binary's inclination increases and/or the mass ratio decreases, the uncertainties associated with the final spin estimates increase, resulting in broader posterior probability distributions. This is because both priors induce a final spin prior that strongly favours binary remnants with $\chi_f \sim 0.6-0.7$, as shown in Fig.~\ref{fig:remnant-prior}. Consequently, the resulting measurement uncertainty is lower when the true spin is within this range, making the likelihood peak near the bulk of the prior, as we observe in the case of the $Q=0.82$ binary system. In contrast, when the true final spin is within the edge of the prior support,  likelihood and prior compete, leading to broader posterior distributions, as in the case of the $Q=0.25$ and $Q=0.11$ binaries. This phenomenon is more pronounced under Prior-4, where a larger \ac{CI} is observed compared to Prior-2, particularly for systems with moderate to high mass ratios and higher inclination angles. The posterior support would progressively move towards the likelihood peak if the SNR of the injection is increased, similarly to what is described in \citet{Leong:2023nuk}. 

We also observe that the inferred posteriors peak closer to the true values when Prior-4 is employed. Additionally, for low $Q$ binaries, we observe that the true value falls outside the 68\% CI of the posteriors generated by Prior-2 as the sampler is unable to probe the higher likelihood regions efficiently.

\begin{figure*}[htb]
    \begin{center}
        \includegraphics[width=0.32\textwidth]{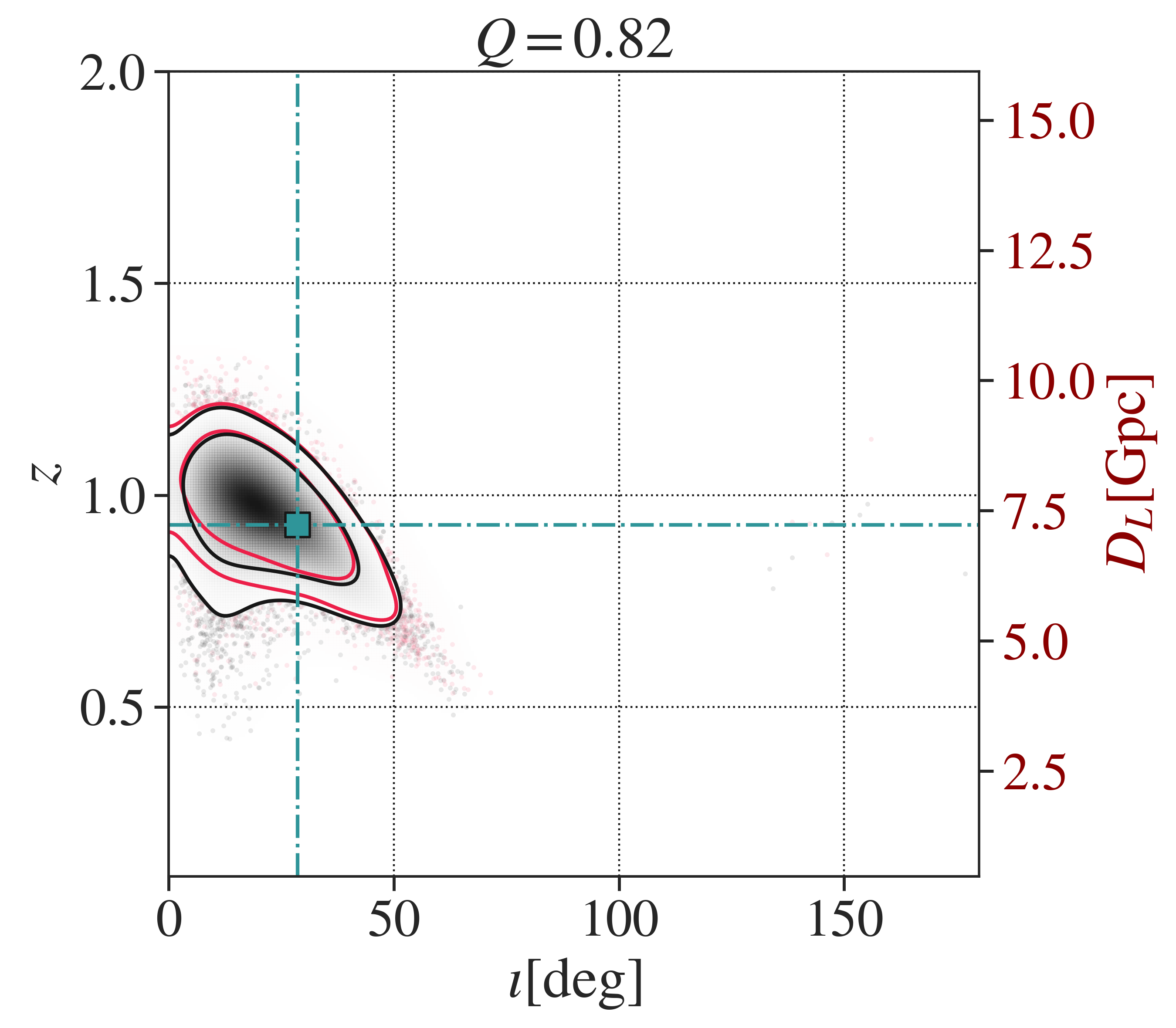}
        \includegraphics[width=0.32\textwidth]{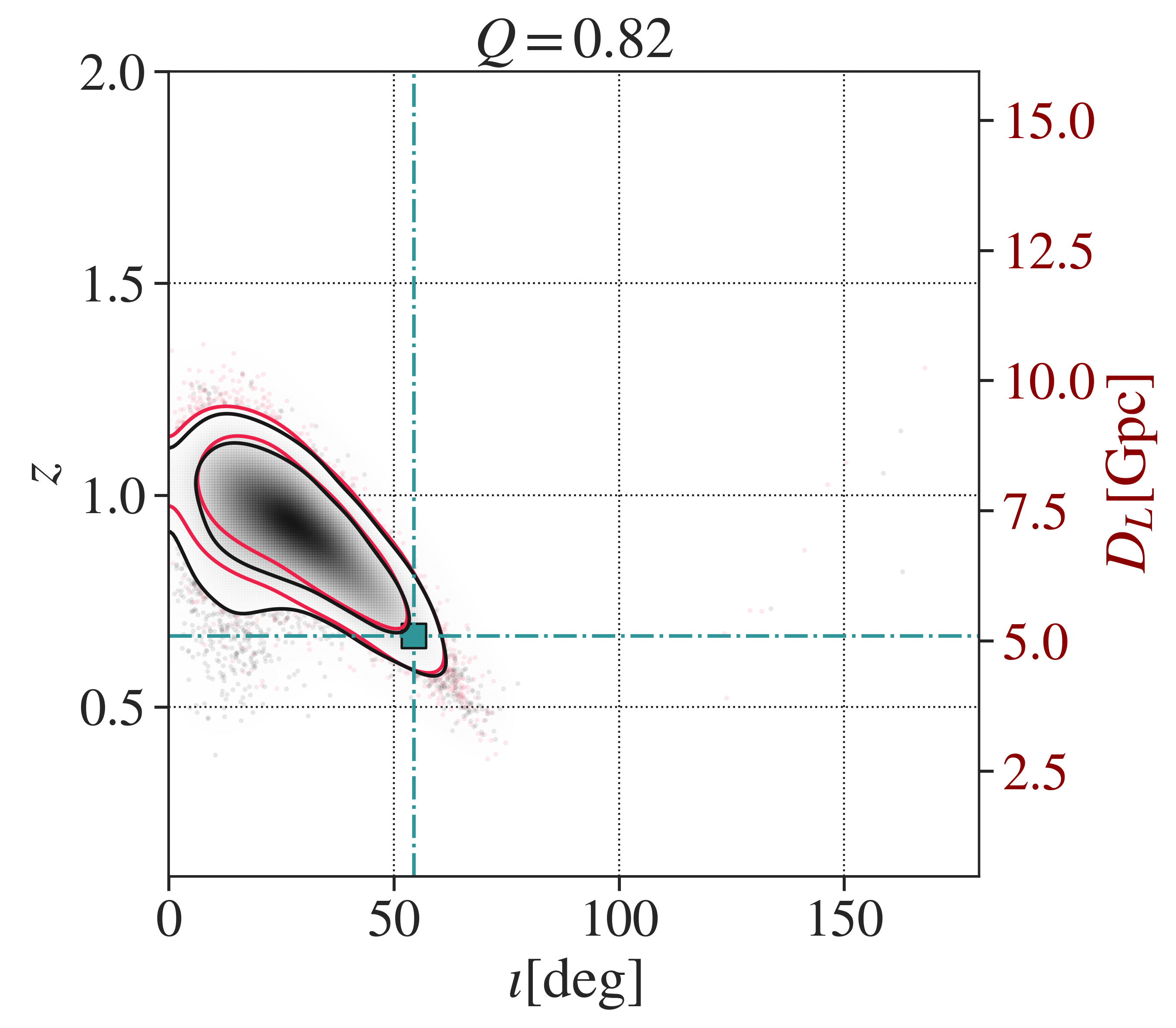}
        \includegraphics[width=0.32\textwidth]{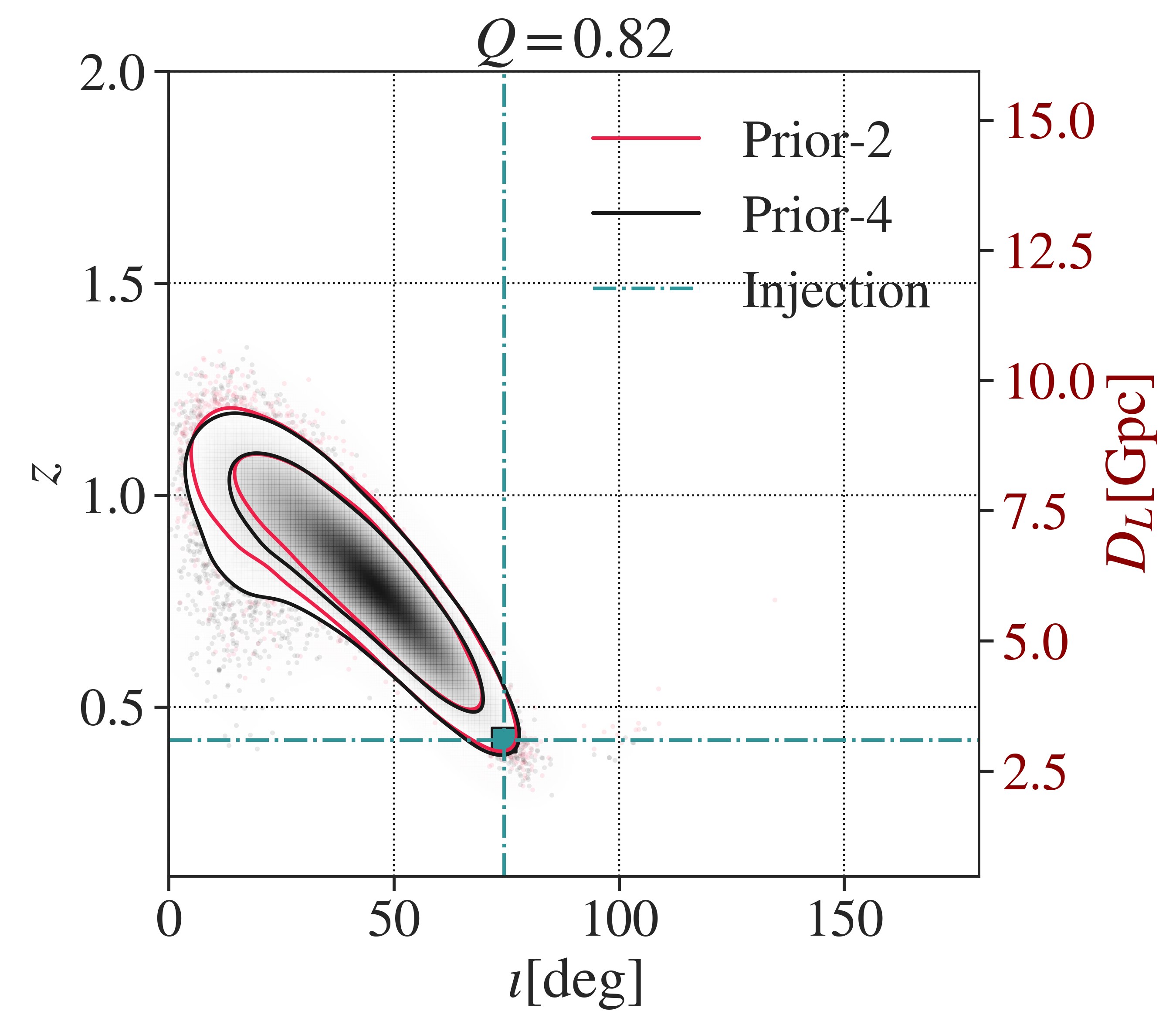} \\
        \includegraphics[width=0.32\textwidth]{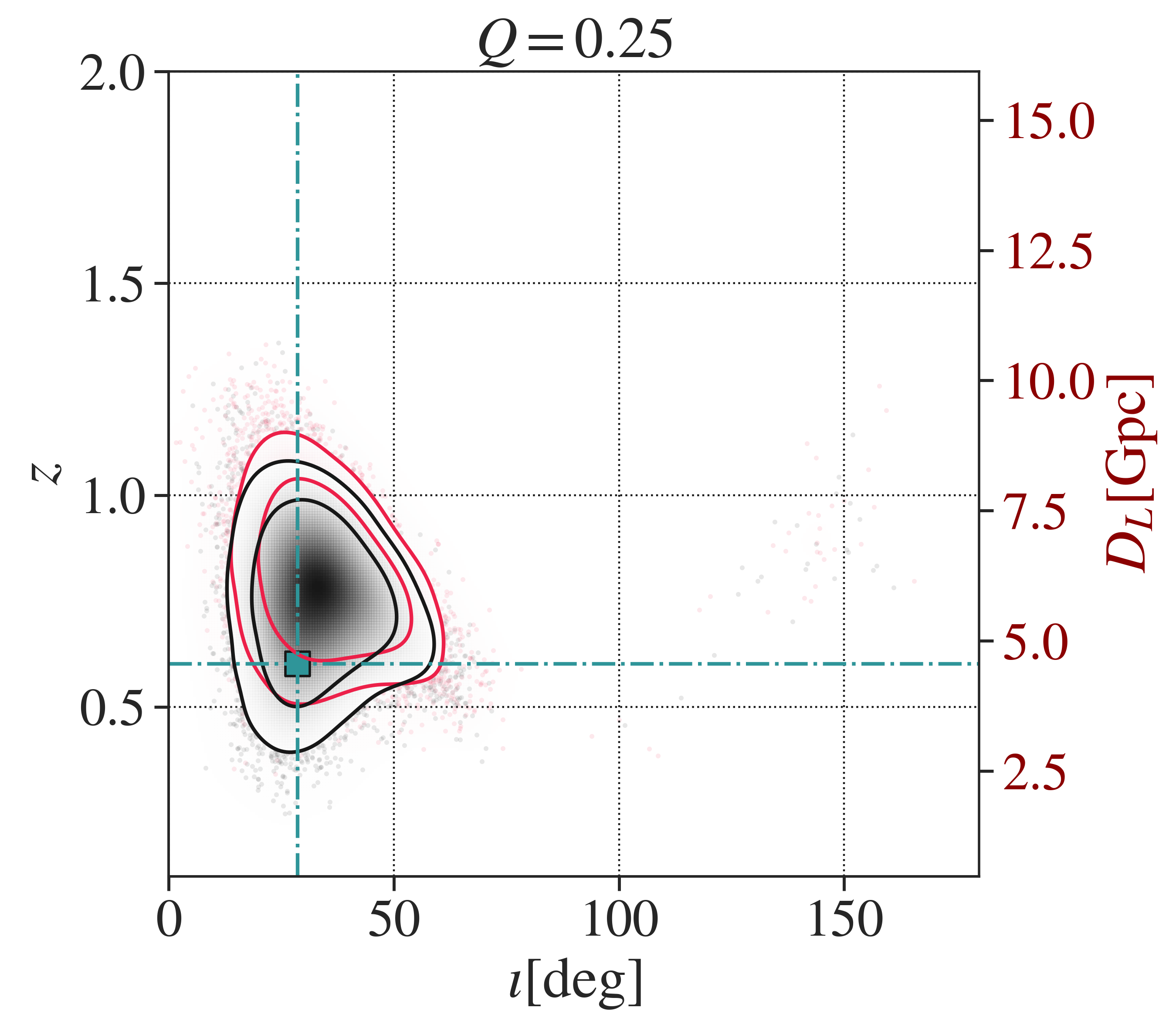}
        \includegraphics[width=0.32\textwidth]{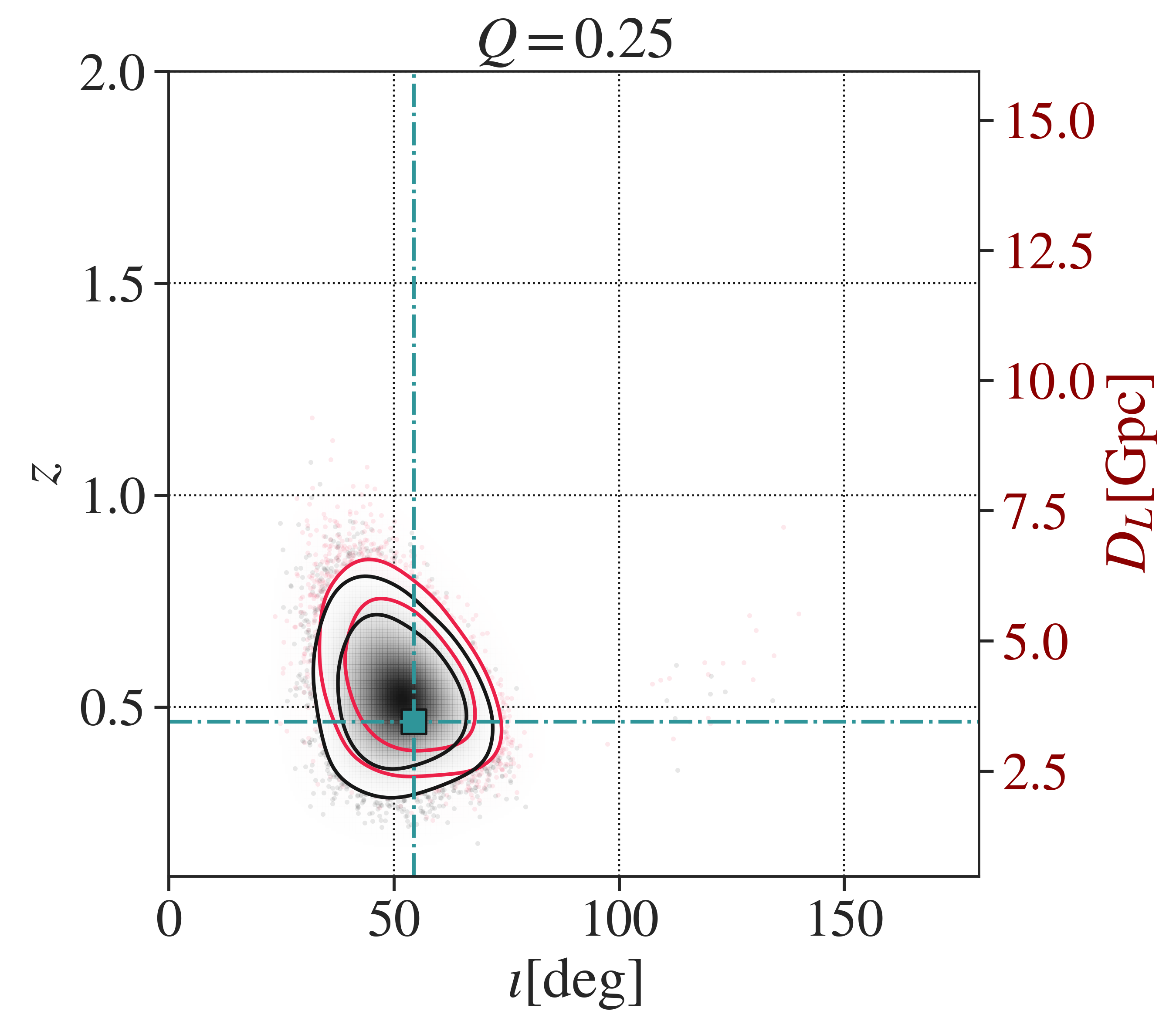}
        \includegraphics[width=0.32\textwidth]{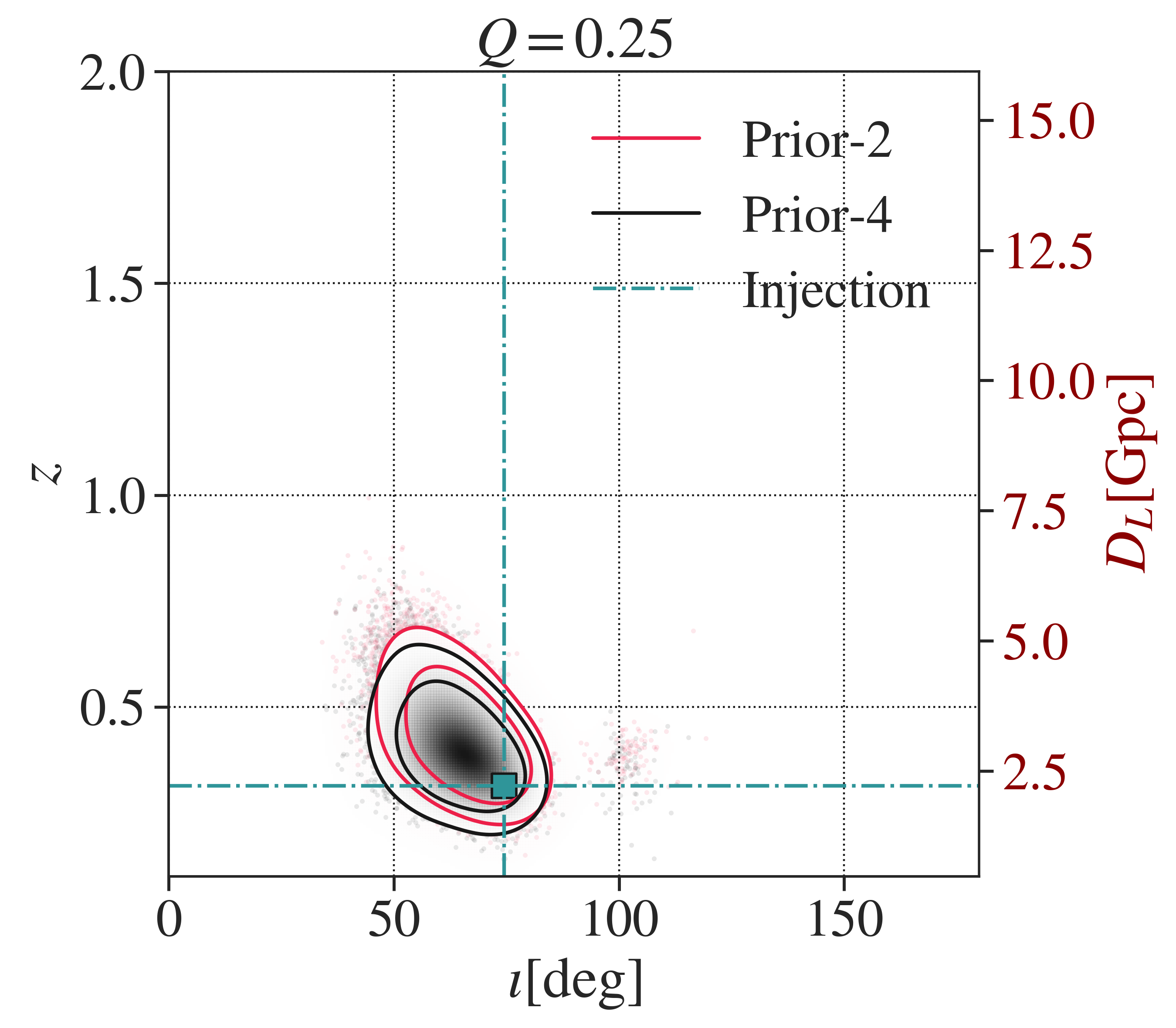} \\
        \includegraphics[width=0.32\textwidth]{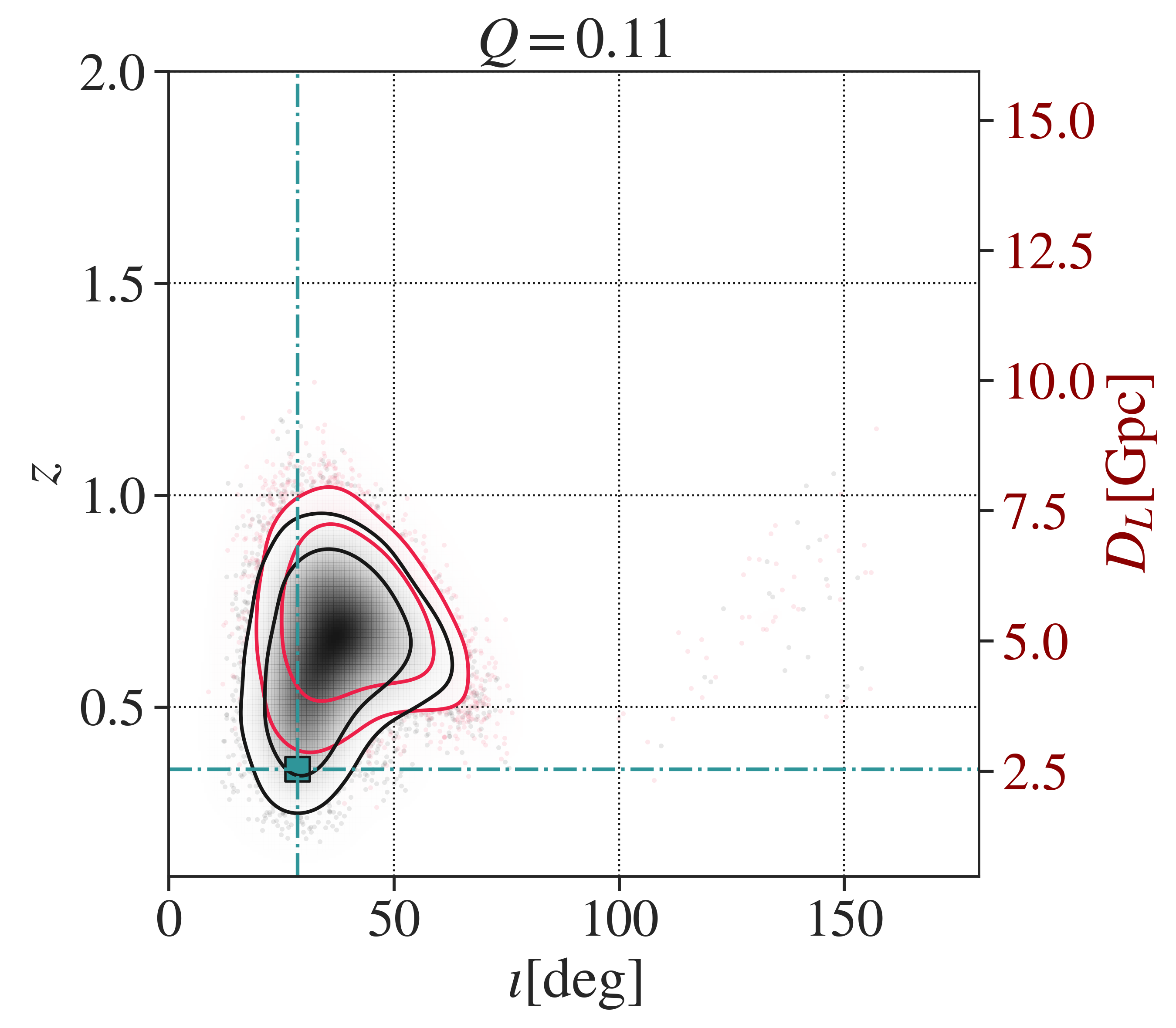}
        \includegraphics[width=0.32\textwidth]{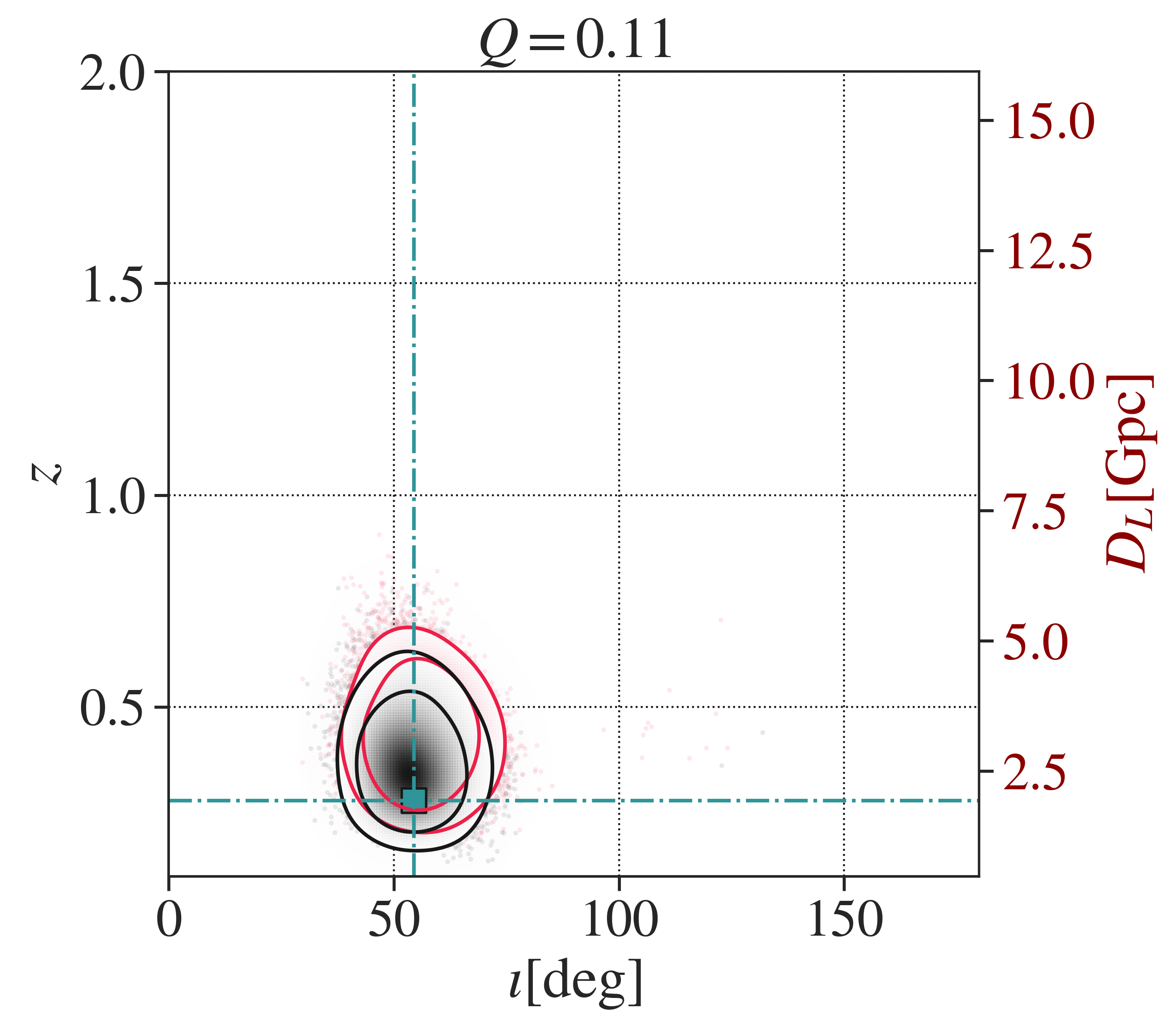}
        \includegraphics[width=0.32\textwidth]{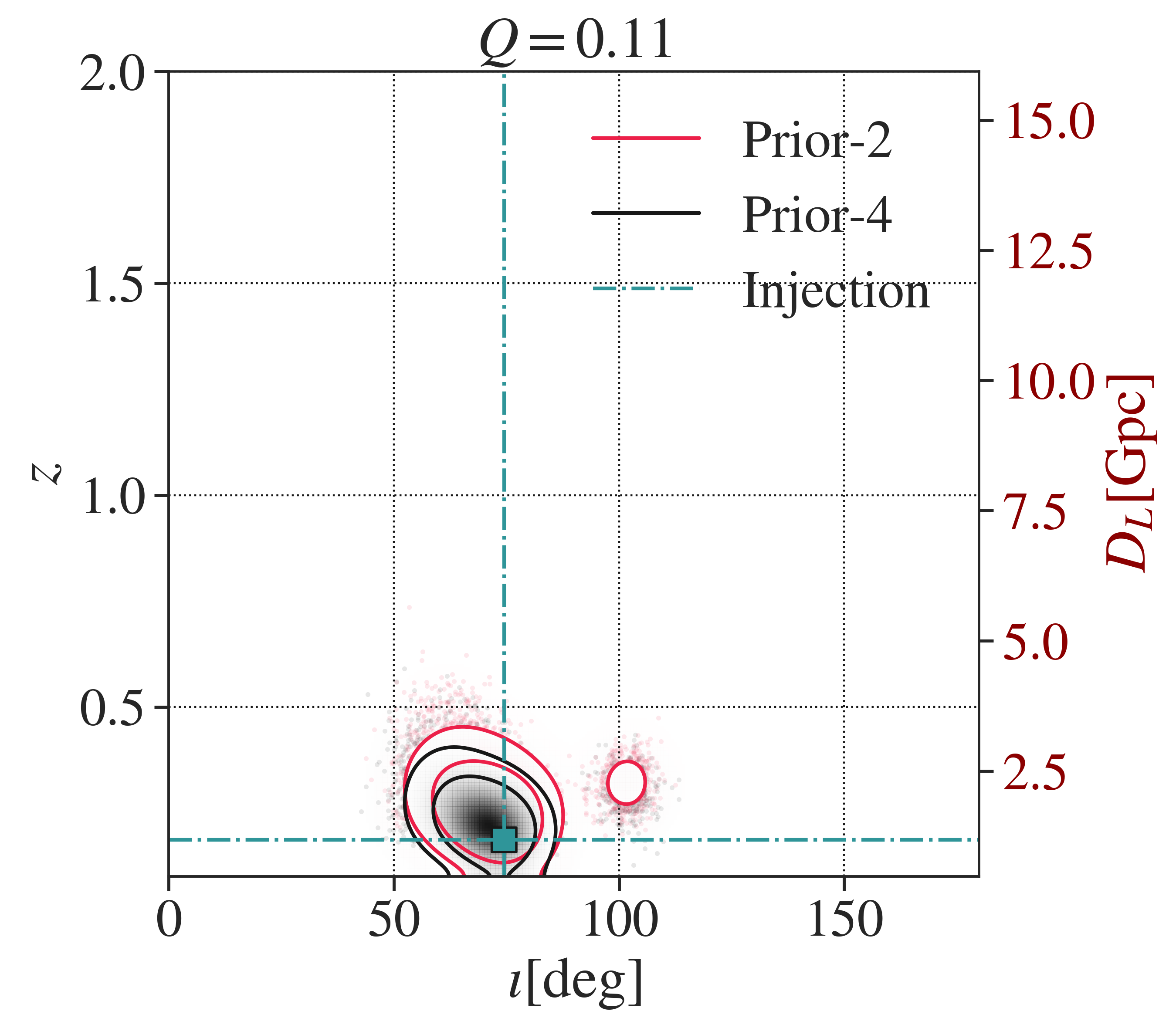}\\
    \caption{The 2-dimensional posterior distribution for the luminosity distance (redshift) and inclination of simulated systems summarised in Table~\ref{table:set-2}.}
    \label{fig:hm-extrinsic}
    \end{center}
\end{figure*}

    \subsection{Impact on inclination and redshift measurement}

\begin{table*}[htb]
    \centering
    \begin{tabular}{|c|c|c|c|c|c|c|c|c|c|}
    \hline \multirow{2}{*}{} & \multicolumn{4}{c|}{Prior-2} & \multicolumn{4}{c|}{Prior-4} \\
    $\iota$ & $m_1~[M_\odot]$ & $m_2~[M_\odot]$ & $\mathbb{P}(m_1 \in \mathrm{PI})$ & $\mathbb{P}(m_2 \in \mathrm{PI})$ & $m_1~[M_\odot]$ & $m_2~[M_\odot]$ & $\mathbb{P}(m_1 \in \mathrm{PI})$ & $\mathbb{P}(m_2 \in \mathrm{PI})$  \\
    \hline
    & & & & & & & & \\[1pt]
    $28.6^\circ$ & $75.9^{+12.3}_{-9.0}$ & $63.6^{+12.2}_{-14.7}$ & 0.99 & 0.43 & $76.6^{+14.4}_{-9.5}$ & $61.7^{+12.9}_{-18.0}$ & 0.99 & 0.34 \\[10pt]
    $54.4^\circ$ & $81.7^{+13.2}_{-11.7}$ & $69.9^{+12.4}_{-15.7}$ & 1.0 & 0.73 & $81.5^{+13.7}_{-11.7}$ & $69.0^{+12.8}_{-18.1}$ & 0.99 & 0.68 \\[10pt]
    $74.5^\circ$  & $84.3^{+14.1}_{-12.7}$ & $71.5^{+13.0}_{-14.9}$ & 1.0 & 0.78 & $88.0^{+13.8}_{-13.2}$ & $73.0^{+13.3}_{-16.0}$ & 1.0 & 0.81 \\
    \hline
    \end{tabular}
    \caption{Component masses and their probability of being within the pair-instability (PI gap) of multipole simulations with $Q=0.82$. We report the median values of the masses with their 90\% symmetric credible intervals.}
    \label{Table:components}
\end{table*}

To measure the source properties of a binary black hole and study its astrophysical implications, it is necessary to determine its source frame masses. Conventionally, we rely on standard cosmology and the gravitational wave measurement of the luminosity distance to estimate the cosmological redshift. This redshift estimate is subsequently used to convert detector-frame masses to source-frame masses. Given this, it is critical to evaluate the impact of our chosen mass priors on the measurement of luminosity distance or, equivalently, redshift. This section presents the results obtained using simulation set-2 as the inclination angle $\iota$ and the distance $D_L$ are degenerate for quadrupole signals. Further, we only discuss the results obtained using Prior-2 and Prior-4 as the results obtained using Prior-1, Prior-2, and Prior-3 exhibit comparable characteristics.

Figure~\ref{fig:hm-extrinsic} displays the two-dimensional posteriors of redshift and inclination for each synthetic binary. We find at $Q=0.82$ and $Q=0.25$, the redshift posterior of low and moderate inclination binaries peaks closer to the true value of the redshift, irrespective of prior choice. However, when the binary inclination becomes large, the true redshift value is relatively poorly recovered for either prior choice.

A subtle disparity emerges for low inclination binaries with $Q=0.11$. Specifically, the true value lies outside the 90\% \ac{CI} when using Prior-2, yet it is within the 68\% \ac{CI} when Prior-4 is used. However, as the binary inclination increases, the measurement bias decreases. We also observe that except for the $Q=0.82$ system at moderate and high inclination, the true inclination angle for most other binaries falls within 68\% CI for both prior choices. Additionally, consistent with other studies~\citep{Graff:2015bba, Usman:2018imj, Shaik:2019dym, Kalaghatgi:2019log}, the correlation between distance and inclination angle weakens as the relative strength of higher harmonics increases with increasing inclination or decreasing mass ratio. We leave the detailed investigation to future work.

\subsection{Pair-instability mass gap}

Stellar physics predicts the existence of a black hole mass gap with no first-generational black holes in the mass range between $65M_\odot$ and $120M_\odot$ owing to the development of pair-instability in massive stars with helium core masses in this range~\citep{Fowler:1964zz, Barkat:1967zz, Bond:1984sn, Woosley:2021xba}. While the precise location of the pair-instability mass gap remains theoretically uncertain~(See for e.g.~\citep{Farmer:2019jed, Fishbach:2020qag} and references therein), we assume that the lower boundary is at $65M_\odot$ and the upper boundary is at $120M_\odot$. The observation of GW190521 by the LVK~\citep{LIGOScientific:2020iuh} challenged this prediction, as the largest merging black hole squarely fits within the mass gap. Given the strong impact that priors can have on mass estimates, it is relevant to ask whether prior choices could induce such observation. We test this idea here by checking if any of our priors falsely places our component masses inside or outside the mass gap. For instance,~\citet{CalderonBustillo:2020xms} showed that such an effect can happen when highly massive and highly eccentric sources are mistakenly taken as precessing  ones.\\ 

Table~\ref{Table:components} shows the inferred source-frame component masses ($m_1,~m_2$) of systems with $Q=0.82$. For increasing inclination, these simulations have $m_1$ of $\sim (75, 85, 105) M_\odot$ and $m_2$ of $\sim (63, 73, 85)M_\odot$. Therefore, except for the lowest inclination, both components lay in the pair-instability mass gap.  

We find that for the lowest inclination, $\iota=28.6^\circ$, Prior-4 places the secondary component outside the mass gap with a slightly higher probability than Prior-2, making it a better choice. Similarly, Prior-4 puts the secondary black hole within the mass gap at the highest inclination with somewhat higher confidence, consistent with the true values. However, at a mid inclination, Prior-2 does a moderately better job. This shows that while prior choices can mildly affect the astrophysical interpretation of our injected sources, none of them lead to false conclusive results. For our remaining injections, for which the true masses straddle the mass gap, both priors place the masses outside the gap with probabilities $\gtrsim 0.97$. 

    \subsection{Model Selection}

\begin{table}[h]
    \centering
    \begin{tabular}{|c|c|c|c|c|c|}
    \hline \multirow{2}{*}{} & \multicolumn{2}{c|}{Prior-2} & \multicolumn{2}{c|}{Prior-4} \\
    & $\ln \mathrm{BF}^{S}_{N}$ & $\ln \mathcal{L}_\mathrm{max}$ & $\ln \mathrm{BF}^{S}_{N}$ & $\ln \mathcal{L}_\mathrm{max}$ \\
    \hline \hline
    $Q=0.82~\iota=28.6^\circ$ & 188.29 & 207.77 & 187.20 & 208.17 \\
    $Q=0.82~\iota=54.4^\circ$ & 186.63 & 207.62 & 185.26 & 207.63 \\
    $Q=0.82~\iota=74.5^\circ$ & 179.08 & 200.93 & 177.15 & 201.75 \\
    \hline
    $Q=0.25~\iota=28.6^\circ$ & 188.45 & 208.82 & 188.47 & 208.94 \\
    $Q=0.25~\iota=54.4^\circ$ & 187.08 & 208.95 & 187.04 & 208.88 \\
    $Q=0.25~\iota=74.5^\circ$ & 175.30 & 197.98 & 176.16 & 199.31 \\
    \hline 
    $Q=0.11~\iota=28.6^\circ$ & 184.76 & 206.12 & 184.85 & 207.14 \\
    $Q=0.11~\iota=54.4^\circ$ & 201.88 & 225.36 & 203.83 & 226.91 \\
    $Q=0.11~\iota=74.5^\circ$ & 196.52 & 221.15 & 198.94 & 223.15 \\
    \hline
    \end{tabular}
    \caption{$\ln \mathrm{BF}^{S}_{N}$ and $\ln \mathcal{L}_\mathrm{max}$ values of simulation set-2 for prior choices, 2 and 4.}
    \label{Table:bayes-factor}
\end{table}

Given gravitational wave data, the support for a given prior is given by the Bayes Factor. This is defined as the prior-averaged value of the log-likelihood $\ln \mathrm{BF}^{S}_{N}$ across the whole parameter space and, therefore, is bounded by the maximum log-likelihood $\ln \mathcal{L}_\mathrm{max}$. In principle, we expect the latter to peak near the true injection parameters, provided that the priors do not prevent a correct sampling of the parameter space. 

Table~\ref{Table:bayes-factor} shows the log Bayes factor for the signal versus noise model for our priors 2 and 4 within the context of the simulation set 2. For $Q=0.82$ systems and the two lowest inclinations, both priors provide similar $\ln \mathrm{BF}^{S}_{N}$ and  $\ln \mathcal{L}_\mathrm{max}$. For larger inclinations and the $Q=0.82$ case, however, we recall that Prior-2 fails to include the right luminosity distance in its $90\%$ \ac{CI} while Prior-4 does. Yet, Prior-2 yields a larger Bayes Factor while finding a worse fit (lower $\ln \mathcal{L}_\mathrm{max}$) to the injection. On the one hand, Prior-2 has very strong support for loud equal-mass sources that emit louder signals, which are, in turn, strongly favoured by the distance prior. Smaller inclinations are also preferred for the same reason. On the other hand, since Prior-2 does not prefer low-$Q$ signals, which generally leads to slightly weaker sources that are then punished by the luminosity closer to the true injection values and, therefore, a larger likelihood. However, this difference in likelihood is not enough to overcome the difference in prior support. 

The above situation changes when $Q$ is decreased, and the increasing contribution of higher-order modes can provide information about the inclination of the source. On the one hand, we note that for decreasing mass ratio, the true injection parameters are progressively displaced from the $90\%$ \ac{CI} of Prior-2, which prefers distant and louder mass-symmetric systems at smaller inclinations. This is particularly true for small inclinations, for which the signal contains weak higher harmonics. In contrast, Prior-4 always returns $90\%$ \ac{CI}, including the true values and larger likelihoods. Even in this situation, the Bayes Factors reveal no clear preference for any of the models, as the larger likelihood obtained by Prior-4 cannot compensate for the strong support that the distance prior shows for the mass ratios preferred by Prior-2. Unlike in the above paragraph, this does not happen for the largest inclinations, for which the higher-order harmonic modes present in the injections do provide strong information about the inclination angle. This way, the maximal likelihood obtained by Prior-4 \textit{is enough to overcome} the preference of Prior-2 for high-$Q$ systems, which are further favoured by the distance prior.\\

Summarising, we have explicitly shown here how a perfectly reasonable prior, a standard one in gravitational-wave data analysis, can lead to biased parameter estimates and still be statistically preferred with respect to a second prior giving unbiased estimates. 

    \section{Summary}
    \label{sec:summary}

This study investigates the influence of various mass prior choices on Bayesian parameter inference of quasi-circular intermediate-mass black hole binaries. We employ two waveform models, IMRPhenomXAS and IMRPhenomXHM, to simulate and infer binary parameters, focusing on their detector-frame total mass and mass ratio. Our results reveal that the choice of mass prior has a substantial influence on the recovery of parameters for systems with a network optimal signal-to-noise ratio of 20. Interestingly, changing the network optimal SNR of the systems to 15 or 30 does not alter the qualitative nature of our results.

When using the quadrupole-only waveform IMRPhenomXAS, we find that Prior-4, which is flat in $M_T(1+z)$ and $1/Q$, recovers the injected mass parameters better when the simulated signal corresponds to a short-lived mass-asymmetric binary. On the other hand, Prior-1, 2 and 3, which are flat in mass ratio, perform relatively better when recovering nearly equal mass binaries with component spins aligned relative to each other and the orbital angular momentum.

We have also shown that when recovering IMRPhenomXHM injections with the same waveform model, Prior-4 yields better recovery of $M_T(1+z)$ and $Q$, especially when the system is mass-asymmetric and is at a low inclination. However, as the relative strength of the higher-order modes increases with inclination or mass asymmetry, the accuracy of inference with conventional prior choices, such as Prior-2, which is flat in component masses, is better. Moreover, the correlation between the mass ratio and detector-frame total mass decreases with increasing higher-order mode content.

Lastly, we highlight a noteworthy observation: the redshift posterior peaks closer to the true value when using Prior-4, irrespective of the binary configuration used in the study, implying that this prior provides a more robust choice. Since accurate luminosity distance and/or redshift inference can impact efforts to measure source properties of black holes, tests of general relativity, and the study of secondary parameters, such as the remnant's recoil kick, it is recommended to use Prior-4 while analysing such short-lived signals. While an exhaustive exploration of distance measurability is beyond the scope of our current study, its importance beckons for future investigations. 

In conclusion, our findings suggest that it is essential to consider multiple mass prior choices when analysing intermediate-mass black hole binary events rather than relying on a single prior. Model selection techniques, such as the Bayes Factor, can then be used with maximum likelihood to evaluate which prior better explains the observation. While previous works have given indications of this either when analysing real events~\citep{Estelles:2021jnz, Olsen:2021qin, bustillo2022searching}, or when using a broader set of injections~\citep{Leong:2023nuk}, we explicitly demonstrate this in a controlled setup.

\section*{Acknowledgements}
\label{sec:acknowledgements}
%
We thank Simon Stevenson, Thomas Dent and Johann Fernandes for their detailed comments and valuable suggestions. KC acknowledges the MHRD, the Government of India, for the fellowship support. AP's research is supported by SERB-Power fellowship grant SPF/2021/000036, DST, India. JCB is funded by a fellowship from the ``la Caixa'' Foundation (ID100010434) and the European Union’s Horizon 2020 research and innovation programme under the Marie Skłodowska-Curie grant agreement No 847648. The fellowship code is LCF/BQ/PI20/11760016. JCB is also supported by the research grant PID2020-118635GB-I00 from the Spain-Ministerio de Ciencia e Innovaci\'{o}n. JAF is supported by the Spanish Agencia Estatal de Investigaci\'on  (PGC2018-095984-B-I00) and by the  Generalitat  Valenciana  (PROMETEO/2019/071).  The authors are grateful for the computational resources provided by the LIGO Laboratory and supported by the National Science Foundation Grants No. PHY-0757058 and No. PHY-0823459. We are grateful for the computational resources provided by Cardiff University and funded by an STFC grant supporting UK Involvement in the Operation of Advanced LIGO. This document has been assigned the LIGO document number LIGO-P2300217.

\bibliography{apssamp}

\begin{figure}[htb]
    \centering
    \includegraphics[width=0.47\textwidth]{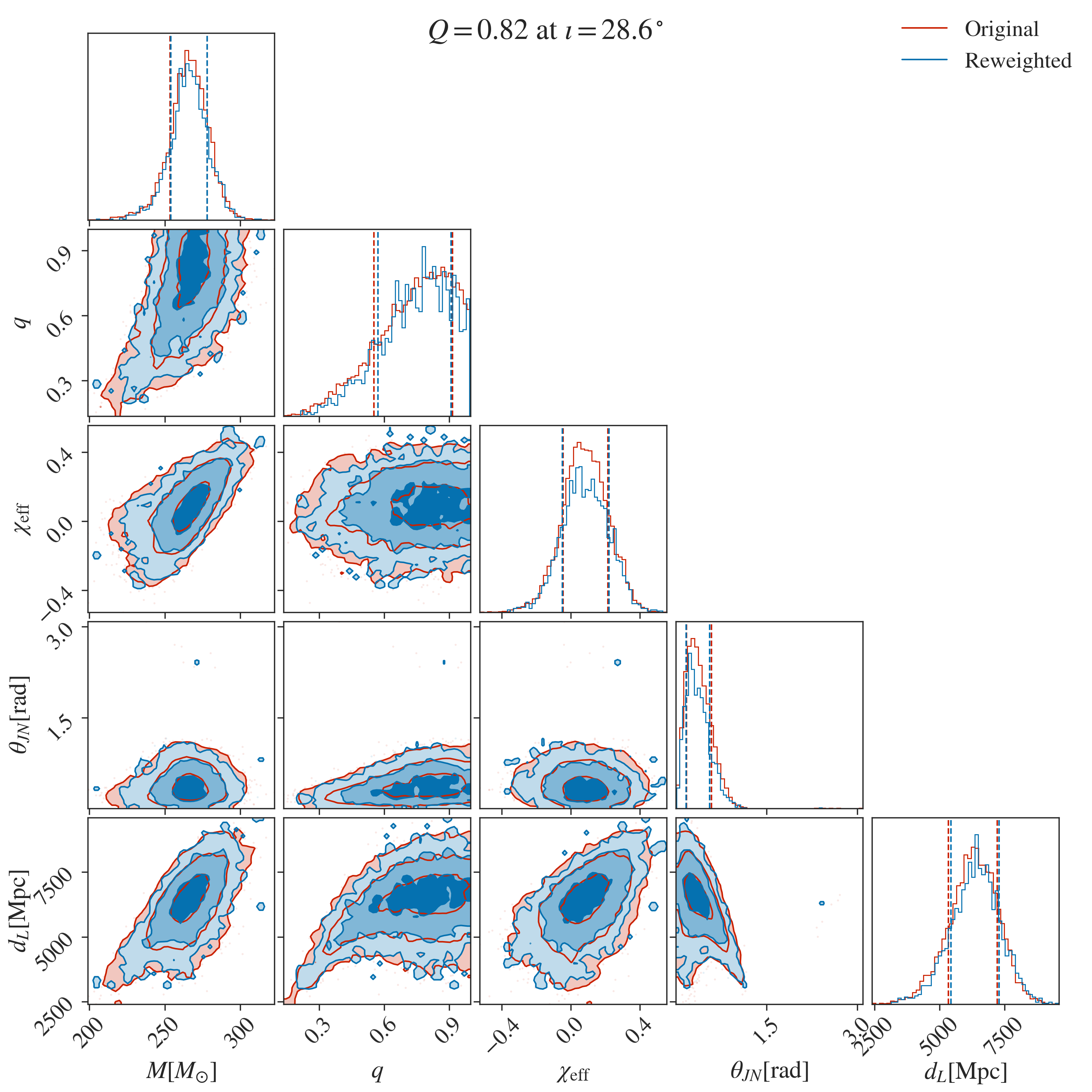}
    \includegraphics[width=0.47\textwidth]{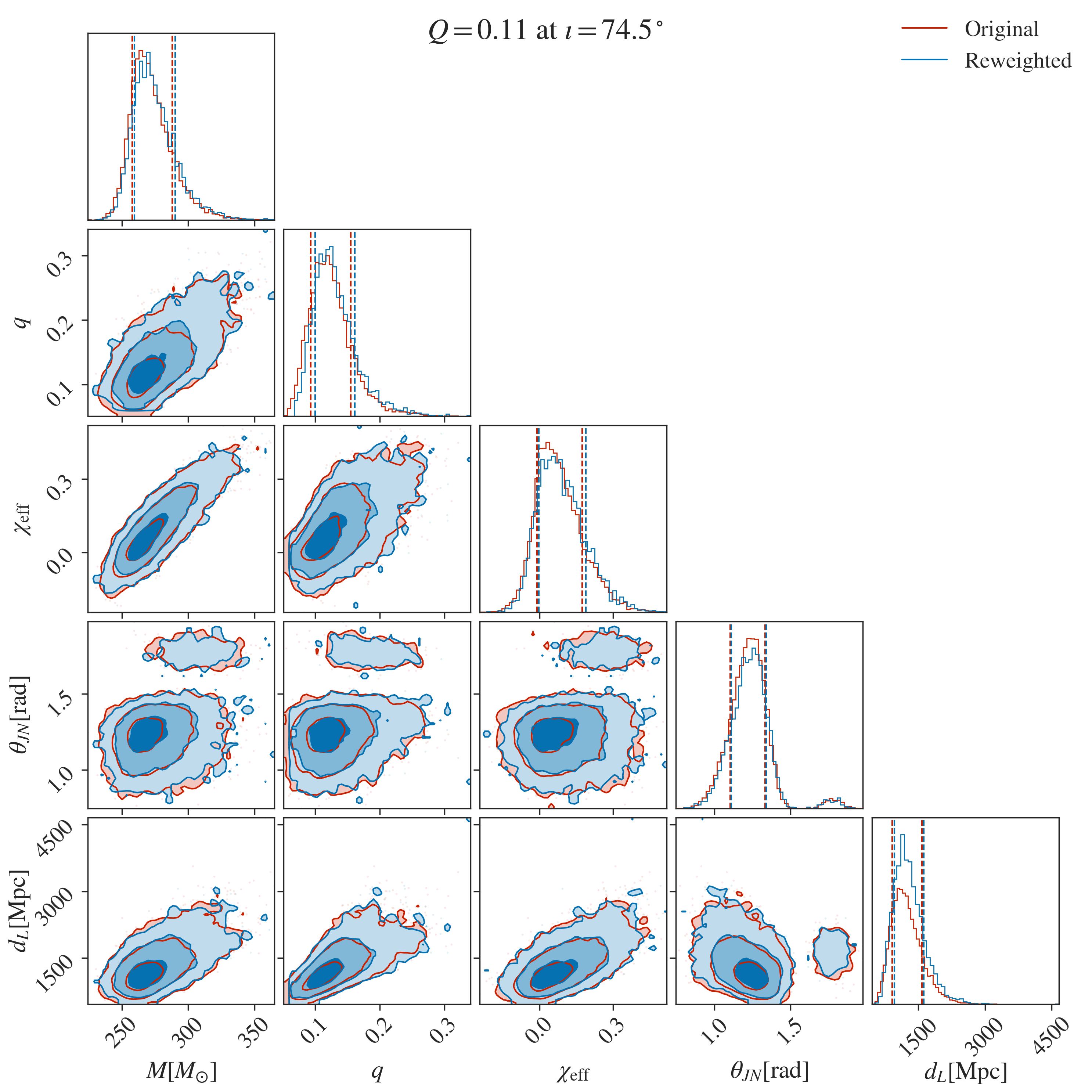}
    \caption{Comparison corner plot for original (samples obtained by directly sampling from Prior-4) and reweighted posterior (samples obtained by reweighing posterior obtained using Prior-2 with Prior-4 [See Eq.~\eqref{eq:prior-reweighing}]).}
    \label{fig:Prior-reweighing}
\end{figure}
\appendix
\section*{Appendix I: Prior Reweighting}\label{appx:reweigh}

\begin{figure}[htb]
    \centering
\includegraphics[width=0.9\columnwidth]{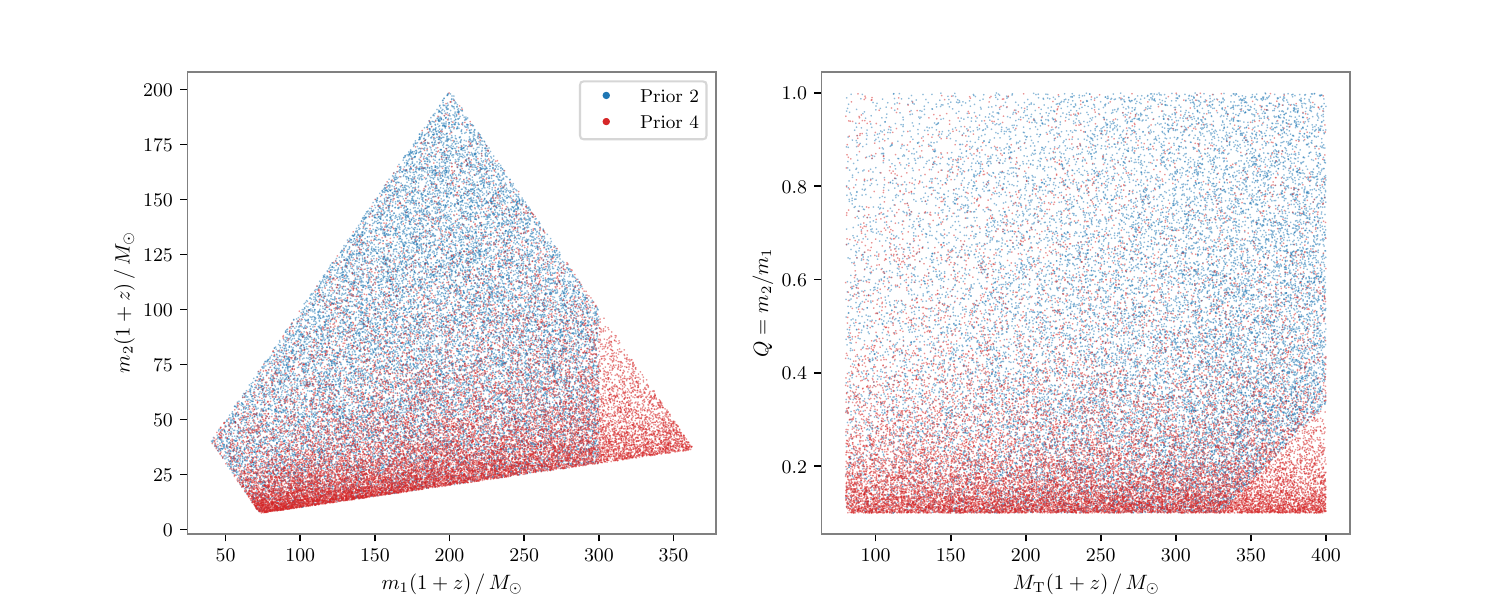}
    \caption{Support regions of the two priors, generated from 20000 sample points from each prior. The blue region (Prior 2) does not extend as much into the high $M_{\rm tot}$ at low $Q$ due to the constraint of $m_{1,2} < 300\,M_\odot$.}
    \label{fig:Prior24-support}
\end{figure}

Fig~\ref{fig:Prior-reweighing} compare the original and reweighted posterior probability distributions of two non-precessing simulations --- [left] $Q=0.82$ system at $\iota=28.6^\circ$ and [right] $Q=0.11$ system at $\iota=74.5^\circ$. The original corresponds to those obtained using Prior-2, while the reweighted distribution is obtained as follows:
\begin{equation}\label{eq:prior-reweighing}
    \bar{p}_4(\boldsymbol{\theta} \mid d) = w(\boldsymbol{\theta}) \cdot p_2(\boldsymbol{\theta} \mid d).
\end{equation}
Here, $p_2(\boldsymbol{\theta} \mid d)$ represents the original posterior probability distribution, and $w(\boldsymbol{\theta})=\pi_4(\boldsymbol{\theta})/\pi_2(\boldsymbol{\theta})$ denotes the weight, with $\pi_4(\boldsymbol{\theta})$ and $\pi_2(\boldsymbol{\theta})$ corresponding to Prior-4 and Prior-2, respectively. Computationally, to calculate $w(\boldsymbol{\theta}_k)$, we evaluate $\pi_4(\boldsymbol{\theta}_k)$ and $\pi_2(\boldsymbol{\theta}_k)$ for each posterior sample $\boldsymbol{\theta}_k \in p_2(\boldsymbol{\theta} \mid d)$ rather than using a Jacobian because the prior support between the two priors, as depicted in Fig.~\ref{fig:Prior24-support}, is different.

Visual inspection reveals a \textit{broad} similarity between the two distributions, consistent with the low JSD (< 0.001) obtained between the two distributions. This shows that it is possible to recompute the posterior using a new prior without stochastic sampling. However, it is worth mentioning that the quality of reweighed samples depends heavily on the posterior samples obtained under the original prior (in this case, Prior-2). The reweighed samples may fail to capture the original posterior distribution if they are sparse or not representative. We prevent this in the main text by sampling under the new prior.

\begin{figure}[htb]
    \centering
\includegraphics[width=0.48\textwidth]{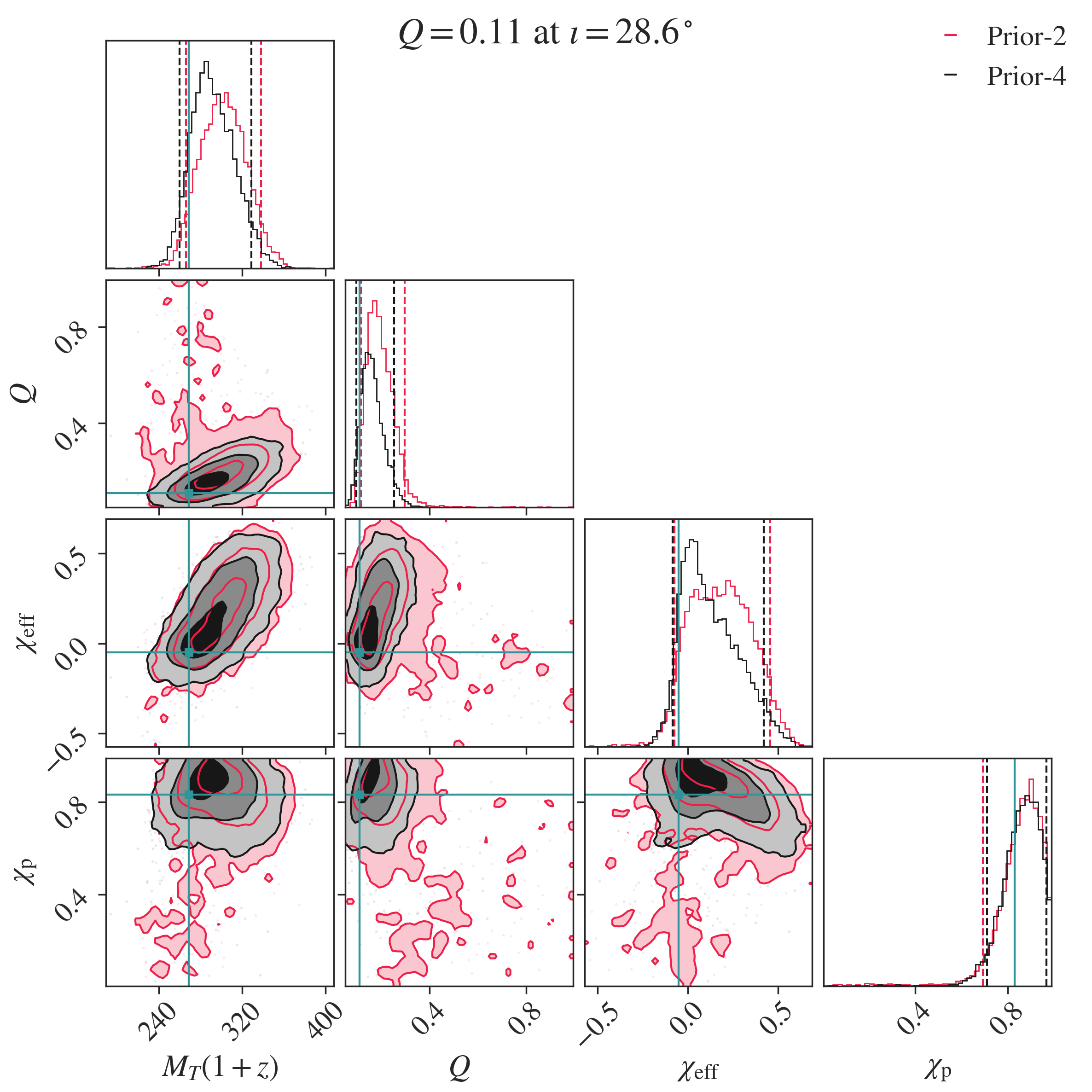}        \includegraphics[width=0.48\textwidth]{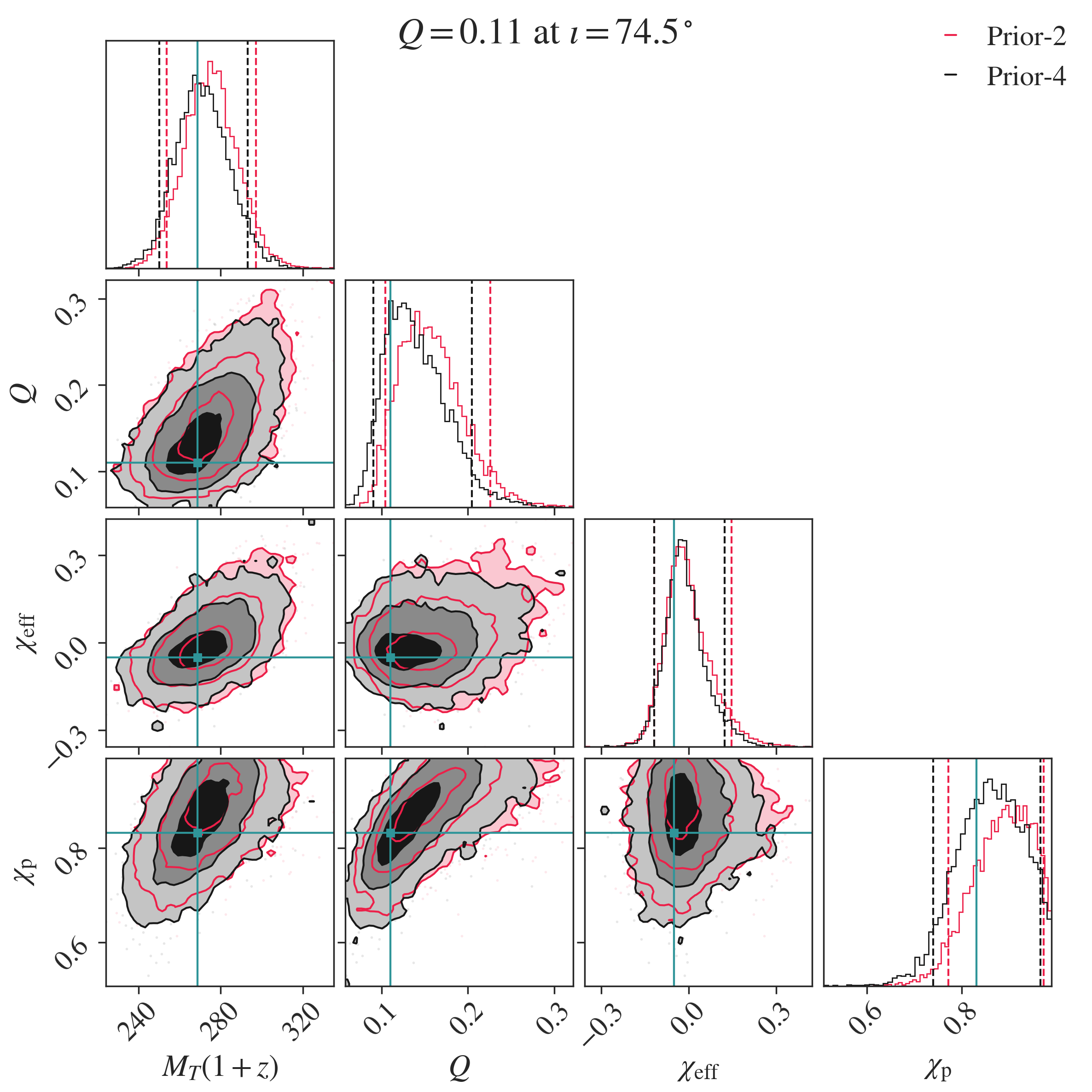} 
    \caption{Comparison corner plot for the two simulated systems at network SNR of 20.}
    \label{fig:corner}
\end{figure}

\section*{Appendix II: Quasi-spherical black hole binaries}\label{appx:quasi-spherical}

While the effect of the masses can be quite degenerate with that of the effective-spin (present already in alighed-spin systems), this is not the case for orbital precession. The reason is that the main impact of precession is an amplitude modulation of the waveform as a function of time. Therefore, the type of impact of the mass priors we describe, must be weakly dependent on whether or not we include precession in the analysis.  To give evidence of the absence we show results for two experiments where we include orbital precession. The binary parameters are fixed at the following values:

\begin{table}[h]
\centering
\caption{Maximum likelihood parameter values used for the simulated system drawn from the GW190521 posterior as given in~\citet{islam2023analysis}.}
\label{table:1}
\begin{tabular}{ll}
\hline 
Parameter & Value \\
\hline 
Mass ratio & 0.11 \\
Detector-frame total mass & $268.83 M_{\odot}$ \\
Primary spin magnitude & 0.83 \\
Secondary spin magnitude & 0.95 \\
Primary tilt & 1.59 \\
Secondary tilt & 1.96 \\
Azimuthal inter-spin angle  & 0.23 \\
Azimuthal precession cone angle & 0.36 \\
Coalescence phase & 1.06 \\
Polarization angle & 1.28 \\
Coalescence GPS time & $1242442967.41$ s \\
Right ascension & 4.39 \\
Declination & 0.85 \\
Network SNR & 20 \\
Inclination angle & $28.6^\circ$, $74.5^\circ$ \\
\hline 
\end{tabular}
\end{table}

These simulations' spin angles and sky location correspond to the maximum-likelihood point of GW190521 from~\citet{islam2023analysis} as listed in Table~\ref{table:1}. 

In line with previous findings, we observe in Fig~\ref{fig:corner} that the actual parameter values for $\iota=28.6^\circ$, represented by the blue lines, fall outside the 68\% CI of Prior-2 but within the 68\% CI of Prior-4. This suggests that in such cases, the prior distribution, rather than the higher harmonic content, plays a more pivotal role. Conversely, for the higher inclination value ($\iota=74.5^\circ$), the actual parameter values lie within the 68\% CI for all prior choices, illustrating the opposite trend. This is anticipated, considering that precession exhibits minimal degeneracy with total mass, as the former introduces frequency modulations while the latter shifts the entire frequency spectrum. Although not the primary focus of this study, we note a similar bias in the $\chi_\mathrm{eff}$ measurement for the $\iota=28.6^\circ$ system. This is due to $\chi_\mathrm{eff}$'s dependence on $Q$.

Thus, while it is true that precession may quantitatively alter our results, we understand that the qualitative impact of the mass priors on the recovered masses would remain unaltered.

\end{document}